\documentclass[universe,review,accept,pdftex,moreauthors]{Definitions/mdpi}

\firstpage{1}
\pubvolume{10}
\issuenum{9}
\articlenumber{339}
\pubyear{2024}
\copyrightyear{2024}
\externaleditor{Academic Editor: Panayiotis Stavrinos}
\datereceived{21 July 2024}
\daterevised{16 August 2024} 
\dateaccepted{21 August 2024}
\datepublished{23 August 2024}
\hreflink{https://\linebreak doi.org/10.3390/universe10090339} 

\usepackage{graphicx}
\usepackage{xcolor}
\usepackage{subcaption}
\newcommand{\be}{\begin{equation}}
\newcommand{\ee}{\end{equation}}

\Title{Energy-Momentum Squared Gravity: A Brief Overview}

\TitleCitation{Energy-Momentum Squared Gravity: A Brief Overview}

\Author{Ricardo 
 A. C. Cipriano $^{1}$,
Nailya Ganiyeva $^{1}$\orcidF{},
Tiberiu Harko $^{2}$\orcidB{}, Francisco S. N. Lobo $^{1,3}$\orcidC{}, Miguel A. S. Pinto $^{1,3,}$*\orcidD{} and João Luís Rosa $^{4,5}$\orcidE{}}

\AuthorNames{Ricardo A. C. Cipriano, Nailya Ganiyeva, Tiberiu Harko, Francisco S. N. Lobo, Miguel A. S. Pinto, João Luís Rosa}

\AuthorCitation{Cipriano, R.A.C.; Ganiyeva, N.; Harko, T.; Lobo, F.S.N.; Pinto, M.A.S.; Rosa, J.L.}

\address{%
$^{1}$ \quad Instituto de Astrofísica e Ciências do Espaço, Faculdade de Ciências da 
 Universidade de Lisboa, \mbox{Campo Grande}, Edifício C8, 1749-016 Lisbon, Portugal; fc51825@alunos.fc.ul.pt
 (R.A.C.C.)
; fc57452@alunos.fc.ul.pt (N.G.); 
fslobo@fc.ul.pt (F.S.N.L.)\\
$^{2}$ \quad Department of Physics, Babes-Bolyai University, Kogalniceanu Street,
	400084 Cluj-Napoca, Romania; tiberiu.harko@aira.astro.ro\\
$^{3}$ \quad Departamento de F\'{i}sica, Faculdade de Ci\^{e}ncias da 
 Universidade de Lisboa, Campo Grande, Edifício C8, 1749-016 Lisbon, Portugal\\
$^{4}$ \quad Institute of Theoretical Physics and Astrophysics, University of Gda\'{n}sk, Jana Ba\.{z}y\'{n}skiego 8, \linebreak 80-309 Gda\'{n}sk, Poland; joao.rosa@ug.edu.pl\\
$^{5}$ \quad Institute of Physics, University of Tartu, W. Ostwaldi 1, 50411 Tartu, Estonia}

\corres{Correspondence: mapinto@fc.ul.pt}

\abstract{In this work, we present a review of Energy-Momentum Squared Gravity (EMSG)---more specifically, $f(R,T_{\mu\nu}T^{\mu\nu})$ gravity, where $R$ represents the Ricci scalar and $T_{\mu\nu}$ denotes the energy-momentum tensor. The inclusion of quadratic contributions from the energy-momentum components has intriguing cosmological implications, particularly during the Universe's early epochs. These effects dominate under high-energy conditions, enabling EMSG to potentially address unresolved issues in General Relativity (GR), such as the initial singularity and aspects of big-bang nucleosynthesis in certain models. The theory's explicit non-minimal coupling between matter and geometry leads to the non-conservation of the energy-momentum tensor, which prompts the investigation of cosmological scenarios through the framework of irreversible thermodynamics of open systems. By employing this formalism, we interpret the energy-balance equations within EMSG from a thermodynamic perspective, viewing them as descriptions of irreversible matter creation processes. Since EMSG converges to GR in a vacuum and differences emerge only in the presence of an energy-momentum distribution, these distinctions become significant in high-curvature regions. Therefore, deviations from GR are expected to be pronounced in the dense cores of compact objects. This review delves into these facets of EMSG, highlighting its potential to shed light on some of the fundamental questions in modern cosmology and gravitational theory.}


\begin{document}

\section{Introduction}

The 
 discovery of late-time cosmic acceleration \cite{SupernovaSearchTeam:1998fmf,SupernovaCosmologyProject:1998vns} has spurred extensive research into modified theories of gravity \cite{Capozziello:2002rd,Nojiri:2006ri,Lobo:2008sg,Sotiriou:2008rp,Nojiri:2010wj,Olmo:2011uz,Capozziello:2011et,Clifton:2011jh,Harko:2018ayt,CANTATA:2021ktz} as an alternative to dark energy \cite{Copeland:2006wr}. Among these, several theories incorporate non-minimal couplings between geometry and matter \cite{Bertolami:2007gv,Harko:2014sja,Harko:2014aja,Harko:2014gwa,Harko:2018gxr}, leading to intriguing modifications in the gravitational dynamics.
A distinctive feature of these theories is the non-conservation of matter's energy-momentum tensor. The explicit coupling between geometry and matter results in a non-zero covariant derivative of the energy-momentum tensor, leading to non-geodesic motion and the emergence of an extra force. This additional force modifies the standard motion of test particles, influencing both local and cosmological scales. Notable examples include $f(R,\mathcal{L}_m)$ \cite{Harko:2010mv}, $f(R,T)$ \cite{Harko:2011kv,Barrientos:2018cnx}, $f(R,T_{\mu\nu}T^{\mu\nu})$ \cite{Katirci:2013okf, Roshan:2016mbt}, and $f(R,T,R_{\mu\nu}T^{\mu \nu})$ \cite{Haghani:2013oma,Odintsov:2013iba} theories of gravity. Here, $R$ and $R_{\mu\nu}$ represent the Ricci scalar and tensor, respectively; $\mathcal{L}_m$ denotes the matter Lagrangian density; and \mbox{$T=T^{\mu}{}_{\mu}=g^{\mu \nu}T_{\mu \nu}$ }is the trace of the energy-momentum tensor ($T_{\mu\nu}$).
In a cosmological context, the geometry--matter couplings introduce compelling phenomenological effects, enabling a unified description of the different cosmological epochs. For instance, these modified gravity models prove invaluable for describing interactions between dark energy and dark matter, providing insights into the mechanisms driving the late-time cosmic speed-up by modifying the gravitational field equations. Moreover, they offer potential explanations for the observed behavior of the galactic flat rotation curves. In this framework, the extra terms in the gravitational field equations alter the equations of motion for test particles, introducing a supplementary gravitational interaction that can account for the rotation curves without invoking dark matter.

\textls[-15]{A novel gravitational framework known as energy-momentum squared gravity} (EMSG) or $f(R, T_{\mu \nu}T^{\mu \nu})$ gravity \cite{Katirci:2013okf} has garnered significant attention in the scientific community since its introduction. The pioneering paper laid the groundwork for this theory, which was later refined \cite{Roshan:2016mbt}. The action was originally expressed as
\begin{equation}
S=\frac{1}{2 \kappa^2} \int \sqrt{-g} f(R, T_{\mu \nu}T^{\mu \nu}) d^4 x + S_m\,,
\label{def:actionEMSG}
\end{equation}
where $\kappa^2 = 8\pi G/c^4$ is a constant; $S_m$ is the action of the matter Lagrangian, i.e.,\linebreak \ $S_m = \int \sqrt{-g} L_m d^4 x$;  $R$ is the Ricci scalar; and, finally, $T_{\mu \nu}T^{\mu \nu}$ is the self-contraction 
of the stress energy tensor that characterizes the aforementioned theory
.
In \cite{Roshan:2016mbt}, the action functional of the theory was simplified to the specific case of $f(R, T_{\mu \nu}T^{\mu \nu}) = R -2\Lambda -\eta T_{\mu \nu}T^{\mu \nu}$, where $\eta$ is a coupling constant and $\Lambda$ is the cosmological constant, which was subsequently dubbed EMSG. This abbreviation has since become the standard term for the theory. However, some authors, such as the authors of \cite{Bahamonde:2019urw}, refer to the original form (as seen in Equation \eqref{def:actionEMSG}) as generalized energy-momentum squared gravity (gEMSG), to differentiate it from other formulations, such as the power law described in the functional form of $f =f\left(R,\left(T_{\mu \nu}T^{\mu \nu}\right)^n\right)$ (see Ref. \cite{Akarsu:2018drb}) or the inclusion of the matter Lagrangian in the functional (see Ref. \cite{Akarsu:2020vii}).
In fact, establishing a consistent notation has been crucial for maintaining coherence and aligning with established conventions in the literature. Thus, in this work, in order to avoid confusion and to simplify definitions, we generically consider the theories described by the action in Equation (\ref{def:actionEMSG}) as energy-momentum squared gravity (EMSG).

Contrary to $f(R,T)$ gravity, EMSG induces quadratic contributions of the energy-momentum components to the right-hand side of the field equations. This self-contraction of the energy-momentum tensor has particularly interesting cosmological consequences during early epochs of the Universe due to its effects being predominant under high-energy regimes, which then allows EMSG to address some of the open questions left in general relativity (GR), such as the initial singularity, dense compact astrophysical objects, or even the big bang nucleosynthesis under some particular models. In fact, an extensive body of literature surrounding the general formulation of this theory includes numerous papers on a plethora of topics, for instance, in a cosmological setting \cite{Board:2017ign,Bahamonde:2019urw,Barbar:2019rfn,Akarsu:2017ohj,Cipriano:2023yhv,Sharif:2023uyv,Canuto:2023gdv,Shahidi:2021lqf} and compact \mbox{objects \cite{Akarsu:2018zxl,Sharif:2021rck, Yousaf:2021ltg, Sharif:2022ibt, Sharif:2022cud, Sharif:2022prm, Sharif:2022mdp, ZeeshanGul:2023ysx, Sharif:2023uac, Sharif:2023gbl, Sharif:2023nrl, Sharif:2023ccr, Chen:2019dip,Kazemi:2020hep,Sharif:2023rcd,HosseiniMansoori:2023mqh,Sharif:2022aei,Nasir:2023pzq,Pretel:2023avv,Naz:2022vvn,Sharif:2022cjv},} among other issues. In fact, energy-momentum squared symmetric teleparallel gravity has also been explored to explain the cosmological dynamics of both the early and the late Universe without resorting to the invocation of dark energy \cite{Rudra:2021ksp}.
Tight constraints have also been obtained on specific parameters of EMSG by binary pulsar observations \cite{Nazari:2022xhv}.

\textls[15]{Indeed, these higher-order quadratic terms lead the field equations to resemble other deeply interesting theories, such as loop quantum gravity \cite{Ashtekar:2011ni} or brane-world cosmologies \cite{Maartens:2003tw,Maartens:2010ar,Brax:2003fv}. In particular, as we will see, EMSG replaces terms such the energy density ($\rho$) with $\rho(1 \pm \mathcal{O}(\rho^2))$, where the negative contribution is associated with loop quantum gravity \cite{Ashtekar:2011ni} and the positive contribution is associated with the brane-world cosmologies \cite{Brax:2003fv} while having an entirely different foundation. It is worth noting that EMSG accomplishes all this without introducing novel forms of fluid stresses (such as extra scalar fields or bulk viscosity) \cite{Akarsu:2018zxl}.}
In regions of high curvature, such as the dense cores of neutron stars or black holes, the energy-momentum tensor becomes significant, causing EMSG corrections to deviate from GR. These deviations manifest in the following several key ways: (i) EMSG changes the internal structure equations of compact objects, affecting density profiles, pressure distributions, and mass--radius relationships; (ii) gravitational waves from compact object mergers may differ under EMSG and be potentially detectable by observatories; (iii) EMSG could influence the formation, evolution, and stability of black holes, neutron stars, and wormholes, as well as the end states of stellar collapse.

The resurgence in popularity of this theory can be attributed to several key developments. Its momentum grew rapidly with the publication of \cite{Roshan:2016mbt}, which demonstrated that under a particular yet simple Lagrangian ($f = R +\gamma T_{\mu \nu}T^{\mu \nu}$), the early epochs of the cosmos had a minimum length and a finite maximum energy density. This finding resolves the primordial singularity without resorting to quantum gravity by generating a bouncing cosmology while maintaining a consistent sequence of cosmic epochs and adequately explaining cosmic behavior. Consequently, EMSG can be considered an emergent universe scenario while still being a subset of the $k$-essence models \cite{Armendariz-Picon:2000ulo}.
It is worth noting that not all EMSG solutions yield these results. However, from a theoretical perspective, it is still a promising indication that EMSG might be an interesting theory to explore further. Although it has been argued that a cosmological evolution that can regularly connect the early-universe bounce to a viable de Sitter late-time universe should not generally exist \cite{Barbar:2019rfn}, this issue is addressed by the existence of a vacuum energy density in EMSG \cite{Nazari:2020gnu}.

Finally, some theoretical remarks regarding EMSG are warranted. Similar to $f(R,T)$ gravity, EMSG does not satisfy the conservation of the energy-momentum tensor. This implies that the paths of test particles may exhibit non-geodesic motion. However, in the special case where $T_{\mu \nu}T^{\mu \nu} = 0$, the standard geodesic path is recovered \cite{Sharif:2021ptz}. Moreover, EMSG may not be equivalent to $f(R)$ gravity when the trace of the energy-momentum tensor ($T = g^{\mu \nu}T_{\mu \nu}$) is zero. For instance, in the case of radiation, although $T=0$, EMSG still contains non-vanishing terms \cite{Akarsu:2017ohj, Cipriano:2023yhv}. These differences are explored below.

This review paper is organized in the following manner: In Section \ref{sec:formalism}, we briefly present the formalism of EMSG
 in the geometric and scalar-tensor representations and consider an extended theory of gravity with an $n$-th-order invariant constructed of contractions of the energy-momentum tensor. In Section \ref{sec:thermodynamics}, we explore the thermodynamics of open systems with the possibility of irreversible matter creation processes due to the non-conservation of the energy-momentum tensor in EMSG. In Section \ref{sec:cosmological_implications}, we briefly review the fundamentals of cosmology in EMSG and explore some general cosmological models. In Section \ref{sec:compactobj}, we explore compact objects in EMSG---in particular, the possibility of black hole solutions---and expand on wormhole geometries. Finally, in Section \ref{sec:conclusion}, we summarize and conclude this research.

Henceforth, to simplify the notation, we denote the self-contraction of the energy-momentum tensor as the following scalar: $\mathcal{T}=T_{\mu \nu} T^{\mu \nu}$. Although this notation is used for brevity, there are instances where displaying the full contraction provides a deeper understanding of the calculations. Whenever $\mathcal{T}$ is mentioned, it refers to this specific contraction. Additionally, it should be noted that most of the analysis presented here is based on the authors' original research. However, readers have access to a substantial body of literature for further exploration and context.

\section{\boldmath{$f(R,\mathcal{T})$} Gravity: Formalism}\label{sec:formalism}

\subsection{Action and Field Equations}

In this section, we provide a brief overview of the geometric formalism of the theory. The action ($S$) that describes $f(R, \mathcal{T})$ gravity is given by
\begin{equation}
S=\frac{1}{2 \kappa^2} \int_{\Omega} \sqrt{-g} f(R, \mathcal{T}) d^4 x+\int_{\Omega} \sqrt{-g} \mathcal{L}_m \,d^4 x ,
\label{eq1_1}
\end{equation}

Varying 
 the action \eqref{eq1_1} with respect to the metric provides the following modified gravitational field equations:
\begin{equation}
f_R R_{\mu \nu}-\frac{1}{2} g_{\mu \nu} f-\left(\nabla_\mu \nabla_\nu - g_{\mu \nu} \square\right) f_R=2\kappa^2 T_{\mu \nu}-f_{\mathcal{T}} \Theta_{\mu \nu} ,
\label{geo_field_simple}
\end{equation}
where $\square=\nabla^{\mu}\nabla_{\mu}$ is the usual D'Alembert operator, $f_R \equiv \partial f / \partial R$ and $f_{\mathcal{T}} \equiv \partial f / \partial \mathcal{T}$, and the auxiliary tensor ($\Theta_{\mu \nu}$) is defined as
\begin{equation}
\Theta_{\mu \nu} \equiv  \frac{\delta \mathcal{T} }{\delta g^{\mu \nu}}.
\label{new_eq4}
\end{equation}

Taking into account the following explicit variation:
\begin{adjustwidth}{-\extralength}{0cm}
\begin{equation} 
\begin{aligned}
\delta \left( T_{\alpha\beta}T^{\alpha\beta}\right) &= \delta \left( g^{\alpha\mu}g^{\beta\nu}T_{\alpha\beta}T_{\mu\nu}\right) = 2 \left(T^{\alpha\beta}\frac{\delta T_{\alpha\beta}}{\delta g_{\mu\nu}}+T^\sigma_\mu T_{\nu\sigma}\right)\delta g^{\mu\nu} \\
&= 2 \left[ T^{\mu\beta} \left( g_{\alpha\beta} \frac{\partial \mathcal{L}_{m}} {\partial g^{\mu\nu}}+\frac{\delta g_{\alpha\beta}} {\delta g^{\mu\nu}} \mathcal{L}_{m}-2 \frac{\partial^{2} \mathcal{L}_{m}} {\partial g^{\mu\nu} \partial
g^{\alpha\beta}} \right)+T_{\mu}^{\nu} T_{\nu\sigma} \right] \delta g^{\mu\nu} \\
&= 2 \left[ T^{\alpha\beta} \left\{-\mathcal{L}_{m} \left( g_{\mu\alpha} g_{\beta\nu}-\frac{1} {2} g_{\mu\nu} g_{\alpha\beta} \right)-\frac{1}{2} g_{\alpha\beta} T_{\mu\nu}-2 \frac{\partial^{2} \mathcal{L}_m} {\partial g^{\mu\nu}\partial g^{\alpha\beta}} \right\} + T_{\mu}^{\sigma} T_{\nu\sigma} \right] \delta g^{\mu\nu} ,
\label{p14}
\end{aligned}
\end{equation}
\end{adjustwidth}
the auxiliary tensor $\Theta_{\mu \nu}$ can be written as
\begin{equation}
\Theta_{\mu \nu}=-2 \mathcal{L}_{m}\left(T_{\mu \nu}-\frac{1}{2} g_{\mu \nu} T\right)-T T_{\mu \nu}+2 T_\mu{ }^\alpha T_{\alpha \nu}-4 T^{\alpha \beta} \frac{\partial^2 \mathcal{L}_{m}}{\partial g^{\mu \nu} \partial g^{\alpha \beta}}.
\label{new_0001}
\end{equation}

The energy-momentum tensor for EMSG is generically not conserved and is transparent according to the following relation
:
\begin{equation}
\nabla^{\mu}T_{\mu\nu}=-f_{\mathcal{T}}g_{\mu\nu}\nabla^{\mu}(T_{\alpha\beta}T^{\alpha\beta})+2\nabla^{\mu}(f_{\mathcal{T}}\theta_{\mu\nu}),
\label{geo_conservation_simple}
\end{equation}
which has implications that are further explored below.

\subsection{Scalar-Tensor Representation}

In addition to the familiar geometrical representation, EMSG can also be described using the scalar-tensor representation. This approach, which has gained popularity recently, separates the scalar and tensor parts of the variational action. It results in second-order equations in the metric, simplifying calculations and potentially addressing issues like Ostrogradsky instabilities \cite{Woodard:2006nt,Ostrogradski}. For a detailed derivation of EMSG in this representation, we refer the reader to \cite{Cipriano:2023yhv}.

Let us start by defining the action in the scalar-tensor representation as
\begin{equation}
S=\frac{1}{2 \kappa^2} \int_{\Omega} \sqrt{-g}\left[\phi R+\psi \mathcal{T}-V(\phi, \psi)\right] d^4 x+\int_{\Omega} \sqrt{-g} \mathcal{L}_m d^4 x ,
\label{eq13}
\end{equation}
where $\phi$ and $\psi$ are the characteristic scalar fields of this representation, which are non-minimally coupled to the Ricci scalar ($R$) and the contraction of the energy-momentum tensor ($\mathcal{T}$), respectively. Moreover, $V(\phi,\psi)$ represents the potential of the two scalar fields.
Varying the action \eqref{eq13} with respect to the metric leads to the following field equations:
\begin{equation}
\phi \left(R_{\mu\nu}-\frac{1}{2} R g_{\mu\nu}\right) +\frac{1}{2} g_{\mu\nu} V - \left(\nabla_\mu \nabla_\nu - g_{\mu\nu} \square\right) \phi =8 \pi T_{\mu\nu}-\psi\left(\Theta_{\mu\nu}-\frac{1}{2} g_{\mu\nu} \mathcal{T}\right)
\label{eq14} ,
\end{equation}
which are equivalent to Equation \eqref{geo_field_simple}. Notably, we have, yet again, introduced the auxiliary tensor (still defined by Equation \eqref{new_0001}), with $f_R$ and $f_\mathcal{T}$ being replaced by the scalar fields of $\phi$ and $\psi$, respectively. Furthermore, the extra terms in Equation \eqref{eq14} result from the proper definition of $V$.

For particular cosmological models, in order to obtain an expression of the potential, we vary the action with respect to both scalar fields ($\phi$ and $\psi$), which yields
\begin{equation}
V_\phi=R ,
\label{eq15}
\end{equation}
\begin{equation}
V_\psi=\mathcal{T},
\label{eq16}
\end{equation}
respectively,
where $V_\phi \equiv \partial V / \partial \phi$ and $V_\psi \equiv \partial V/ \partial \psi$. Indeed, to find the analytical expression for $V(\phi,\psi)$, one can integrate Equations 
 \eqref{eq15} and \eqref{eq16}. Lastly, by taking the covariant derivative of \eqref{eq14}, we obtain the following corresponding conservation equation:
\begin{equation}
8 \pi \nabla_\nu T^{\mu\nu}=\nabla_\nu\left(\psi \, \Theta^{\mu\nu}\right)-\frac{1}{2} g^{\mu\nu}\left[R \nabla_\nu \phi+\nabla_\nu (\psi \mathcal{T}-V)\right].
\label{eq17}
\end{equation}

\subsection{Geometrical and Scalar-Tensor $f(R, \mathcal{T}^{(n)})$ Gravity}\label{sec:gEMSPG}

It is possible to generalize the concept of EMSG presented above to an arbitrary finite power of the scalar ($\mathcal{T}$), i.e., $\mathcal{T}^{(n)}$. This work was developed in \cite{Abedi:2022hzq} by considering an extension of the action \eqref{eq1_1} that includes a function of the scalar curvature and the higher-order invariant, defined as
\begin{equation}
\begin{aligned}
	{{\mathcal{T}}^{\mu}_{\phantom{\mu}\nu}}^{(n)}\equiv& T^{\mu}_{\phantom{\mu}\alpha_1} \, T^{\alpha_1}_{\phantom{\alpha_1}\alpha_2} \, \ldots \, T^{\alpha_{n-1}}_{\phantom{\alpha_{n-1}}\nu},
	\\
	{\mathcal{T}}^{(n)}\equiv&{{\mathcal{T}}^{\mu}_{\phantom{\mu}\mu}}^{(n)}.
\end{aligned}
\end{equation}

The potential physical significance of this expansion is still under investigation as research continues to progress. However, for the present, it serves as a valuable mathematical generalization. With this in mind, to derive the field equations for the model, we consider a general function ($f(R, \mathcal{T}^{(n)})$) that depends on the exponent $n$. The starting point for the geometrical representation is now the following higher-order gravitational action:
\begin{equation}
S= \frac{1}{2 \kappa^2}  \int \mathrm{d}^4 x \sqrt{-g}\,f\left(R, \mathcal{T}^{(n)}\right) +\int_{\Omega} \sqrt{-g} \mathcal{L}_m d^4 x.
\label{p4}
\end{equation}

We assume that the matter Lagrangian ($\mathcal{L}_{m}$) only depends on the metric tensor and not on its derivatives
. By varying Equation (\ref{p4}) with respect to the metric, we obtain the following field equations:
\begin{equation}
f_R R_{\mu \nu}-\frac{1}{2} g_{\mu \nu} f+\left(g_{\mu \nu} \square-\nabla_\mu \nabla_\nu\right) f_R=\kappa^2 T_{\mu \nu}-f_{\mathcal{T}^{(\mathbf{n})}} \Theta_{\mu \nu}^{(n)} ,
\label{geo_field_n}
\end{equation}
where $f_{\mathcal{T}^{(n)}} \equiv \partial f / \partial \mathcal{T}^{(n)} $ and $ f_R \equiv \partial f / \partial R$.
The auxiliary tensor of the $n$-th order ($\Theta_{\mu \nu}^{(n)} \equiv \delta \mathcal{T}^{(n)} / \delta g^{\mu \nu}$), which can be explicitly obtained, similarly to \eqref{new_0001}, takes the following form:
\begin{equation}
\Theta_{\mu \nu}^{(n)} =n \mathcal{T}_{\mu \nu}^{(n)}-n \mathcal{L}_{m} \mathcal{T}_{\mu \nu}^{(n-1)}+n\left(\frac{1}{2} g_{\mu \nu} \mathcal{L}_m-T_{\mu \nu}\right) \mathcal{T}^{(n-1)}-2 n g^{\alpha \gamma} \frac{\partial^2 \mathcal{L}_{m}}{\partial g^{\mu \nu} \partial g^{\beta \gamma}} \mathcal{T}_\alpha^\beta{ }^{(n-1)} ,
\label{p15}
\end{equation}
which allows us to describe the geometrical field equations for $f(R, \mathcal{T}^{(n)})$ gravity.

On the other hand, in the scalar-tensor representation, in a similar manner as above, one can start from a generalization of Equation (\ref{eq13}) as follows:
\begin{equation}
S=\frac{1}{2 \kappa^2} \int_{\Omega} \sqrt{-g}[\phi R+\psi \mathcal{T}^{(n)}-V(\phi, \psi)] d^4 x+\int_{\Omega} \sqrt{-g} \mathcal{L}_m d^4 x ,
\end{equation}
which yields a similar set of field equations given by
\begin{equation}
\phi \left(R_{\mu\nu}-\frac{1}{2} R g_{\mu\nu}\right) +\frac{1}{2} g_{\mu\nu} V - \left(\nabla_\mu \nabla_\nu - g_{\mu\nu} \square\right) \phi =8 \pi T_{\mu\nu}-\psi\left(\Theta_{\mu\nu}^{(n)}-\frac{1}{2} g_{\mu\nu} \mathcal{T}^{(n)}\right) ,
\end{equation}
which results in the now more general conservation equation given by
\begin{adjustwidth}{-\extralength}{0cm}
\begin{equation}
\left(\kappa^2-\psi\right) \nabla^\mu T_{\mu \nu}=\left(T_{\mu \nu}+\Theta_{\mu \nu}^{(n)}\right) \nabla^\mu \psi+\psi \nabla^\mu \Theta_{\mu \nu}^{(n)}-\frac{1}{2} g_{\mu \nu}\left[R \nabla^\mu \varphi+\nabla^\mu\left(\psi \mathcal{T}^{(n)}-V\right)\right] .
\label{general}
\end{equation}
\end{adjustwidth}

Taking into account the definition of Equation (\ref{p15}), we have our field equations of a scalar-tensor gravity ($f(R, \mathcal{T}^{(n)})$). In particular, for the case of $n=2$, the general $2n$-th auxiliary tensor takes the following form:
\begin{equation}
\Theta_{\mu \nu}^{(2)}=-2 \mathcal{L}_{m}\left(T_{\mu \nu}-\frac{1}{2} g_{\mu \nu} T\right)-T T_{\mu \nu}+2 T_\mu{ }^\alpha T_{\alpha \nu}-4 T^{\alpha \beta} \frac{\partial^2 \mathcal{L}_{m}}{\partial g^{\mu \nu} \partial g^{\alpha \beta}} ,
\end{equation}
recovering the expression for the conservation equation previously derived in Equation \eqref{eq14}.
\begingroup\makeatletter\def\f@size{9.6}\check@mathfonts
\begin{equation}
\left(\kappa^2-\psi\right) \nabla^\mu T_{\mu \nu}=\left(T_{\mu \nu}+\Theta_{\mu \nu}^{(2)}\right) \nabla^\mu \psi+\psi \nabla^\mu \Theta_{\mu \nu}^{(2)}-\frac{1}{2} g_{\mu \nu}\left[R \nabla^\mu \varphi+\nabla^\mu\left(\psi T^{(2)}-V\right)\right] .
\end{equation}
\endgroup

As mentioned in \cite{Abedi:2022hzq}, one may determine the extra force that appears from this non-conservation of the energy-momentum tensor in a manner analogous to that reported in~\cite{Bertolami:2007gv}, and it was shown to be related to MOND theory and other anomalies in test particle motions. Indeed, this fact may generically constitute a reliable test bed for these theories at local scales and at the Galactic level.

\section{Thermodynamics of Open Systems}\label{sec:thermodynamics}

Large-scale matter creation is of the utmost importance in cosmology, as it can provide a description of the origin of the large-scale structure, as well as for the evolution of the various components that constitute the cosmological fluid. Motivated by the relevance of this topic, in this section, we summarize some of the existent literature regarding matter creation in the cosmological context, with an emphasis on the irreversible thermodynamics of open systems. Afterward, we apply the aforementioned framework to a spatially flat, homogeneous, and isotropic universe and briefly discuss its consequences.

\subsection{Particle Production in Cosmology}

The investigation of particle production processes in expanding universes started with the pioneering works of Erwin Schrödinger in 1939 and 1940 \cite{Schrodinger:1939, Schrodinger:1940}. While studying a quantum wave packet propagating in an expanding Friedmann--Lemaitre--Robertson--Walker (FLRW) universe, Schrödinger reached a peculiar conclusion, namely that a scalar particle could have a non-zero probability of stimulating the creation of a pair of scalar particles spontaneously. As pointed out by Leonard Parker in Ref. \cite{Parker:2012at},
instead of having scalar particles, if one considers photons,
then a single one could induce the creation of a pair of them. Indeed, Schrödinger referred to this occurrence as an “alarming” phenomenon. Unfortunately, since quantum field theory in curved spacetime was in its early stages at the time, some technical details regarding Schrödinger's approach were not sufficiently robust. Thus, the idea of exploring particle creation in expanding universes was abandoned for a while. In addition, it is important to mention that in curved, expanding spacetime, the question ``what is matter?'' is more difficult to answer because of the lack of a global time symmetry, a problem intrinsically related to the definition of energy in gravitation.

It was only in the late 1960s that the same Leonard Parker brought this research program back (although unaware of Schrödinger's works at the beginning), formulating the first successful mechanism for the production of particles by gravity \cite{Parker:1968mv,Parker:1969au,Parker:1971pt,Parker:1972kp}. Using the more developed mathematical apparatus of quantum field theory in curved spacetime, Parker demonstrated that there is a deep relationship between an expanding FLRW geometry and particle production. By considering a quantized massive scalar field in an expanding FLRW geometry, whose initial vacuum state (with no particles) is the Minkowski vacuum, Parker verified that the particle number associated with the final vacuum state of such an evolution was not zero. The physical process responsible for this peculiar result was attributed to the expansion of the Universe itself, meaning that a massive scalar field evolving in an FLRW geometry can give rise to particles spontaneously. Furthermore, it is also worth mentioning that this matter creation can be seen as the process responsible for originating the power spectrum of the inflation field perturbations during inflation \cite{Parker:2012at}, demonstrating possible imprints on physical observations.

Later, in the 1980s, Nobel Prize winner Ilya Prigogine and collaborators \cite{Prigogine:1988jax,Prigogine:1986,Prigogine:1989zz} searched for an alternative framework to accommodate cosmological particle production with the entropy evolution of the Universe. Since the semiclassical Einstein field equations used by Parker (and subsequently by others \cite{Grib:1976pw,Brout:1979qe,Ford:1986sy}) are both adiabatic and reversible, Prigogine et al. argued that these could not provide a natural explanation for the increase in cosmological entropy accompanied by the creation of particles due to the non-reversible nature of this process. As such, they proposed an alternative cosmology based on the irreversible thermodynamics of open systems, in which the explanation for macroscopic matter and entropy production relies on a reinterpretation of the matter energy-momentum tensor that includes an irreversible matter creation term.

Although this framework was successful at the time, because it was employed in GR, a physical interpretation of this irreversible process was missing. The main problem was, again, the adiabatic and reversibility character of the Einstein field equations or, equivalently, the covariant conservation of the energy-momentum tensor. However, in the last two decades, the appearance of classical modified gravity theories that contain previously discussed non-minimal geometry--matter couplings gave a new meaning to Prigogine's approach to cosmology. Since in all these theories, the matter energy-momentum tensor is not conserved, this feature, in the context of the thermodynamics of open systems, allows one to physically interpret such a non-conservation as an irreversible energy flow from the gravitational field to the matter sector that could result in particle creation \cite{Harko:2014pqa}.
The effects and implications of the irreversible matter creation processes on late cosmological evolution have been studied in some of these modified gravity theories by the authors of the present paper \cite{Harko:2014pqa,Pinto:2022tlu,Harko:2015pma,Cipriano:2023yhv} (see \cite{Pinto:2023phl} for a review). In addition, particle creation may also result from the following different processes:
\begin{itemize}

\item Vacuum instability in the presence of both gravitational and gauge fields, possibly resulting from the conformal trace anomaly, as shown in \cite{Chernodub:2023pwf};

\item The existence of quadratic curvature terms in the action of Weyl gravitational theory and the direct interaction of the perfect fluid particles. Hence, in such models, particles may also be created directly from the vacuum \cite{Berezin:2022phu};

\item  Cosmological models such as the one presented in \cite{Xue:2020tpf}, in which there is an interaction between dark energy and massive particle pairs that can produce both stable and unstable particle pairs.
\end{itemize}

In summary, many different physical mechanisms exist that may generate particles. Nonetheless, we emphasize that since in EMSG, the energy-momentum tensor is not conserved, and the natural framework to study particle production is through the lens of the thermodynamics of open systems \cite{Harko:2014pqa}.

\subsection{Thermodynamic Interpretation of Irreversible Matter Creation}

In this subsection, we introduce the basics of this formalism by considering a spatially flat FLRW universe. Generally speaking, the application of the irreversible thermodynamics of open systems in cosmology relies on the basic assumption that the Universe can be seen as an effective thermodynamic open system \cite{Pinto:2023phl}. From this premise, it follows that the bulk of the system corresponds to the observable Universe and the boundary to the apparent horizon 
(which, in a flat FLRW geometry, coincides with the Hubble radius), and the surroundings correspond to the unobservable universe \cite{Pinto:2023phl}.

This framework consists of two main equations, namely the first and second laws of thermodynamics.

\subsubsection{First Law of Thermodynamics: Temperature Evolution}

In particular, for a homogeneous universe, the heat ($dQ$) is negligible, so the first law can be expressed as
\begin{equation}
\label{1st law}
d(\rho V)+p d V=\frac{h}{n} d(n V),
\end{equation}
where $\rho$ is the energy density of the cosmological fluid, $V$ the volume, $p$ is the usual thermodynamic pressure of the fluid, $h=\rho+p$ is the enthalpy per unit volume, and $n=N / V$ is the particle number density. In addition, it is assumed that $p$ and $\rho$ describe non-exotic matter contents. Therefore, these must satisfy the following various \mbox{energy conditions:}
\begin{equation}\label{def:ECs}
\rho \geq 0, \qquad p\geq 0,  \qquad \rho+p \geq 0,  \qquad \rho+3p\geq 0,  \qquad \rho \geq \left|p\right|.
\end{equation}

To further develop Equation \eqref{1st law}, one can take the time derivative and describe the volume of the Universe in terms of the scale factor ($a(t)$) as $V(t) = a^3(t)$, which yields
\begin{equation}
\dot{\rho}+3 H(\rho+p)=\frac{\rho+p}{n}(\dot{n}+3 H n),
\label{uau2}
\end{equation}
with $H=\dot{a}/a$ being the Hubble function. In order to further develop the above equation, it is convenient to introduce two quantities, namely the number current, defined as
\begin{equation}
\label{number_current}
N^\mu\equiv n u^\mu,
\end{equation}
and the particle production rate ($\Gamma$). This quantity expresses the number of particles that are being created in comparison with the total number of particles at a specific instant, i.e.,
\begin{equation}
\Gamma =\frac{1}{N}\frac{dN}{dt}=\frac{1}{nV}\frac{d}{dt}(nV).
\label{gamma}
\end{equation}

From Equations \eqref{number_current} and \eqref{gamma}, it follows that the covariant derivative of the number current in the flat FLRW geometry has the following form:
\begin{equation}
\label{cov_nd}
\nabla_\mu N^\mu =  \dot{n}+3Hn \equiv  n \Gamma,
\end{equation}

By inserting Equation \eqref{cov_nd} into Equation \eqref{uau2}, we obtain the particle production rate in terms of the Hubble function, pressure, and energy density.
\begin{equation}\label{31}
\Gamma =3H+\frac{\dot{\rho}}{\rho +p}.
\end{equation}

Additionally, in the context of the thermodynamics of open systems, the total pressure makes two contributions
, namely the usual thermodynamic pressure ($p$) and the creation pressure ($p_c$), i.e.,\ $\tilde{p} = p + p_c$. The creation pressure can be viewed as the pressure responsible for describing, in an effective way, the irreversible matter creation processes that occur within the bulk of the open system, which, in this case, is the observable Universe. Hence, \mbox{Equation \eqref{1st law}} can be expressed as
\begin{equation}
d\left(\rho V\right)+\left(p+p_c\right)dV=0.
\end{equation}

With all the previous considerations in mind, one can find a relation between the creation pressure and the particle production rate.
\begin{equation}
p_c=-\left(\frac{\rho+p}{3 H} \right) \, \Gamma.
\label{pc}
\end{equation}

With this framework established, it is feasible to obtain a cosmological temperature evolution. To do so, we start by assuming that the energy density ($\rho$) and the pressure ($p$) are both functions of the particle number density ($n$) and of the temperature ($\mathbb{T}$), that is,
\begin{equation}
\rho=\rho(n, \mathbb{T}), \qquad p=p(n, \mathbb{T}).
\end{equation}

This allows us to express the first law as
\begin{equation}
\left(\frac{\partial \rho}{\partial n}\right)_{\mathbb{T}} \dot{n}+\left(\frac{\partial \rho}{\partial \mathbb{T}}\right)_n \dot{\mathbb{T}}+3(\rho+p) H=(\rho+p) \Gamma.
\label{cons_new}
\end{equation}

From the equation above and using some useful thermodynamic relations, one can arrive at the following equation for the temperature evolution:
\begin{equation}
\frac{\dot{\mathbb{T}}}{\mathbb{T}}=c_s^2(\Gamma-3 H),
\label{mehhh}
\end{equation}
where we naturally define the speed of sound as $c_s=\sqrt{(\partial p / \partial \rho)_n}$ (for a more in depth derivation, we refer the reader to \cite{Pinto:2023phl}). Then , the general solution of Equation \eqref{mehhh} can be expressed as
\begin{equation}
\mathbb{T}(t)=\mathbb{T}_0 \exp \left\{c_s^2 \int_0^{t^{\prime}}\left[ \Gamma\left(t^{\prime}\right)-3 H\right] d t^{\prime}\right\},
\end{equation}
where $\mathbb{T}_0=\mathbb{T}(0)$ is the constant initial temperature.

\subsubsection{Second Law of Thermodynamics: Entropy Evolution}

Moreover, the second law of thermodynamics for an open system is usually stated as
\begin{equation}
d \mathcal{S}=d_e \mathcal{S}+d_i \mathcal{S} \geq 0,
\label{dif_entro}
\end{equation}
where $d_e \mathcal{S}$ is called the entropy flow and $d_i \mathcal{S}$ represents entropy creation. It was demonstrated in \cite{Pinto:2022tlu} that these two entropy terms can be respectively written as 
\begin{subequations}
\begin{align}
d_e \mathcal{S} &= \frac{d Q}{\mathbb{T}}, \label{eq9a}\\
d_i \mathcal{S} &= \frac{s}{n} d(n V) \label{eq9b}.
\end{align}
\end{subequations}

As stated above, we are considering a homogeneous universe ($dQ=0$); therefore, the entropy flow vanishes ($d_e \mathcal{S}=0$). This implies that the variation of entropy is only dependent on the entropy of creation. Taking into account these considerations, the second law of thermodynamics is reduced to
\begin{equation}
d \mathcal{S}=\frac{s}{n} d(n V) \geq 0.
\label{dS_final}
\end{equation}

\textls[-25]{By recalling the definition of the creation rate in Equation \eqref{31}, we can rewrite \mbox{Equation \eqref{dS_final}} as}
\begin{equation}
\dot{\mathcal{S}}=\mathcal{S} \Gamma \geq 0,
\label{nhe1}
\end{equation}
which has the following solution:
\begin{equation}\label{33}
\mathcal{S}(t)=\mathcal{S}_0 \exp \left[\int_0^t \Gamma\left(t^{\prime}\right) d t^{\prime}\right],
\end{equation}
where $\mathcal{S}_0=\mathcal{S}(0)$ is the constant initial entropy. It is important to note that the derivations presented above are not specific to EMSG but general solutions for modified theories of gravity working under the following assumptions:
\begin{enumerate}
    \item Universe locally considered as an open system;
    \item Geometry being flat FLRW.
\end{enumerate}

Furthermore it is noteworthy that for theories where the energy-momentum tensor is conserved, the particle production rate also vanishes.

\section{Cosmology of \boldmath{$f(R,\mathcal{T})$} Gravity}
\label{sec:cosmological_implications}

In this section, we briefly review the fundamentals of the cosmology of $f\left(R,\mathcal{T}\right)$ gravity. We introduce the generalized Friedmann equations and discuss some of their implications. We examine the behavior of an important observational quantity, the deceleration parameter, in detail. Additionally, we cover the de Sitter evolutionary phases and explore some general cosmological models.

\subsection{The Generalized Friedmann Equations}

The first step in investigating the cosmological implications of $f\left( R,\mathcal{T}\right) $ gravity is to consider the dynamical evolution within the flat, homogeneous, and isotropic Friedmann--Lemaître--Robertson--Walker metric given by
\vspace{-6pt}
\begin{equation}
ds^{2}=-dt^{2}+a^{2}(t)\left[ dr^{2}+r^{2}\left( d\theta ^{2}+\sin
^{2}\theta d\varphi ^{2}\right) \right] ,  \label{eq18}
\end{equation}%
written in spherical coordinates as $(t,r,\theta,\varphi)$, where $a(t)$ is the scale factor, the homogeneity of cosmological spacetime requires that the scalar fields $\phi \equiv \phi(t)$ and $\psi \equiv \psi(t)$, and all thermodynamic quantities describing the matter content of the Universe depend only on the time coordinate ($t$). Additionally, we introduce the Hubble parameter ($H \equiv \dot{a}/a$, where $(\dot{} \equiv d/dt)$ denotes the derivative with respect to time).

We assume that the matter content of the Universe consists of an isotropic perfect fluid, with the energy-momentum tensor given by
\begin{equation}
T_{\mu \nu }=(\rho +p)u_{\mu }u_{\nu }+pg_{\mu \nu },  \label{eq19}
\end{equation}%
where $\rho (t)$ is the energy density of the Universe and $p(t)$ is the isotropic pressure. Moreover, $u^{\mu }$ is the velocity four-vector of the
fluid, satisfying the normalization condition of \mbox{$u_{\mu }u^{\mu }=-1$}.

 The matter energy-momentum tensor (\ref{eq19}) can be obtained from two different forms of the matter Lagrangian after variation with respect to the tensor component metric 
  \cite{Schutz, Brown}. First, we note that the variation of the Lagrangian of a perfect fluid must be constrained by the requirements of the conservation of the rates of the entropy ($s$) and of particle production. The baryon number flux vector density is defined according to $n^{\mu }=nu^{\mu }\sqrt{-g}$, where the baryon number density $n$ is \mbox{given by}
\begin{equation}
n=\sqrt{\frac{1}{g}n^{\mu }n^{\nu }g_{\mu \nu }}.  \label{n}
\end{equation}

We require that the variation of the thermodynamic variables satisfies the
constraints of $\delta s=0$ and $\delta n^{\mu }=0$ \cite{Schutz, Brown}. For the
variation of the particle flux ($nu^{\mu }$), we obtain the following relationship:
\vspace{-6pt}
\begin{equation}
\nabla _{\mu }\left( nu^{\mu }\right) =\frac{1}{\sqrt{-g}}\frac{\partial }{%
\partial x^{\mu }}n^{\mu }=\frac{1}{\sqrt{-g}}\partial _{\mu }n^{\mu }.
\end{equation}

Now, we take the variation of the gradient of $\delta n^{\mu }$, and we take
into account the relation of $\partial _{\mu }\delta =\delta \partial _{\mu }$. Thus, we obtain

\begin{equation}
    \delta \left( \partial _{\mu }n^{\mu }\right) = \sqrt{-g} \left[ \delta \left( \nabla _{\mu }\left( nu^{\mu
}\right) \right) +\frac{1}{2}\nabla _{\mu }\left( nu^{\mu }\right) g^{\alpha
\beta }\delta g_{\alpha \beta }\right]= 0
\end{equation}

This relation guarantees that the rate of particle production in the fluid is preserved under the variation ($\delta $). This is a weaker demand than
imposing the condition that the number density be
conserved ($\nabla _{\mu }\left( nu^{\mu }\right) =0$) from the beginning. A similar relation for the entropy ($\delta \left( n^{\mu }\partial _{\mu }s\right) =0$) can also be easily obtained \cite{Schutz, Brown}.

Now, we consider the assumption that the fluid matter Lagrangian is $\mathcal{L}_{m}=-\rho
$
. We also assume that the fluid obeys an equation of state ($\rho =\rho
\left( n,s\right) $). From the thermodynamic equality ($\left( \partial \rho
/\partial n\right) _{s}=w$, where $w$ is the specific enthalpy $w=\left(
\rho +p\right) n$), we obtain the relation of $\delta \rho =w\delta n$ \cite{Schutz, Brown}.

With the above choice of the Lagrangian, the matter action is
\begin{equation}
S_{m}=-\int \rho \sqrt{-g}d^{4}x.  \label{mact}
\end{equation}

For the variation of the action (\ref{mact}), we first obtain
\begin{equation}
\delta S_{m}=-\int \left[ \delta \rho \sqrt{-g}+\rho \delta \sqrt{-g}\right]
d^{4}x.  \label{varm}
\end{equation}

By taking the variation of Equation (\ref{n}), we find
\begin{equation}
\delta n=\frac{n}{2}\left( -g\right) u^{\mu }u^{\nu }\left( \frac{\delta
g_{\mu \nu }}{g}-\frac{g_{\mu \nu }}{g^{2}}\delta g\right) =-\frac{n}{2}%
\left( u^{\mu }u^{\nu }+g^{\mu \nu }\right) \delta g_{\mu \nu }.
\end{equation}

Hence, Equation (\ref{varm}) becomes
\begin{eqnarray}
\delta S_{m} &=&-\frac{1}{2}\int \left[ wn\left( u^{\mu }u^{\nu }+g^{\mu \nu
}\right) -\rho g^{\mu \nu }\right] \sqrt{-g}d^{4}x \\
&=&\frac{1}{2}\int \left[ \left( \rho +p\right) u^{\mu }u^{\nu }+pg^{\mu \nu
}\right] d^{4}x.  \nonumber
\end{eqnarray}
This immediately yields the relation expressed by (\ref{eq19}) for the matter energy-momentum tensor
.

The same expression for the matter energy-momentum tensor is obtained by assuming that $\mathcal{L}_{m}=p$. We now express the equation of state in the form of $%
p=p\left( w^{\prime },s\right) $, where $w^{\prime }=\left( \rho +p\right)
/\rho _{0}$, with $\rho _{0}$ being the rest mass density. In this case, we have the following thermodynamic relation \cite{Schutz, Brown}:
\begin{equation}
dp=\rho _{0}dw^{\prime }-\rho _{0}\mathbb{T}ds.
\end{equation}

Generally, the four-velocity of the fluid can be represented as \cite{Schutz, Brown}
\begin{equation}
u^{\mu }=\frac{1}{w^{\prime }}\left( \partial _{\nu }\phi +\alpha \partial
_{\nu }\beta +\theta \partial _{\nu }s\right) g^{\mu \nu },
\end{equation}%
where $\phi $, $\alpha $, $\beta $, $\theta $, and $s$ are the velocity
potentials (scalar fields), which, together with $w^{\prime }$ and $g_{\mu
\nu }$, comprise the dynamical variable of the fluid to be varied freely. Varying
with respect to the metric and using the normalization condition of $u^{\mu
}$, we obtain
\begin{equation}
\delta w^{\prime }=-\frac{1}{2w^{\prime }}\left( \partial _{\nu }\phi
+\alpha \partial _{\nu }\beta +\theta \partial _{\nu }s\right) \left(
\partial _{\mu }\phi +\alpha \partial _{\mu }\beta +\theta \partial _{\mu
}s\right) \delta g^{\mu \nu }.
\end{equation}

By taking into account that $\delta g^{\mu \nu }=-g^{\mu \alpha }g^{\nu
\beta }g_{\alpha \beta }$, we obtain
\begin{equation}
\delta w^{\prime }=\frac{w^{\prime }}{2}u^{\mu }u^{\nu }\delta g_{\mu \nu }.
\end{equation}

By adopting the action expressed as
\begin{equation}
S_{m}=\int p\sqrt{-g}d^{4}x,  \label{actp}
\end{equation}%
for the fluid and by assuming that the fluid is isothermal  ($\mathbb{T}=cons\tan t$) and that $%
\delta s=0$, the variation of Equation (\ref{actp}) yields
\begin{equation}
\delta S_{m}=\frac{1}{2}\int \left( \rho _{0}w^{\prime }u^{\mu }u^{\nu
}+pg^{\mu \nu }\right) \sqrt{-g}\delta g_{\mu \nu }d^{4}x,
\end{equation}%
a relation from which we immediately reobtain Equation (\ref{eq19}).

 In the following, for the
matter Lagrangian, we adopt the expression  of $\mathcal{L}_{m}=p$
(see Ref. \cite{Bertolami:2008ab}), which, for the auxiliary tensor, yields $\Theta _{\mu
\nu }$ \mbox{the following expression}:
\begin{equation}
\Theta _{\mu \nu }=-\left( \rho ^{2}+4\rho p+3p^{2}\right) u_{\mu }u_{\nu }.
\label{eq20}
\end{equation}%

To investigate the cosmological implications of the EMSG theory, we adopt the comoving reference frame, with $u_{\mu
}=(1,0,0,0)$. Then, we obtain the system of two generalized Friedmann
equations in $f\left( R,\mathcal{T}\right) $ gravity given by 

\vspace{-6pt}
\begin{subequations}
\begin{adjustwidth}{-\extralength}{0cm}
\begin{align}
3H^{2}& =\frac{1}{\phi }\left[ \kappa ^{2}\rho +\frac{1}{2}V+\frac{1}{2}\psi
\left( \rho ^{2}+8\rho p+3p^{2}\right) -3H\dot{\phi}\right] =\kappa
_{\rm eff}^{2}\;\rho +\rho _{\rm eff},  \label{eq21}
	\\
2\dot{H}& =\frac{1}{\phi }\left[ -\kappa ^{2}(\rho +p)-\psi \left( \rho
^{2}+4\rho p+3p^{2}\right) +H\dot{\phi}-\ddot{\phi}\right] =-\kappa
_{\rm eff}^{2}\,\left( \rho +p\right) -\left( \rho _{\rm eff}+p_{\rm eff}\right) ,
\label{eq22}
\end{align}%
\end{adjustwidth}
where $\kappa _{\rm eff}^{2}=\kappa ^{2}/\phi $, and we denote
\end{subequations}
\begin{equation}\label{rhoeff}
\rho _{\rm eff}=\frac{1}{\phi }\left[ \frac{1}{2}V+\frac{1}{2}\psi \left( \rho
^{2}+8\rho p+3p^{2}\right) -3H\dot{\phi}\right] ,
\end{equation}%
and
\begin{equation}\label{peff}
p_{\rm eff}=\frac{1}{\phi }\left[ -\frac{1}{2}V+\frac{1}{2}\psi \left( \rho
^{2}+8\rho p+3p^{2}\right) -3H\dot{\phi}+\ddot{\phi}\right] ,
\end{equation}%
where $\rho_{\rm eff}$ and $p_{\rm eff}$ represent the effective scalar-tensor contributions to the cosmological evolution equations resulting from the coupling between matter and geometry via the square of the matter energy-momentum tensor. Moreover, the gravitational coupling constant becomes time-dependent, indicating the time variation of the gravitational constant in this model. The equations of motion for the scalar fields of $\phi $
and $\psi $ are given by 
\begin{subequations}
\begin{align}
V_{\phi }& =R=6\left( \dot{H}+2H^{2}\right) ,  \label{eq23} \\
V_{\psi }& =\mathcal{T}=\rho ^{2}+3p^{2}.  \label{eq24}
\end{align}

\subsubsection{The Energy Balance Equation}

From Equations (\ref{eq21}) and (\ref{eq22}), one can obtain the generalized
energy-conservation equation in the following form:
\end{subequations}
\begin{equation}
\frac{d}{dt}\left( \kappa _{\rm eff}^{2}\;a^{3}\rho \right) +p\frac{d}{dt}\left(\kappa
_{\rm eff}^{2}\,a^{3}\right)+\frac{d}{dt}\left( a^{3}\rho _{\rm eff}\right) +p_{\rm eff}\frac{d}{%
dt}a^{3}=0.
\end{equation}

Explicitly, we obtain
\begin{eqnarray}
\dot{\rho}+3H\left( \rho +p\right) &=&-\frac{1}{\kappa _{\rm eff}^{2}}\left(
\rho +p\right) \frac{d}{dt}\kappa _{\rm eff}^{2}-\frac{1}{\kappa _{\rm eff}^{2}}%
\left[ \dot{\rho}_{\rm eff}+3H\left( \rho _{\rm eff}+p_{\rm eff}\right) \right] \\
&=&\left( \rho +p\right) \Gamma ,
\end{eqnarray}
which yields the following expression for the particle creation rate:
\begin{equation}
\Gamma =-\frac{1}{\kappa _{\rm eff}^{2}}\frac{d}{dt}\kappa _{\rm eff}^{2}-\frac{1}{%
\left( \rho +p\right) \kappa _{\rm eff}^{2}}\left[ \dot{\rho}_{\rm eff}+3H\left(
\rho _{\rm eff}+p_{\rm eff}\right) \right] .
\end{equation}

\subsubsection{The Deceleration Parameter}

An important cosmological parameter, the deceleration parameter, is defined
as
\begin{equation}
q=\frac{d}{dt}\frac{1}{H}-1=-\frac{\dot{H}}{H^{2}}-1.
\end{equation}

With the use of the gravitational field equations, the deceleration parameter
is obtained in the following form:
\begin{equation}
q=\frac{3}{2}\frac{\kappa _{\rm eff}^{2}\left( \rho +p\right) +\left( \rho
_{\rm eff}+p_{\rm eff}\right) }{\kappa _{\rm eff}^{2}\rho +\rho _{\rm eff}}-1,
\end{equation}
or explicitly,
\begin{equation}
q=\frac{3}{2}\frac{\kappa ^{2}(\rho +p)+\psi \left( \mathcal{T} +4\rho
p\right) -H\dot{\phi}+\ddot{\phi}}{\kappa ^{2}\rho +\frac{1}{2}V+\frac{1}{2}%
\psi \left( \mathcal{T} +8\rho p\right) -3H\dot{\phi}}-1.
\end{equation}

Thus, the condition for accelerating expansion ($q<0$) yields
\begin{equation}
\kappa _{\rm eff}^{2}\left( \frac{\rho }{3}+p\right) <-\left( \frac{\rho _{\rm eff}}{%
3}+p_{\rm eff}\right) .
\end{equation}

Explicitly, we obtain the condition for the accelerated expansion as
\begin{equation}
\kappa _{\rm eff}^{2}\left( \frac{\rho }{3}+p\right) <\frac{V}{3}-\frac{2}{3}%
\psi \left( \mathcal{T} -2\rho p\right) +H\dot{\phi}-\ddot{\phi}.
\end{equation}

For a dust Universe, the late time-accelerated phase occurs for
\begin{equation}
\kappa _{\rm eff}^{2}\rho <\frac{V}{3}-2\psi \rho ^{2}+3H\dot{\phi}-3\ddot{\phi}.
\end{equation}

\subsubsection{Dark Matter and Dark Energy}

\textls[-15]{The EMSG theory generates, in its two scalar field representations,
an effective energy and an effective pressure of an essentially geometric
nature, as expressed by Equations (\ref{rhoeff}) and (\ref{peff}), respectively.
These effective geometric quantities can be interpreted as describing both dark matter and dark energy at the same time. The presence of an
effective gravitational constant ($\kappa ^{2}/\phi $) could already have some
important implications for the behavior of massive particles gravitating around galaxies---behavior usually explained by assuming the presence of dark matter. However, an effective gravitational constant appearing in Newton's
law ($F_{grav}=-G_{eff}(r)mM/r^{2}$) can already provide some insights into the dark matter problem, since, for the rotational velocity, the modified Newtonian--Keplerian law ($%
v^{2}/r=G_{eff}M/r^{2}$) yields $%
v\approx \sqrt{GM/\phi r}$, which could explain the flatness of the rotation
curves once $\phi$$\sim$$1/r$
. The effective energy density ($\rho _{DM}$) and
pressure ($p_{DM}$) of dark matter can be inferred from Equations (\ref{rhoeff})
and (\ref{peff}) by assuming that in the case of pressureless matter with $%
p=0$,  $\rho _{DM}=p_{DM}=\left( \psi /2\phi \right) \rho ^{2}$. Hence, in
this model, there is a close relationship between (effective) dark matter and ordinary matter. Interestingly enough, in this model, dark matter satisfies the stiff (causal) equation of state, with $\rho =p$.}

From a cosmological point of view, $\rho _{\mathrm{eff}}$ and $p_{\mathrm{eff%
}}$ can be interpreted as globally describing both dark components of the Universe, and in our analysis we do not make an explicit distinction between dark matter and dark energy, treating the dark components of the Universe as a single physical entity. However, by considering the above interpretation of dark matter, we may assume that the dark energy term is described by time-dependent energy density and pressure expressed as $\rho
_{DE}=\left( V/2-3H\dot{\phi}\right) /\phi $ and $p_{DE}=\left( -V/2-3H\dot{%
\phi}+\ddot{\phi}\right) /\phi $, respectively. The equation of state of the dark energy depends on the explicit choice of $V$ and on the dynamical
evolution of the two scalar fields. Then, a necessary condition for the late accelerated expansion of the Universe is that the parameter of the dark energy equation of state ($w=$ $p_{DE}/\rho _{DE}$) must be negative, i.e., $w<0$.

\subsection{de Sitter Expansion}

The fact that the Universe experiences an accelerated expansion, which may end in a de Sitter-type phase of the Universe, is an important result of observational cosmology. Hence, viable theoretical cosmological models must have an explicitly exponentially expanding solution. We now consider the de
Sitter-type expansion of the Universe with \mbox{$H=H_0=\mathrm{constant}$} in $%
f\left(R,\mathcal{T}\right)$ gravity. The theory admits
exponential expansion for both vacuum and dust universes.

\subsubsection{Self-Interacting Potential and Constant-Density de Sitter Expansion
}

We first consider the self-interacting potential during de Sitter evolution in the constant-density dust case, which is obtained in the de Sitter phase with $H=H_{0}=\mathrm{constant}$, $p=0$, and $\rho =\rho _{0}=\mathrm{%
constant}$
. Then, by integrating \eqref{eq23} and \eqref{eq24},
we immediately find
\begin{equation}
V(\phi ,\psi )=12H_{0}^{2}\phi +\rho _{0}^{2}\psi +\Lambda _{0},
\label{vphipsi}
\end{equation}%
where $\Lambda _{0}$ is the resulting constant of integration. For this specific form of the potential, the energy balance equation becomes
\begin{align}
\kappa ^{2}[\dot{\rho} +3H(\rho +p)]+(\rho +p)(\rho +3p)(\dot{\psi}+3H\psi )
 +\psi \lbrack \dot{\rho}(\rho +4p)+\dot{p}(4\rho +3p)]=0.  \label{eq25}
\end{align}

The matter creation rate  for the constant-density de Sitter phase is given by
\begin{equation}
\Gamma =-\dfrac{1}{\kappa ^{2}}\bigg[(\dot{\psi}+3H\psi )(\rho +3p)+\psi
\frac{\dot{\rho}(\rho +4p)+\dot{p}(4\rho +3p)}{(\rho +p)}\bigg].
\label{creation_rate}
\end{equation}%

The creation pressure can be obtained as
\begin{equation}
p_{c}=\dfrac{\rho +p}{3H\kappa ^{2}}\bigg[(\dot{\psi}+3H\psi )(\rho
+3p)+\psi \frac{\dot{\rho}(\rho +4p)+\dot{p}(4\rho +3p)}{(\rho +p)}\bigg].
\end{equation}

\subsubsection{The Vacuum de Sitter Solution}

For the vacuum case, with $\rho =p=0$, for the de Sitter expansion, the potential ($V$) becomes $V(\phi,\psi)=12H_0^2\phi$.  The evolution equation of the scalar field ($\phi$) that follows from Equation~(\ref{eq22}) is
\begin{equation}
H_0\dot{\phi}-\ddot{\phi}=0,
\end{equation}
with the general solution of
\begin{equation}
\phi(t)=C_1\frac{e^{H_0t}}{H_0}+C_2,
\end{equation}
where $C_1$ and $C_2$ are arbitrary constants of integration. For this solution, Equation~(\ref{eq21}) gives $3C_2H_0^2=0$, a condition that is satisfied by $C_2=0$. Hence, the exact vacuum de Sitter solution corresponds to an exponentially increasing $\phi =C_1
e^{H_0t}/H_0 $ in the presence of an arbitrary field ($\psi$). The particle creation rate ($\Gamma$) is zero ($\Gamma \equiv 0$) for the vacuum Universe, and the energy balance equation (Equation (\ref{eq25})) is identically satisfied. It is important to note that in the present model, matter creation takes place only if matter already exists.

\subsubsection{de Sitter Solution with Constant Matter Density}

We now assume that the Universe consists of pressureless dust, with $p=0$, and that the matter density is constant ($\rho=\rho_0$). From Equation \eqref{eq25}, we obtain the following differential equation \mbox{for $\psi$}:
\begin{equation}
\dot\psi(t) + 3H_0\psi + \dfrac{3\kappa^2H_0 }{\rho_0}= 0,  \label{coc}
\end{equation}
with the general solution of
\begin{equation}
\psi(t)=e^{-3 H_0 t}\left[\psi_0+\dfrac{\kappa^2}{\rho_0}\left(1-e^{3 H_0
t}\right)\right] , \label{psi}
\end{equation}
where  $\psi_0=\psi(0)$. From Equation~\eqref{eq21}, it follows that the differential equation that describes the scalar field ($\phi$) is given by
\begin{equation}
\dot\phi(t) -H_0\phi -\dfrac{1}{6H_0}\left[\psi\left(\rho_0^2-\rho_0%
\right)+2\kappa^2\rho_0+\Lambda_0\right] = 0,  \label{def_phi}
\end{equation}
with the general solution of 
\begin{eqnarray}
\phi(t) &=& \dfrac{1}{12 H_0^2}\Bigg\{e^{-3 H_0 t}\Bigg[-2 e^{3 H_0 t}
\Lambda_0-\rho\left(k^2+\rho \psi_0\right)  \notag \\
&& +e^{4 H_0 t}\left(2 \Lambda_0+\kappa^2 \rho+12 H_0^2 \phi_0+\rho^2
\psi_0\right)\Bigg]\Bigg\},
\end{eqnarray}
where $\phi_0=\phi(0)$. The creation pressure and the creation rate %
can now be obtained as 
\begin{equation}
\Gamma = 3H_0,\quad p_c =-\rho_0.
\end{equation}

\subsubsection{de Sitter Solution with Arbitrary Matter Density}

 For a de Sitter Universe with time-varying, non-zero matter density, the generalized Friedmann equations (Equations (\ref{eq21}) and (\ref{eq22})) take the following form:
\begin{equation}
3H_0^2\phi=\kappa ^2\rho +\frac{1}{2}V+\frac{1}{2}\psi \rho^2-3H_0\dot{\phi},
\end{equation}
and
\begin{equation}  \label{eq52}
-\kappa ^2\rho-\psi \rho^2+H_0\dot{\phi}-\ddot{\phi}=0,
\end{equation}

The solutions of the generalized Friedmann equations,
as well as the cosmological dynamics, depend on the  form of the potential ($V$). In the following, we assume that $V$
can be \mbox{represented as}
\begin{equation}  \label{V1}
V(\phi,\psi)=12H_0^2\phi-\frac{\alpha ^2}{\psi}.
\end{equation}

Then from Equation~(\ref{eq24}), we first obtain $V_\psi=\alpha^2 /\psi
^2=\rho ^2$,   $\psi =\alpha/\rho $, and $\dot{%
\psi}=-\alpha \dot{\rho}/\rho^2$.

Equation \eqref{eq25} takes the following form:
\begin{equation}
\dot{\rho}\left(\kappa^2+\psi\rho\right) + \rho^2\left(\dot{\psi}%
+3H_0\psi\right)+3\kappa^2H_0\rho = 0,
\end{equation}
and
\begin{equation}
\dot{\rho}+3H_0\left(1+\frac{\alpha}{\kappa ^2}\right)\rho=0.
\end{equation}

Hence, the matter density is obtained as
\begin{equation}
\rho (t)=\rho _0e^{-3H_0\left(1+\frac{\alpha}{\kappa ^2}\right)t},
\end{equation}
where $\rho_0=\rho (0)$. From Equation~(\ref{eq52}), for $\phi$, we obtain the following equation:
\begin{equation}
\ddot{\phi}-H_0\dot{\phi}+\left(\kappa ^2+\alpha\right)\rho
_0e^{-3H_0\left(1+\frac{\alpha}{\kappa ^2}\right)t}=0,
\end{equation}
the general solution of which is expressed as
\begin{equation}
\phi (t)=\frac{c_1 e^{H_0 t}}{H_0}+c_2-\frac{\kappa ^4 \rho_0 e^{-\frac{3
H_0 t \left(\alpha +\kappa ^2\right)}{\kappa ^2}}}{3 H_0^2 \left(3 \alpha +4
k^2\right)},
\end{equation}
where $c_1$ and $c_2$ are arbitrary constants of integration. The first
Friedmann Equation~(\ref{eq21}) yields $3H_0^2c_2=0$, or $c_2=0$. Thus, this solution satisfies all cosmological evolution equations.

In the present model, the particle creation rate, given by
\begin{equation}
\Gamma =-\frac{1}{\kappa ^2}\left[\left(\dot{\psi}+3H\psi\right)\rho+\psi
\dot{\rho}\right],
\end{equation}
becomes
\begin{equation}
\Gamma =-\frac{3\alpha}{\kappa ^2}H_0.
\end{equation}

Thus, for the de Sitter-type solution corresponding to the potential in Equation (\ref{V1}), the thermodynamics of open systems require a negative particle creation rate, corresponding to a decrease in the density of ordinary matter
.

\subsection{Matter and Radiation Domination Phases}

When the Universe was about 47,000 years old, it entered into the matter-dominated phase, in which the energy density of matter largely exceeded the energy densities of radiation and dark energy (cosmological constant). The matter-dominated phase lasted until the Universe was about 9.8 billion years old \cite{Dod}. During this phase, the standard Big Bang cosmological scenario predicts that the scale factor of the Universe evolved as $t\propto t^{2/3}$, while the deceleration parameter took a constant value of $q=1/2$. The matter density varied during this period as $\rho(t)\propto 1/a^3(t)=1/t^2$. In the following, we investigate the conditions under which a similar matter-dominated era could exist in the energy-momentum squared gravity theory, assuming a more general form for the scale factor than that of standard general relativity ($a=t^{n}$, $n>0$)
.

\subsubsection{Models with a Quadratic Additive Potential}

We consider a case in which during the matter-dominated phase, the potential ($V(\phi,\psi)$) has the following quadratic form:
\begin{equation}
V(\phi,\psi)=\frac{\alpha}{2}\phi^2+\frac{\beta}{2}\psi ^2,
\end{equation}
where $\alpha$ and $\beta$ are constants. Then, from Equation~(\ref{eq23}), we obtain
\begin{equation}
\alpha \phi=\frac{6n(2-n)}{t^2},
\end{equation}
while Equation~(\ref{eq24}) yields
\begin{equation}
\beta \psi=\rho ^2.
\end{equation}

With these choices for $V$, $\phi$, and $\psi$, the generalized Friedmann equations of the EMSG theory take the following forms:
\be
\label{94}
27 \beta  n^2 (2 n-1)+\alpha  \beta  \kappa ^2 t^4 \rho (t)+\alpha  t^4 \rho ^4(t)=0
\ee
and
\be\label{95}
36 \beta  n(n-1)+\alpha  \beta  \kappa ^2 t^4 \rho (t)+\alpha  t^4 \rho ^4(t)=0.
\ee

\textls[20]{For $n$, the required equivalence of Equations~(\ref{94}) and (\ref{95}) yields the following algebraic equation}:
\be
-54 \beta  n^3+99 \beta  n^2-36 \beta  n=0,
\ee
which fixes $n$ as
\be
n=0, n=\frac{1}{2}, n=\frac{4}{3}.
\ee

The matter density can be obtained by solving any of the algebraic equations \linebreak(Equation (\ref{94}) or (\ref{95})).

For $n=1/2$, the density takes a constant value, i.e., $\rho (t)=\rho_0=\left(-\beta \right)^{1/3}$. For $n=4/3$, the density can be obtained as a solution of the following algebraic equation:
\be\label{98}
\alpha t^4\rho ^4(t)+ \kappa ^2\alpha \beta t^4 \rho (t)+80 \beta =0, n=\frac{4}{3}.
\ee

In the first order of approximation, we neglect the term containing the fourth power of the density, obtaining
\be
\kappa ^2\rho (t)\approx -\frac{80 }{\alpha}\frac{1}{t^4}, n=\frac{4}{3}.
\ee

In order to assure the physical viability of the density, the ratio of $1 /\alpha $ must be negative, i.e., $1 /\alpha <0$.

Hence, we have obtained two distinct solutions of the EMSG theory in the matter-dominated phase of cosmological evolution in the presence of an additive quadratic potential ($V$). In the first solution, $a(t)=t^{1/2}$, $H(t)=1/2t$, $\rho (t)={\rm constant}$, and $q=1$. The density of the Universe is constant, and the expansion is decelerating. The constancy of density is maintained by the particle creation rate, given by $\Gamma=3H=3/2t$, which decreases over time. But, even so, it can maintain the constancy of the matter density. In the second solution, $a=t^{4/3}$, $H(t)=4/37$, $\rho (t)\propto 1/t^4$, and $q=-1/4$. In this model, the Universe is already accelerating in the matter-dominated phase but with a relatively low value of the deceleration parameter. By taking the derivative of the density equation (Equation~(\ref{98})), we obtain
\be
\frac{\dot{\rho}}{\rho}=\frac{320\beta}{\alpha \rho\left(\rho^3+\kappa ^2 \beta\right)},
\ee
a relation that allows for the expression of the particle creation rate as
\be
\Gamma =3H+\frac{320\beta}{\alpha \rho\left(\rho^3+\kappa ^2 \beta\right)}, n=\frac{4}{3}.
\ee

In the approximation of $\rho \propto t^{-4}$, the particle creation rate is $\Gamma =3H+\dot{\rho}/\rho\approx 0$. However, when using the exact solution of the algebraic density equation, the particle creation rate is non-zero, indicating the presence of matter creation during an accelerating expansionary phase.

\subsubsection{Radiation-Dominated Models}

In the very early stages of cosmological evolution, most of the total energy density was in the form of radiation, which  was the major constituent of the Universe. With the expansion of the Universe and the cooling down of radiation, matter became the dominant 
component. Although we are presently in a phase dominated by dark energy, an understanding of the roles of radiation and matter is fundamental for the description of the evolution of the early Universe.

We now consider the effects of the EMSG theory on radiation-dominated cosmological models. The generalized Friedmann equations in the presence of radiation with energy density ($\rho_r$) and the equation of state ($p_r=\rho_r/3$) take the following forms:
\be \label{eq21n}
3H^{2} =\frac{1}{\phi }\left[ \kappa ^{2}\rho_r +\frac{1}{2}V+2\psi \rho_r^2-3H\dot{\phi}\right]
\ee
and
\be\label{eq22n}
2\dot{H} =\frac{1}{\phi }\left[ -\kappa ^{2}\frac{4}{3}\rho_r-\frac{8}{3}\psi \rho_r^2 +H\dot{\phi}-\ddot{\phi}\right].
\ee

For the potential ($V$), we assume a simple, multiplicative form so that
\be
V(\phi,\psi)=\alpha \phi \psi,
\ee
where $\alpha $ is a constant. Hence, we immediately obtain
\be
V_\phi=\alpha \psi=6\left(\dot{H}+2H^2\right), \alpha \phi=\rho_r ^2.
\ee

Then, the generalized Friedmann equations of the EMSG theory become
\be\label{g1}
15 \dot{H}-\frac{6 H \dot{\rho} }{\rho }+27 H^2+\frac{\alpha  \kappa ^2}{\rho }=0,
\ee
\be\label{g2}
18 \dot{H}+\frac{-6 H \dot{\rho} +4 \alpha  \kappa ^2+6 \ddot{\rho} }{3 \rho }+32 H^2+\frac{2
   \dot{\rho} ^2}{\rho ^2}=0.
\ee

We again look for solutions of the above generalized Friedmann equations with $a(t)=t^n$, where $n$ is a constant. With this choice of the scale factor, \mbox{Equations~(\ref{g1}) and (\ref{g2})} take the following forms
\be\label{g11}
-\alpha  k^2 t^2-27 n^2 \rho +15 n \rho +6 n t \dot{\rho} =0,
\ee
and
\begin{equation}
\frac{\ddot{\rho}}{\rho }+\frac{\dot{\rho}^{2}}{\rho ^{2}}-\frac{n\dot{\rho}%
}{t\rho }+\frac{2\alpha k^{2}}{3\rho }+\frac{n(16n-9)}{t^{2}}=0,  \label{g22}
\end{equation}
respectively. From Equation (\ref{g11}), we obtain $\dot{\rho}$ as
\be
\dot{\rho}=\frac{\alpha  \kappa ^2 t}{6 n}+\frac{(9 n-5) \rho }{2 t}.
\ee

\textls[10]{By taking the time derivative of the above equation and substituting the result into Equation~(\ref{g22}), we obtain the following for the time variation of the density the first-order \mbox{differential equation}}:
\be\label{gfin}
\dot{\rho} =\frac{3 n \rho  \left\{8 \alpha  k^2 (1-3 n) t^2-3 n [n (127 n-134)+35] \rho
   \right\}-\alpha ^2 k^4 t^4}{18 n^2 (9 n-5) t \rho }.
\ee

Equation~(\ref{gfin}) cannot be generally solved for arbitrary values of $n$. However, some simple particular solutions can be obtained for some particular values of $n$. Thus, for $n=1/3$, we obtain the following expression for the variation of matter density in the radiation-dominated phase of EMSG theory cosmology:
\be
\rho (t)=\sqrt{c_1 t^{20/9}+\frac{9}{32} \alpha ^2 k^4 t^4},
\ee
where $c_1$ is an arbitrary constant of integration. $\rho(t)$  has the property of $\rho (0)=0$. The density of the Universe is increasing during the expansionary phase, and a large amount of radiation is produced. This qualitative behavior is similar to that considered in the warm inflationary scenario \cite{wi1,wi2,wi3,wi4,wi5,wi6},  which represents an interesting and  successful alternative to standard inflation and reheating
. In warm inflation, not only the dynamics of the scalar field but also the creation and effects of radiation during the process of accelerated expansion are considered. Radiation is produced via the decay of the scalar field as a result of a dissipative process. A similar result can be obtained in the framework of the EMSG theory. The non-conservation of the matter energy-momentum tensor, which is associated with particle creation, allows for the production of photons during the early expansion of the Universe. Hence, radiation creation could take place as triggered by the presence of the two essentially geometric scalar fields ($\phi$ and $\psi$), and the interplay between these two fields and gravity may be responsible for the creation of the matter content of the early Universe. It is also important to note that in the present model, matter creation takes place during a decelerating phase of evolution, with $q=2$. Of course, other values of $n$ could lead to power law accelerating models, which may also explain other observational features of the early Universe.

\subsection{General Cosmological Models}

We now investigate cosmological models with the Hubble function ($%
H $), an arbitrary function of time whose behavior must be obtained from the generalized Friedmann equations. We also require that $H$ tend towards a constant value in the final stages of cosmological evolution, that is, that the Universe end in a de Sitter-type phase.

\subsubsection{The Dimensionless Representation}

A significant simplification of the mathematical formalism can be obtained by introducing a set of the dimensionless variables $\left(h,\tau,r,U,\Psi\right)$,
defined as
\vspace{-6pt}
\begin{equation}
H=H_0h, \qquad \tau =H_0t,  \qquad  \rho=\frac{3H_0^2}{\kappa ^2}r, \qquad  V=3H_0^2U,  \qquad  \psi=\frac{\kappa ^4}{9H_0^2}\Psi.
\end{equation}

Then, the system of differential equations describing cosmological evolution in $f\left(R,\mathcal{T}\right)$ gravity takes the following form:
\vspace{-6pt}
\begin{equation}  \label{F1}
h^2=\frac{1}{\phi}\left(r+\frac{1}{2}U+\frac{1}{6}\Psi r^2-h\frac{d\phi}{%
d\tau}\right),
\end{equation}
\begin{equation}  \label{F2}
\frac{dh}{d\tau}=\frac{1}{\phi}\left(-\frac{3}{2}r-\frac{1}{2}\Psi r^2+\frac{%
1}{2}h\frac{d\phi}{d\tau}-\frac{1}{2}\frac{d^2\phi}{d\tau ^2}\right),
\end{equation}
\begin{equation}  \label{F3}
\left(1+\frac{1}{3}r\Psi\right)\frac{dr}{d\tau}+3hr+\frac{1}{3}r^2\left(%
\frac{d\Psi}{d\tau}+3h\Psi\right)=0.
\end{equation}

The system of Equations (\ref{F1})--(\ref{F3}) contains two independent dynamical equations but four unknown functions $%
(h,r,\phi,\Psi)$. To close the system, one must first specify the potential ($U$); then, using Equations~(\ref{eq23})
and (\ref{eq24}), one can obtain two relations for $\phi$ and $\psi$, given by
\begin{equation}
U_\phi=2\left(\frac{dh}{d\tau}+2h^2\right), \qquad U_\Psi=\frac{1}{3}r^2.
\end{equation}

The potential can now be represented as
\begin{equation}
U_\phi=\frac{2}{\phi}\left(\frac{1}{2}r-\frac{1}{6}\Psi r^2-\frac{3}{2}h%
\frac{d\phi}{d\tau}-\frac{1}{2}\frac{d^2\phi}{d\tau^2}\right).
\end{equation}

\subsubsection{The Redshift Representation}

A comparison with the observational data of cosmological models can be performed in the easiest way by considering the evolution of the models and adopting redshift ($z$) as an independent variable
. Redshift is defined as
\begin{equation}
1+z=\frac{1}{a},
\end{equation}
yielding $\frac{d}{d\tau}=-(1+z)h\frac{d}{dz}$.

After introducing the new variable ($d\phi/d\tau =\theta$), the cosmological evolution equations in the two scalar field representations of $f\left(R,\mathcal{T}\right)$ gravity can be formulated in a redshift space as a first-order dynamical system given by
\begin{equation}  \label{d1}
-(1+z)h\frac{d\phi}{dz}=\theta,
\end{equation}
\begin{equation}  \label{d2}
-(1+z)h\frac{dh}{dz}=\frac{1}{\phi}\left[-\frac{3}{2}r-\frac{1}{2}\Psi r^2+%
\frac{1}{2}h\theta +\frac{1}{2}(1+z)h\frac{d\theta}{dz}\right],
\end{equation}
\begin{eqnarray}  \label{d3}
-(1+z)\left(1+\frac{1}{3}r\Psi\right)h\frac{dr}{dz}+3hr
+\frac{1}{3}r^2\left[-(1+z)h\frac{d\Psi}{dz}+3h\Psi\right]=0,
\end{eqnarray}
\begin{equation}  \label{d4}
(1+z)h\frac{d\theta}{dz}=\phi U_\phi-r+\frac{1}{3}\Psi r^2+3h\theta,
\end{equation}
\begin{equation}  \label{d5}
(1+z)h\frac{dh}{dz}=2h^2-\frac{1}{2}U_\phi,
\end{equation}
\begin{equation}  \label{d6}
U_\Psi=\frac{1}{3}r^2.
\end{equation}

The system of Equations (\ref{d1})--(\ref{d6}) must be solved together with
the initial conditions of $h(0)=1$, $r(0)=r_0$, $\phi (0)=\phi_0$, and $\theta
(0)=\theta _0$. No dynamical evolution equation for $\psi$ exists, but  the evolution of $h$ is described by two equivalent equations. Thus, one can consider only one of Equations~(\ref{d2}) and (\ref{d5}) when solving the
cosmological field equations.

To describe the accelerating/decelerating phases of the evolution of the Universe, we use the deceleration parameter ($q$),
which takes the following form in the redshift space:
\begin{equation}
q=\frac{d}{d\tau}\frac{1}{h}-1=(1+z)\frac{1}{h}\frac{dh}{dz}-1.
\end{equation}

The negative sign of $q$ indicates an accelerated evolution of the Universe, whereas a positive sign describes the decelerating stages of evolution.

The particle creation rate is obtained from the following relation:
\begin{equation}
\Gamma =H_0\left[3h-(1+z)\frac{1}{r}\frac{dr}{dz}\right]=H_0\Sigma,
\end{equation}
while the creation pressure is given by
\begin{equation}
P_c=\frac{p_c\kappa ^2}{H_0^2}=-\frac{r}{h}\Sigma.
\end{equation}

\subsubsection{Specific Cosmological Models: A Qualitative Discussion}

In the following, we briefly discuss, , the general properties of two distinct cosmological models corresponding to two independent functional forms of the potential ($U$) from a qualitative point of view.

\begin{figure}[H]
	\includegraphics[scale=0.54]{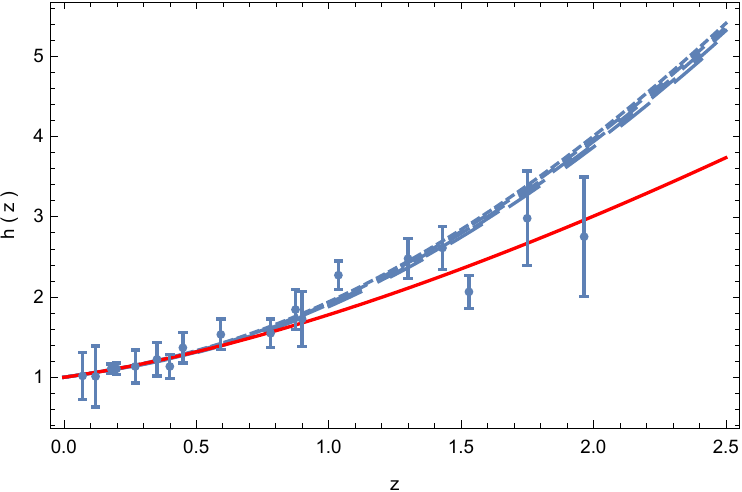}
\includegraphics[scale=0.54]{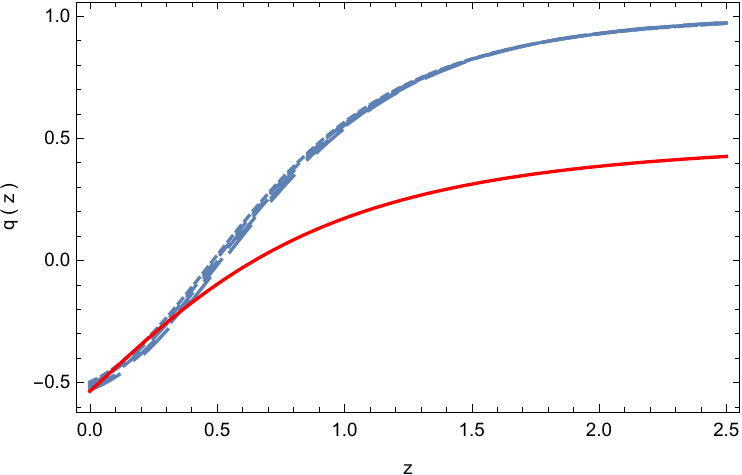}
	\caption{\label{fig1_U} The 
 evolution of the dimensionless Hubble function ($h(z)$) (\textbf{left panel}) and of the deceleration parameter ($q(z)$) (\textbf{right panel}) as a function of the redshift ($z$) in the EMSG cosmological model with $U(\phi,\Psi)=2\alpha \phi^{n+1}+\frac{1}{3}\beta \Psi ^{m+1}$ for $n=0.975$, $m=2$, $\beta =1$, $r(0)=0.30$, and different values of $\alpha$ ($\alpha =0.368$, solid curve; $\alpha =0.370$, dotted curve; $\alpha =0.372$, short dashed curve; $\alpha =0.374$, dashed curve; $\alpha =0.376$, long dashed curve). Predictions of the standard $\Lambda$CDM model are represented by the solid red curve, while the observational data are shown with their error bars.}
\end{figure}

{Additive} 
 power law 
 potential: $U(\phi,\Psi)=2\alpha \phi^{n+1}+\frac{1}{3}\beta \Psi ^{m+1}$. The first case corresponds to a potential ($U(\phi, \psi)$) with an additive algebraic structure of the following form:
\begin{equation}
U(\phi,\Psi)=2\alpha \phi^{n+1}+\frac{1}{3}\beta \Psi ^{m+1},
\end{equation}
where $\alpha$, $\beta$, $n$, and $m$ are constants. The system of differential equations describing cosmological evolution must be solved with initial conditions or $h(0)=1$, $r(0)=r_0$, $\phi (0)=\phi_0$, and $%
\theta (0)=\theta_0$. In this model, cosmological evolution depends on the model parameters, as well as on the initial conditions for the scalar field ($\phi$) and its
derivative. The variations of the dimensionless Hubble function ($h(z)$) and of the deceleration parameter of this model are represented in Figure~\ref{fig1_U}.

For $n\approx 1$, $m=2$, $\beta =1$, and $\alpha $ in the range of $\alpha \in (0.38,0.42)$, this model provides a very good description of the observational
data for the Hubble function up to a redshift (\mbox{$z\approx 1$) \cite{Cipriano:2023yhv}.} However, at higher redshifts ($z>1$), important differences appear with respect to the observational data and the $\Lambda$CDM predictions, at least for the considered
range of model parameters. At redshifts greater than $z=1$, the Hubble function increases faster as compared to the $\Lambda$CDM model. The evolution of the deceleration parameters of the two models also shows important differences.

At redshifts of $z>1.5$, the deceleration parameter in
$f\left(R,\mathcal{T}\right)$ gravity takes higher values close to $%
q\approx 1$. The transition to the accelerated expansion phase occurs at redshifts of $z<0.5$. On the other hand, the present-day value of the deceleration parameter is dependent on the model parameters.

\begin{figure}[H]
	\includegraphics[scale=0.54]{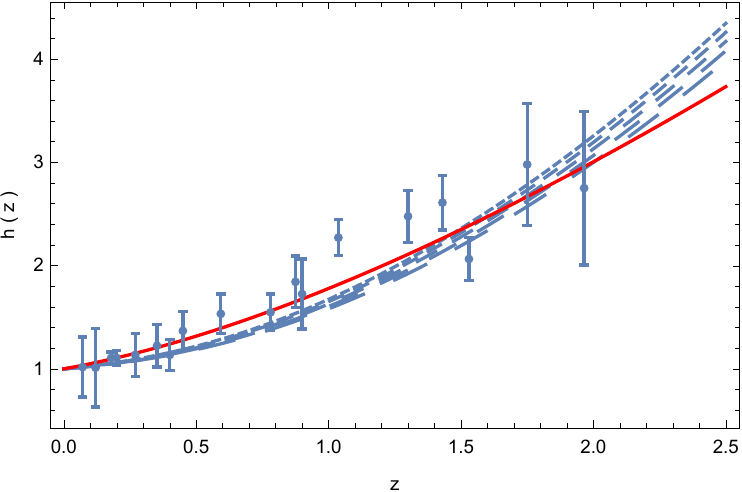}
\includegraphics[scale=0.54]{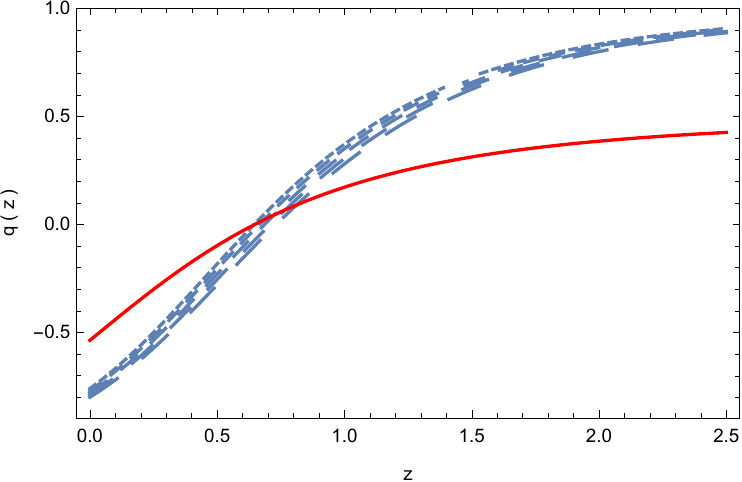}
	\caption{\label{fig2_U} The 
 evolution of the dimensionless Hubble function ($h(z)$) (\textbf{left panel}) and of the deceleration parameter ($q(z)$) (\textbf{right panel}) as a function of the redshift ($z$) in the EMSG cosmological model with $U(\phi,\Psi)=U(\phi,\psi)=\alpha \phi +\frac{1}{3}\beta \Psi +\frac{\gamma }{3}\phi ^n\Psi ^{m+1}$ for $n=10^{-8}$, $m=1.05$, $\beta =0.02$, $\gamma =0.04$, $r(0)=0.30$, and different values of $\alpha$ ($\alpha =0.352$, solid curve; $\alpha =0.354$, dotted curve; $\alpha =0.356$, short dashed curve; $\alpha =0.358$, dashed curve; $\alpha =0.360$, long dashed curve). Predictions of the standard $\Lambda$CDM model are represented by the red solid curve, while the observational data are shown with their error bars. The initial conditions used to integrate the generalized Friedmann equations are $\phi (0)=0.10$, and $\theta (0)=-0.01$.}
\end{figure}

{Additive}
--multiplicative potential: $U(\phi,\psi)=\alpha \phi +(1/3)\beta \Psi +(\gamma /3)\phi ^n\Psi ^{m+1}$. As a second cosmological model, one can consider the model corresponding to potential ($U$) with a non-additive structure, given by
\begin{equation}
U(\phi,\psi)=\alpha \phi +\frac{1}{3}\beta \Psi +\frac{\gamma }{3}\phi ^n\Psi ^{m+1},
\end{equation}
where $\alpha$, $\beta$, $\gamma$, $n$, and $m$ are constants. The variations of the dimensionless Hubble function ($h(z)$) and of the deceleration parameter for this model are represented in Figure~\ref{fig2_U}.

For $n=10^{-8}$, $m=1.05$, $\beta =0.02$, $\gamma =0.04$, and $\alpha \in (3.50,3.6)$, the model provides a good description of the observational data and of the Hubble function up to a redshift of $z\approx 2$ \cite{Cipriano:2023yhv}. Differences in the behavior of the deceleration parameter appear at both low and high redshifts. The model also predicts a lower matter density as compared to \mbox{$\Lambda$CDM \cite{Cipriano:2023yhv}.}

\subsection{Summary and Discussion}

Our preliminary results on the cosmology of $f\left(R,\mathcal{T}\right)$ gravity point towards the possibility that the considered theory and the cosmological models that can be built using it can provide qualitative and quantitative descriptions of the observational data once the optimal values of the model parameters have been obtained either by a trial-and-error method or by systematic fitting of the observational data. The models we discussed above can provide both a qualitative and quantitative descriptions of the observational Hubble data for a limited range of redshifts, and they can also reproduce the predictions of the $\Lambda$CDM model.

An interesting issue related to $f\left(R,\mathcal{T}\right)$ gravity and to the cosmological models based upon it  is the  possibility of at least alleviating, if not solving, the Hubble tension. One can see from the first Friedmann equation (Equation~(\ref{eq21})) that the present-day value of the Hubble function is fully determined by the present-day values of the energy density of matter, potential, and the two scalar fields
. These quantities satisfy the constraint of the \mbox{(closure relation)}
.
  \begin{equation}
  H_0^2\frac{\dot{\phi}(0)}{\phi(0)}H_0=\kappa ^2\frac{\rho (0)}{3\phi (0)}+\frac{1}{6}\frac{V(0)}{\phi (0)}+\frac{1}{6}\frac{\psi (0)}{\phi (0)}\rho ^2 (0).
  \end{equation}

By appropriately fixing the initial values of the matter energy density, the potential ($V$), the $\phi$ and $\psi$ fields, and their derivatives
, the present-day value of the Hubble function can be obtained in concordance with supernova data.

 Moreover, the function $H(z)$ can generally be represented as
 \begin{equation}
 H(z)=H_0 \Phi (z),
 \end{equation}
 where
 \begin{equation}
 \Phi (z)=\sqrt{\frac{r(z)+U(z)/2+\Psi (z)r^2(z)}{1-(1+z)\frac{d}{dz}\ln \phi (z)}}.
 \end{equation}

If the $\Lambda$CDM model is correct for all redshifts, then $H(0)=H_0\Phi(0)$, $\Phi (0) = 1$, and $H_0$ can be determined from the value obtained from supernova observations ($H_0=H_0^{sup}=H(0)/\Phi (0)$). But if the $\Lambda$CDM model is not correct for all $z$, then $\Phi (0)$ may be different from one or have values of the order of one that minimally deviate from the $\Lambda$CDM value
. On the other hand, as estimated from the early Universe, $\Phi (z)$ yields a different value for $H_0$. Let us assume that at a redshift of $z_r$, the function $\Phi (z)$ takes a value of $\Phi \left(z_r\right)=0.92$. Then, the present-day value of the Hubble function is $H_0=H\left(z_r\right)/\Phi \left(z_r\right)=H_0^{Planck}$. Therefore, we have
\begin{equation}
\frac{H_0^{Planck}}{H_0^{sup}}=\frac{H\left(z_r\right)}{H(0)}\times \frac{\Phi (0)}{\Phi\left(z_r\right)}=\zeta.
\end{equation}

For $\zeta =0.92$,  $H_0^{Planck}=\zeta\times H_0^{sup}=0.92\time 73=67.16$ km/s/Mpc \cite{Cipriano:2023yhv}. Numerical values of this order of magnitude for $\zeta$ can be obtained in the $f\left(R,T_{\protect\mu \protect\nu}T^{\protect\mu\protect\nu}\right)$ gravity theory for appropriate initial conditions of the two scalar fields. Of course, the form of the potential ($U$) must also be known. As an additional constraint on the theoretical models, one can also use the condition of the existence of small deviations only from the $\Lambda$CDM model, a condition that must be valid, especially in the period of formation of galaxies and stars, that is, during the reionization phase, and the cosmic dawn era, corresponding to $z <11$.

The behavior of the cosmological models of $f\left(R,\mathcal{T}\right)$ gravity theory is strongly influenced by the choice of the potential ($V$) of the two scalar fields. This choice is relatively arbitrary; however,  the form of the potential can be determined by comparison of the theoretical predictions with observational data. Constraints on the potential may be found from other astrophysical and cosmological observations, including the study of gravitational waves, black holes, gravitational lensing, or structure formation.

$f\left(R,\mathcal{T}\right)$ gravity provides an interesting approach to gravitational force, and it certainly has the potential to explain the yet-unknown physical aspects of cosmology. In this brief presentation, we have outlined the basic theoretical concepts and mathematical tools necessary to further develop the theory.

\section{Compact Objects in \boldmath{$f(R,\mathcal{T})$} Gravity}\label{sec:compactobj}

It is interesting to note that EMSG reduces to GR in vacuum, where differences only arise in the presence of an energy-momentum distribution. These effects become significant in regions of high curvature. Thus, deviations from GR are expected within compact objects. In this context, $f\left(R,\mathcal{T}\right)$ gravity has been studied in \cite{nari1,Akarsu:2018zxl,singh1,sharif1}, including black holes in~\cite{Khodadi:2022xtl,Roshan:2016mbt,Chen:2019dip,Rudra:2020rhs} and wormholes in \cite{Moraes:2017dbs,Sharif:2021ptz,ZeeshanGul:2023ysx,Rosa:2023guo,Rosa:2023tph}.
More specifically, the deviations from the predictions of GR due to EMSG are expected to become pronounced in the high-density cores of neutron stars \cite{Akarsu:2018zxl}. Here, the hydrostatic equilibrium equations in EMSG were derived and solved numerically to obtain the neutron star mass--radius relations for several realistic equations of state. The existing observational measurements of the masses and radii of neutron stars were then used to constrain the free parameter that characterizes the coupling between matter and spacetime in EMSG.

In the following sections, we outline several strategies to obtain black hole solutions and briefly review specific wormhole geometries obtained by several of the authors.

\subsection{Black Holes in EMSG Coupled with Electrodynamics
}
\label{subsec:blackh_hole}

In this section, we present the formalism of EMSG coupled with 
electrodynamics and consider a strategy to obtain black hole solutions.

Consider and action \eqref{eq1_1} expressed in the following form:
\begin{equation}
S=\frac{1}{2 \kappa^2} \int_{\Omega} \sqrt{-g} f(R, \mathcal{T}) d^4 x+\int_{\Omega} \sqrt{-g} \mathcal{L}_m^{(em)} \,d^4 x ,
\label{em_action_1}
\end{equation}
where $\mathcal{L}_m^{(em)}$ is the Maxwell Lagrangian given by
\begin{equation}
\mathcal{L}_m^{(em)} = -\frac{1}{4} F_{\mu\nu}F^{\mu\nu} = -\frac{1}{4}\mathcal{F},
\label{MaxLag}
\end{equation}
where $\mathcal{F} = F_{\mu\nu}F^{\mu\nu}$, and the electromagnetic tensor is defined as $F_{\mu\nu}=\partial_\mu A_\nu-\partial_\nu A_\mu$.
In the case of linear electrodynamics (Maxwell), the energy-momentum tensor is given by
\begin{equation}
T_{\mu\nu}^{(em)}=F_{\mu\sigma}{F_{\nu}}^\sigma-\frac{1}{4}g_{\mu\nu}\mathcal{F} \,.
\label{Tab_em}
\end{equation}

To obtain the gravitational field equation in the presence of the electromagnetic field. we vary Equation \eqref{em_action_1} with respect to the $g_{\mu\nu}$ metric, which yields
\begin{equation}
f_R R_{a b}-\frac{1}{2} g_{a b} f-\left(\nabla_a \nabla_b-g_{a b} \square\right) f_R=2\kappa^2 T_{a b}^{(em)}-f_{\mathcal{T}} \Theta_{a b}^{(em)} ,
\label{em_field_eq}
\end{equation}
where $\Theta_{\mu \nu}^{(em)}$ is given by
\begin{equation}
\Theta_{\mu \nu}^{(em)} = \frac{1}{16\pi^2}\left[4 F_{\gamma}^{~ \beta} F^{\gamma\alpha} F_{\alpha\mu} F_{\beta\nu}- {F^{\rho}}_{\mu} F_{\rho\nu} \mathcal{F}\right] .
\label{theta_em}
\end{equation}

Finally incorporating 
both Equations \eqref{Tab_em} and \eqref{theta_em} into Equation \eqref{em_field_eq} yields {the following}:
\begin{equation}
\begin{aligned}
f_R R_{a b}-\frac{1}{2} g_{a b} f&-\left(\nabla_a \nabla_b-g_{a b} \square\right) f_R= \\ &2\kappa^2 \left(F_{\mu\sigma}{F_{\nu}}^\sigma-\frac{1}{4}g_{\mu\nu}\mathcal{F}\right)-\frac{1}{16\pi^2}f_{\mathcal{T}}\left[4 F_{\gamma}^{~ \beta} F^{\gamma\alpha} F_{\alpha\mu} F_{\beta\nu}- {F^{\rho}}_{\mu} F_{\rho\nu} \mathcal{F}\right] ,
\label{field_eqations_em}
\end{aligned}
\end{equation}
which are the gravitational field equations assuming that the only matter in the Universe is described by $T_{\mu\nu}^{(em)}$. Specific solutions for $f(R,\mathcal{T})$ were obtained in
\cite{Chen:2019dip,Sharif:2022ibt, Sharif:2022prm, Sharif:2022cud, Pretel:2023avv}, and we refer the reader to these articles for more details.

In principle, one can also consider baryonic matter contributions ($\mathcal{L}_m^{(B)}$) to the matter total Lagrangian density so that $\mathcal{L}_m = \mathcal{L}_m^{(B)} + \mathcal{L}_m^{(em)}$, and the gravitational field equations generalize to
\begin{equation}
\begin{aligned}
f_R R_{a b}-\frac{1}{2} g_{a b} f&-\left(\nabla_a \nabla_b-g_{a b} \square\right) f_R =2\kappa^2\left\{T_{a b}^{(B)} + F_{\mu\sigma}{F_{\nu}}^\sigma-\frac{1}{4}g_{\mu\nu}\mathcal{F}\right\} \\ &-f_{\mathcal{T}}\left\{\Theta_{a b}^{(B)} + \frac{1}{16\pi^2}\left[4 F_{\gamma}^{~ \beta} F^{\gamma\alpha} F_{\alpha\mu} F_{\beta\nu}- {F^{\rho}}_{\mu} F_{\rho\nu} \mathcal{F}\right]\right\} \,.
\end{aligned}
\end{equation}

However, we do not explore this topic further at this stage, as it is still a work in progress, and developments are ongoing. We look forward to sharing more detailed insights once the work has reached a more finalized stage.

As one may expect, the standard Maxwell equations for EMSG are also different. It is more useful to use the Euler--Lagrange equations applied to the vector potential ($A_\mu$)\mbox{ given by}
\begin{equation}
\nabla_\mu\left[\frac{\partial\mathcal{L}}{\partial\left(\nabla_\mu A_\nu\right)}\right] - \frac{\partial\mathcal{L}}{\partial A_\nu} = 0 ,
\end{equation}
which yield
\begin{equation}
\begin{aligned}
\nabla_\mu\left[\frac{\partial}{\partial\left(\nabla_\mu A_\nu\right)}\left\{ \frac{1}{2 \kappa^2}\sqrt{-g} f(R, \mathcal{T}) + \sqrt{-g} \mathcal{L}_m^{(em)} \right\}\right] &= 0, \\
\nabla_\mu\left[\frac{\partial}{\partial\left(\nabla_\mu A_\nu\right)}\left\{ f(R, \mathcal{T}) + 2 \kappa^2 \mathcal{L}_m^{(em)} \right\}\right] &= 0.
\end{aligned}
\end{equation}

Using Equation (\ref{MaxLag}), these take the following form:
\begin{equation}
\begin{aligned}
\nabla_\mu\left[\frac{\partial f}{\partial\left(\nabla_\mu A_\nu\right)}\right] - \frac{\kappa^2}{2} \nabla_\mu\left[\frac{\partial \mathcal{F}}{\partial\left(\nabla_\mu A_\nu\right)} \right] &= 0 ,
	\\
\nabla_\mu\left[f_\mathcal{T}\, \left\{ \frac{1} {2 \pi^{2}} \left( F_{\gamma} {}^{\nu} F^{\mu\rho} F^{\gamma} {}_{\rho}-\frac{1} {4} F^{\mu\nu} \mathcal{F} \right) \right\}\right] - 2\kappa^2 \nabla_\mu F^{\mu\nu} &= 0, \\
\end{aligned}
\end{equation}
which yield the modified Maxwell equations for EMSG as follows:
\begin{equation}
\nabla_\mu F^{\mu\nu} =
\frac{1} {4 \kappa^2\pi^{2}}\nabla_\mu\left[f_\mathcal{T}\,  \left( F_{\gamma} {}^{\nu} F^{\mu\rho} F^{\gamma} {}_{\rho}-\frac{1} {4} F^{\mu\nu} \mathcal{F} \right)\right].
\end{equation}

One may also generalize the action (\ref{em_action_1}) to EMSG coupled with 
nonlinear electrodynamics (NLED), where the Maxwell Lagrangian ($\mathcal{L}_m^{(em)}$) is substituted with an NLED Lagrangian density ($\mathcal{L}_m^{(\rm N)}(\mathcal{F})$).
Varying the action (\ref{em_action_1}) with respect to the vector potential, one obtains the following equation of motion:
$\nabla_\mu ({\cal L}_{\cal F} \, F^{\mu\nu})=\sqrt{-g}\, \partial_\mu (\sqrt{-g} \, {\cal L}_{\cal F} \, F^{\mu\nu})=0$, where we denote ${\cal L}_{\cal F}=\partial \mathcal{L}_m^{(\rm N)}(\mathcal{F})/\partial {\cal F}$. Now, we can explore novel black hole solutions (including regular black holes and black bounces) with static and spherical symmetry by coupling EMSG with NLED, as carried out extensively in the recent literature \cite{Lobo:2020ffi,Rodrigues:2023vtm,Junior:2023qaq,Fabris:2023opv,Junior:2023ixh,Junior:2024xmm,Junior:2024vrv}. This is work in progress and will shortly be submitted for publication.

\subsection{Wormhole Geometries}\label{subsec:wh_metric}

In this section, we focus on wormholes, which are structures that connect two different spacetime manifolds or two distinct regions within the same spacetime manifold. Wormholes have been extensively studied within the framework of general relativity \mbox{(GR) \cite{morris1,Morris:1988tu,visser1,lemos1,Visser:2003yf,Kar:1995ss}.} A crucial requirement for a wormhole to be traversable is the flaring-out condition \cite{morris1}, which, when applied through the Einstein field equations, leads to the violation of the null energy condition (NEC) and, consequently, all other energy \mbox{conditions \cite{visser1,Hawking:1973uf,Sajadi:2016hko}.} Matter that violates the NEC is referred to as \textit{exotic
} matter, and it lacks physical relevance due to the scarcity of experimental evidence supporting \mbox{its existence.}

\textls[-10]{One way to avoid the use of exotic matter to sustain wormhole geometries is to consider these objects within the context of modified theories of gravity \cite{Sharif:2021ptz,ZeeshanGul:2023ysx,Rosa:2023guo,agnese1,nandi1,camera1,lobo1,garattini1,lobo2,garattini2,lobo3,MontelongoGarcia:2011ag,garattini3,myrzakulov1,lobo4}. In these theories, the additional curvature components of the gravitational sector preserve the geometry of the wormhole throat, making it traversable while keeping the matter components non-exotic. This outcome can be achieved in various modifications of GR, from $f\left(R\right)$ gravity and its extensions \cite{lobo5,capozziello1,rosa1,rosa2,rosalol,rosalol2,kull1} to couplings between curvature and matter \cite{garcia1,garcia2}, theories with additional fundamental fields \cite{harko1,anchordoqui1}, Einstein--Cartan gravity \cite{DiGrezia:2017daq}, Gauss--Bonnet \mbox{gravity \cite{bhawal1,dotti1,mehdizadeh1},} and brane-world scenarios \cite{bronnikov1,lobo6}. Regarding wormhole physics in the context of $f\left(R,\mathcal{T}\right)$ gravity, specific solutions were initially found using a Noether symmetry approach; however, these lacked physical relevance, as they violated the NEC \cite{Sharif:2021ptz,ZeeshanGul:2023ysx}. More recently, physically relevant traversable wormhole spacetimes that satisfy all energy conditions have been found \cite{Rosa:2023guo}, and these are the solutions presented in this section.}

\subsubsection{Metric and Field Equations}

We start by considering a static and spherically symmetric traversable wormhole metric, which can be written in spherical coordinates $\left(t,r,\theta,\varphi\right)$ as
\begin{equation}\label{def_metric}
ds^2=-e^{\zeta(r)}dt^2+\left[1-\frac{b(r)}{r}\right]^{-1}dr^2+r^2d\Omega^2,
\end{equation}
where $\zeta(r)$ is the redshift function, $b(r)$ is the shape function, and $d\Omega^2=d\theta^2+\sin^2\theta d\varphi^2$ is the solid-angle surface element. The necessary conditions to ensure the traversability of the wormhole are the finiteness of the redshift function throughout the entire spacetime, i.e., $|\zeta(r)|<\infty$, to avoid event horizons, therefore allowing an observer to cross the wormhole's interior without being trapped, and the flaring-out condition at the wormhole throat ($r=r_0$)\endnote {The flaring-out condition in the neighborhood of the throat takes the form of $(b - b'r)/b^2 > 0$ \cite{morris1}
	.}, which is given by the following boundary conditions:
\begin{equation}\label{def_flaring}
b(r_0)=r_0, \qquad b'(r_0)<1.
\end{equation}

Given these requirements, we consider the following two broad families of solutions for the functions of $\zeta(r)$ and $b(r)$:
\begin{equation}\label{zbfunctions}
\zeta(r) = \zeta_0 \left(\frac{r_0}{r}\right)^\alpha, \qquad b(r) = r_0 \left(\frac{r_0}{r}\right)^\beta,
\end{equation}
\textls[-25]{where $\zeta_0$ is an arbitrary constant to be specified and $\alpha$ and $\beta$ are arbitrary positive exponents.}

Regarding the matter sector, we assume that the matter distribution is well described by an anisotropic perfect fluid. Thus, the energy-momentum tensor ($T_{\mu \nu}$) is given by
\begin{equation}\label{def_matter}
T_\mu^\nu = \text{diag}(-\rho, p_r, p_t, p_t),
\end{equation}
where $\rho \equiv \rho(r)$ represents the energy density, $p_r \equiv p_r(r)$ denotes the radial pressure, and $p_t \equiv p_t(r)$ is the tangential pressure, which depend only on the radial coordinate ($r$) to maintain the spherical symmetry of the wormhole. Under these assumptions, the matter Lagrangian is given by $\mathcal{L}_m = \frac{1}{3}(p_r + 2p_t)$, and consequently, the auxiliary tensor ($\Theta_{\mu \nu}$) results in
\begin{equation}\label{def_theta_2}
\Theta_{\mu \nu} = -\frac{2}{3}(p_r + 2p_t) \left(T_{\mu \nu} - \frac{1}{2} g_{\mu \nu} T\right) - T T_{\mu \nu} + 2 T_{\mu}^\sigma T_{\sigma \nu}.
\end{equation}

We developed the methodology to obtain wormhole solutions by considering that the function of $f\left(R,\mathcal T\right)$ is separable and linear in both $R$ and $\mathcal{T}$, that is,
\begin{equation}
    f(R,\mathcal{T}) = R + \gamma \mathcal{T},
\end{equation}
and briefly generalized it to cases where the function includes highers powers of $\mathcal{T}$ along the way, which can be seen in detail in \cite{Rosa:2023guo}.

Under this assumption, the field equations in Equation \eqref{geo_field_simple} and the conservation equation in Equation \eqref{geo_conservation_simple} become
\begin{equation}\label{field}
G_{\mu \nu} = 8\pi T_{\mu \nu} - \gamma \left( \Theta_{\mu \nu} - \frac{1}{2} g_{\mu \nu} \mathcal{T} \right),
\end{equation}
\begin{equation}\label{conservation}
8\pi \nabla_b T^{\mu \nu} = \gamma \nabla_b \left( \Theta^{\mu \nu} - \frac{1}{2} g^{\mu \nu} \mathcal{T} \right),
\end{equation}
\textls[-15]{respectively. Now, with the assumptions from Section \ref{subsec:wh_metric}, the field equations in \mbox{Equation \eqref{field}} yield three independent components, which are expressed as follows:}
\begin{eqnarray}\label{eqrho}
8\pi\rho &=& \frac{\gamma}{6}\left(p_r^2-2p_t^2-3\rho^2-8p_rp_t-8p_r\rho-16p_t\rho\right) - \frac{\beta}{r^2}\left(\frac{r_0}{r}\right)^{\beta+1},
\end{eqnarray}
\begin{eqnarray}\label{eqpr}
8\pi p_r &=& \frac{\gamma}{6}\left(p_r^2+2p_t^2-3\rho^2-12p_rp_t+4p_r\rho-4p_t\rho\right)
	\nonumber \\
&& \hspace{-0.5cm} -\frac{1}{r^2}\left(\frac{r_0}{r}\right)^{\beta+1}-\frac{\alpha\zeta_0}{r^2}\left(\frac{r_0}{r}\right)^\alpha\left[1-\left(\frac{r_0}{r}\right)^{\beta+1}\right],
\end{eqnarray}
\begin{eqnarray}\label{eqpt}
8\pi p_t &=&-\frac{\gamma}{6}\left(p_r^2+6p_t^2+3\rho^2+2p_rp_t+2p_r\rho-2p_t\rho\right)
	\nonumber \\
&& +\frac{1+\beta}{2r^2}\left(\frac{r_0}{r}\right)^{\beta+1}
+\frac{\alpha^2\zeta_0^2}{4r^2}\left(\frac{r_0}{r}\right)^{2\alpha}\left[1-\left(\frac{r_0}{r}\right)^{\beta+1}\right]
\nonumber \\
&&	+\frac{\alpha\zeta_0}{4r^2}\left(\frac{r_0}{r}\right)^\alpha\left[2\alpha-\left(1+2\alpha+\beta\right)\left(\frac{r_0}{r}\right)^{\beta+1}\right].
\end{eqnarray}

Equations \eqref{eqrho}--\eqref{eqpt} form a system of three equations with three unknowns, namely $\rho$, $p_r$, and $p_t$. Each of these equations is quadratic in its respective unknown, implying that the system can potentially yield up to eight independent solutions. However, the nature of these solutions may vary, including the possibility of some solutions being complex, depending on the specific values of the parameters involved.

\subsubsection{Wormhole Solutions}\label{subsec:whsolutions}

Due to the complexity of the system of Equations \eqref{eqrho}--\eqref{eqpt}, analytical solutions for $\rho$, $p_r$, and $p_t$ cannot be obtained explicitly, even when considering specific choices for the free parameters of $r_0$, $\alpha$, $\beta$, $\gamma$, and $\zeta_0$. However, analytical solutions for these quantities can still be obtained using a recursive approach. We begin by choosing specific values for the free parameters and solve the system to find $\rho(r_0)$, $p_r(r_0)$, and $p_t(r_0)$ at an initial radius of $r = r_0$. This provides an initial set of values $\{\rho_0^i, p_{r0}^i, p_{t0}^i\}$ for $i \in \{1, \ldots, 8\}$ corresponding to the eight independent solutions of the system. Then, for each solution, the radius ($r$) is incremented in small steps, such as $r_{n+1} = r_n + \epsilon$, where $\epsilon$ is a small increment, to compute $\rho(r_{n+1})$, $p_r(r_{n+1})$, and $p_t(r_{n+1})$. By recursively applying this process up to a sufficiently large radius ($r$), one can analytically determine the behavior of the solutions. The same method is applicable to any form of the function of $f\left(R,\mathcal T\right)$ containing higher powers of $\mathcal{T}$ and no crossed $R\mathcal T$ terms, given that the relation between the matter fields remains algebraic, although the set of solutions is larger.

Among the obtained solutions, we focus only on those that are astrophysically relevant, meaning that their matter components satisfy the energy conditions (see Equation \eqref{def:ECs}). For a diagonal energy-momentum tensor ($T_{\mu \nu}$) as given in Equation \eqref{def_matter}, these energy conditions are expressed as follows:
\begin{equation}
    \rho+p_r \geq 0,\quad \rho+p_t \geq 0,\quad \rho\geq0,\quad \rho+p_r+2p_t \geq 0,\quad \rho \geq |p_r|, \quad \rho \geq |p_t|.
\end{equation}

From the set of eight solutions for the matter quantities within the theory, solutions that violate any of these energy conditions are discarded, and only those that satisfy all of the energy conditions are considered.

\begin{figure}[H]
  \centering
  \begin{subfigure}[b]{0.48\textwidth}
    \includegraphics[width=\textwidth]{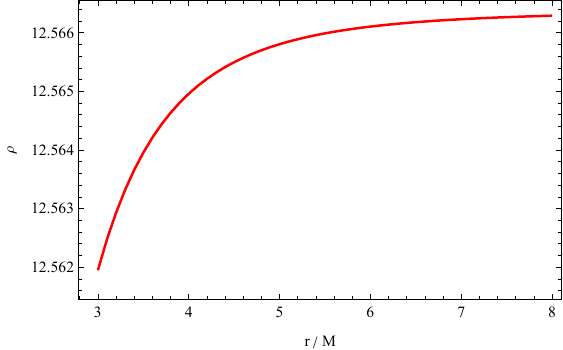}
  \end{subfigure}
  \begin{subfigure}[b]{0.48\textwidth}
    \includegraphics[width=\textwidth]{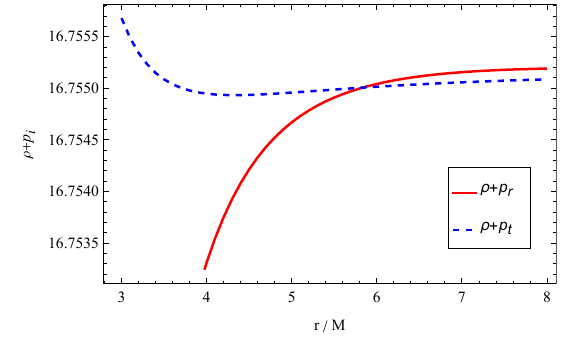}
  \end{subfigure}
  \begin{subfigure}[b]{0.48\textwidth}
    \includegraphics[width=\textwidth]{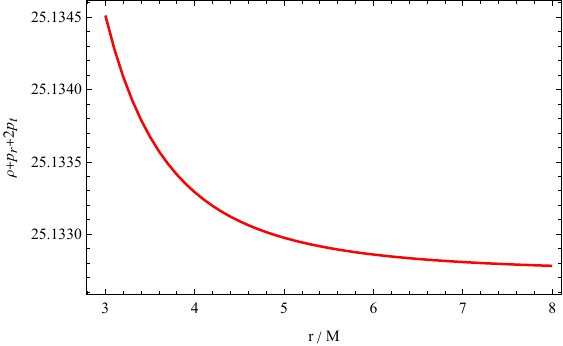}
  \end{subfigure}
  \begin{subfigure}[b]{0.48\textwidth}
    \includegraphics[width=\textwidth]{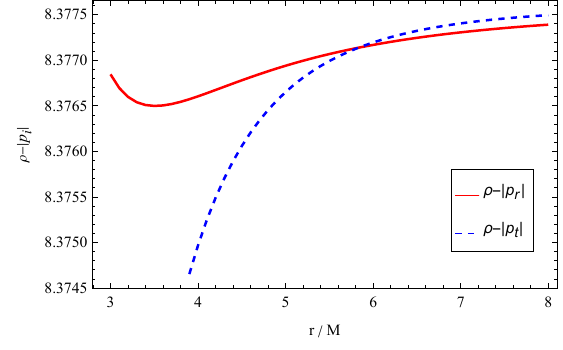}
  \end{subfigure}
  \caption{Considering the parameters $\alpha = \beta = -\gamma = 1$, $r_0 = 3M$, and $\zeta_0 = -\frac{6}{5}$, the matter density $\rho$ and the combinations of $\rho + p_r$, $\rho + p_t$, $\rho + p_r + 2p_t$, $\rho - |p_r|$, and $\rho - |p_t|$ are presented.}
  \label{fig:solution}
\end{figure}

As a specific example, let us consider the parameter combination of $\alpha = \beta = -\gamma = 1$, $r_0 = 3M$, and $\zeta_0 = -\frac{6}{5}$.\endnote{
	While the choice of $\zeta_0$ at this point is somewhat arbitrary, we chose this particular value for reasons that we clarify in the subsequent section. Various other values of $\zeta_0$, including positive values, would yield qualitatively similar solutions.} The matter density ($\rho$), as well as the combinations of $\rho + p_r$, $\rho + p_t$, $\rho + p_r + 2p_t$, $\rho - |p_r|$, and $\rho - |p_t|$, are plotted in Figure \ref{fig:solution} for the solution that satisfies all previously mentioned energy conditions.
	
	Similar to findings in linear $f\left(R,T\right)$ gravity \cite{kull1}, from the example above and from all the solutions satisfying the energy conditions that we have found for the linear case of the function of $f(R,\mathcal{T})$, we concluded that solutions satisfying the energy conditions throughout the entire spacetime could only be obtained by taking negative values of $\gamma$, as shown in Figure \ref{fig:solution}. However, when considering higher powers of $\mathcal{T}$, i.e., \mbox{$f(R,\mathcal{T})=R+\gamma \mathcal{T}+\sigma \mathcal{T}^n$}, we found that solutions satisfying all the energy conditions can be achieved even for positive values of $\gamma$, provided that $\sigma$ remains negative. Furthermore, for both cases, a notable characteristic stood out, which is the fact that despite the spacetime being asymptotically flat, the matter components do not vanish across the entire radial coordinate range, indicating non-localized behavior, as seen in Figure \ref{fig:solution}. Therefore, to improve the physical relevance of these solutions, we must match them with an exterior vacuum spacetime with 
a finite radius. We address this issue in the next section.

\subsubsection{Junction Conditions and Matching}\label{sec:matching}

\textls[10]{To obtain physically relevant spacetime solutions that describe localized objects, one has to employ the junction conditions to perform matching between the interior and exterior spacetimes at a finite radius. In the context of GR, these equations were derived long ago \cite{israel1} and have been applied in various astrophysical scenarios, such as in the analysis of traversable wormholes \cite{visser2,visser3,Lobo:2004rp,Lobo:2004id,Lobo:2005us,Lobo:2005yv}, fluid stars \cite{schwarzschild1,rosafluid,rosafluid2}, and gravitational collapse \cite{oppenheimer1,rosa112}. Since these conditions are theory-dependent, several studies have analyzed them in the context of different modified theories of gravity (we refer to \cite{Rosa:2023tph} for a review), from $f(R)$ gravity and its extensions \cite{senovilla1,Vignolo:2018eco,Reina:2015gxa,Deruelle:2007pt,Olmo:2020fri,rosafrt,rosafrt2} to theories with additional fundamental fields \cite{suffern,Barrabes:1997kk,Padilla:2012ze,Casado-Turrion:2023omz}; metric-affine gravity \cite{delaCruz-Dombriz:2014zaa,Arkuszewski:1975fz,amacias}; and, more recently, in the context of $f\left(R,\mathcal{T}\right)$ gravity \cite{Rosa:2023guo}. In this section, we present the latter work, including both the derivation of the junction conditions and the matching performed between the interior wormhole spacetime and an exterior vacuum spacetime
	.}

\paragraph{Junction Conditions 
}
\label{subsec:jc}

We begin the derivation of the junction conditions by first considering the linear case of the function, that is, $f(R,\mathcal{T}) = R + \gamma \mathcal{T}$, and later briefly extend the analysis to cases where the function includes higher powers of $\mathcal{T}$, namely $f(R,\mathcal{T}) = R + \gamma \mathcal{T} + \sigma \mathcal{T}^n$.

Let us consider a spacetime manifold ($\Omega$) consisting of two distinct and complementary regions ($\Omega^\pm$), each described by metric tensors ($g_{\alpha \beta}^\pm$) in their respective coordinate systems ($x^\alpha_\pm$). We define $\Omega^+$ as the exterior spacetime and $\Omega^-$ as the interior spacetime. The interface between $\Omega^\pm$ is a three-dimensional hypersurface ($\Sigma$) with a metric ($h_{ab}$) expressed in terms of a coordinate system ($y^a$), where Latin indices exclude the direction orthogonal to $\Sigma$. The projection tensors from the four-dimensional spacetime ($\Omega$) onto the hypersurface ($\Sigma$) are expressed as $e^\alpha_a = \partial x^\alpha / \partial y^a$. The normal vector on $\Sigma$ is defined as $n_\alpha = \epsilon \partial_\alpha l$, where $l$ is the affine parameter along geodesics orthogonal to $\Sigma$, and $\epsilon$ takes values $1$, $-1$, and $0$ for space-like, time-like, and null geodesic congruences, respectively. By construction, note that $n^\alpha e_\alpha^a = 0$ holds true. Using this notation, the induced metric ($h_{ab}$) and the extrinsic curvature ($K_{ab}$) of the hypersurface ($\Sigma$) are expressed as
\begin{equation}\label{defhK}
h_{ab}=e^\alpha_a e^\beta_b g_{\alpha \beta}, \qquad K_{ab}=e^\alpha_a e^\beta_b\nabla_\alpha n_\beta.
\end{equation}

To derive the junction conditions, we employ the distribution formalism. In this approach, any quantity ($X$) and its derivative ($\nabla_\alpha X$) are expressed in terms of distribution functions \mbox{as follows}:
\begin{equation}\label{distX}
X=X^+\theta\left(l\right)+X^-\theta\left(-l\right),
\end{equation}
\begin{equation}\label{distdX}
\nabla_\alpha X=\nabla_\alpha X^+\theta\left(l\right)+\nabla_\alpha X^-\theta\left(-l\right)+\epsilon n_\alpha\left[X\right]\delta\left(l\right),
\end{equation}
where $X^\pm$ represents the quantity ($X$) in the spacetimes ($\Omega^\pm$)
; $\theta\left(l\right)$ is the Heaviside distribution function defined as $\theta\left(l\right)=0$ for $l<0$, $\theta\left(l\right)=1$ for $l>0$, and $\theta\left(l\right)=\frac{1}{2}$ for $l=0$; $\delta\left(l\right)=\partial_l\theta\left(l\right)$ is the Dirac delta distribution; and we have introduce the definition of $\left[X\right]$ to represent the jump of $X$ across $\Sigma$, that is,

\begin{equation}
\left[X\right]=X^+|_\Sigma - X^-|_\Sigma.
\end{equation}

If the quantity ($X$) is continuous across $\Sigma$, then $\left[X\right]=0$. Additionally, notice that by definition, we have $\left[n^\alpha\right]=\left[e^\alpha_a\right]=0$.

To obtain the junction conditions, we start by writing every quantity appearing in the field equations (see Equation \eqref{field}) in the distribution formalism. Let us begin with the metric  of $g_{\alpha \beta}$, which, in the distributional formalism, becomes
\begin{equation}\label{eq:def_metric}
g_{\alpha \beta}=g_{\alpha \beta}^+\theta\left(l\right)+g_{\alpha \beta}^-\theta\left(-l\right).
\end{equation}

\textls[10]{With the metric expression in Equation \eqref{eq:def_metric}, we proceed to compute the Christoffel symbols ($\Gamma^\gamma_{\alpha \beta}$) associated with the metric of $g_{\alpha \beta}$, which involves calculating the derivatives ($\partial_\lambda g_{\alpha \beta}$). Using \mbox{Equation \eqref{distdX},} these derivatives are expressed as $\partial_\lambda g_{\alpha \beta} = \partial_\lambda g_{\alpha \beta}^+ \theta(l) + \partial_\lambda g_{\alpha \beta}^- \theta(-l) + \epsilon n_\lambda [g_{\alpha \beta}] \delta(l)$. However, the presence of the term proportional to $\delta(l)$ poses a challenge when defining the Riemann tensor ($R^\alpha_{\beta \lambda \gamma}$) in the distributional formalism. Since the Riemann tensor ($R^\alpha_{\beta \lambda \gamma}$) depends on products between the Christoffel symbols ($\Gamma^\gamma_{\alpha \beta}$), it results in terms proportional to $\delta^2(l)$, which are singular in the distributional formalism. The junction conditions are introduced precisely to eliminate such singular terms from the field equations when expressed in the distribution formalism. Thus, to remove the singular term arising in the Riemann tensor ($R^\alpha_{\beta \lambda \gamma}$), we impose that the metric of $g_{\alpha \beta}$ must be continuous across $\Sigma$, i.e., $\left[g_{\alpha \beta}\right]=0$. Considering $\left[e^\alpha_a\right]=0$, this condition can be expressed in a coordinate-independent manner by projecting both indices onto the hypersurface ($\Sigma$), resulting in}
\begin{equation}\label{junction1}
\left[h_{ab}\right]=0.
\end{equation}

Equation \eqref{junction1} represents the first junction condition, stipulating that the induced metric at $\Sigma$ must be continuous. Using this result, the derivatives of $g_{\alpha \beta}$ become
\begin{equation}\label{dmetric}
\partial_\lambda g_{\alpha \beta}=\partial_\lambda g_{\alpha \beta}^+\theta\left(l\right)+\partial_\lambda g_{\alpha \beta}^-\theta\left(-l\right).
\end{equation}

We can now compute the Christoffel symbols in the distributional formalism and subsequently derive the Riemann tensor, along with its contractions, namely the Ricci tensor ($R_{ab}$) and the Ricci scalar ($R$), which are well-defined. These quantities are expressed as follows:
\begin{equation}\label{eq:dist_Rab}
R_{\alpha \beta}=R^+_{\alpha \beta}\theta\left(l\right)+R^-_{\alpha \beta}\theta\left(-l\right)-\left(\epsilon\left[K_{ab}\right]e^a_\alpha e^b_\beta+n_\alpha n_\beta \left[K\right]\right)\delta\left(l\right),
\end{equation}
\begin{equation}\label{eq:dist_R}
R=R^+\theta\left(l\right)+R^-\theta\left(-l\right)-2\epsilon\left[K\right]\delta\left(l\right),
\end{equation}
where $K=h^{ab}K_{ab}$ represents the trace of the extrinsic curvature.

Considering the matter sector, it is useful to link the existence of a thin shell at the hypersurface ($\Sigma$) with any terms proportional to $\delta\left(l\right)$ in the gravitational sector of the modified field equations. Therefore, in the distribution formalism, we express the energy-momentum tensor as follows:
\begin{equation}\label{eq:dist_tab}
T_{\alpha \beta}=T_{\alpha \beta}^+\theta\left(l\right)+T_{\alpha \beta}^-\theta\left(-l\right)+S_{\alpha \beta}\delta\left(l\right),
\end{equation}
where $S_{\alpha \beta}=S_{ab}e^a_\alpha e^b_\beta$, and $S_{ab}$ denotes the three-dimensional energy-momentum tensor of the thin shell. To obtain the scalar ($\mathcal{T}$) in the distributional formalism, one can simply contract $T_{\alpha \beta}$ with itself using the previous expression \eqref{eq:dist_tab}, resulting in
\begin{equation}\label{eq:dist_tab2}
\mathcal T=\mathcal T^+\theta\left(l\right)+\mathcal T^-\theta\left(-l\right)+\left(T_{\alpha \beta}^++T_{\alpha \beta}^-\right)S^{\alpha \beta}\delta\left(l\right)+S_{\alpha \beta}S^{\alpha \beta}\delta^2\left(l\right),
\end{equation}

Note that the term proportional to $\delta^2\left(l\right)$ in Equation \eqref{eq:dist_tab2} is singular in the distributional formalism and needs to be removed. The only possible approach to eliminate the term proportional to $\delta^2\left(l\right)$ in $\mathcal{T}$ is to require the energy-momentum tensor of the thin shell to vanish, that is,
\begin{equation}\label{junctionS}
    S_{\alpha \beta}=0.
\end{equation}

When Equation \eqref{junctionS} holds true, the matching is referred to as a smooth matching. Unlike in several other theories of gravity, where this kind of matching is considered a specific case of a broader thin-shell matching, in $f\left(R,\mathcal{T}\right)$ gravity, smooth matching between two spacetimes is the only method allowed to maintain the regularity of the action.

\textls[15]{Based on the definitions provided above and under the restriction in \mbox{Equation \eqref{junctionS}}, we project the field equations for the function of $f\left(R,\mathcal{T}\right)= R + \gamma \mathcal{T}$ that are presented in \mbox{Equation \eqref{field}} onto the hypersurface ($\Sigma$) with $e^\alpha_a e^\beta_b$, for which they take the form of \mbox{\textls[-5]{$\left[K_{ab}\right] - \left[K\right]h_{ab}=0$.}} Taking the trace of this result with $h^{ab}$ yields $\left[K\right]=0$. Substituting this back into the original equation results in}
\begin{equation}
\left[K_{ab}\right]=0.
\end{equation}

Therefore, the second junction condition implies that the extrinsic curvature ($K_{ab}$) must be continuous across $\Sigma$.

In summary, the matching between two spacetimes in linear $f\left(R,\mathcal T\right)$ gravity must always be smooth, meaning it must occur without a thin shell. The two junction conditions that must be satisfied are the same as in GR---the induced metric ($h_{ab}$) and the extrinsic curvature $K_{ab}$ must be continuous across the hypersurface ($\Sigma$) as follows:
\begin{equation}\label{junction}
\left[h_{ab}\right]=0, \qquad \left[K_{ab}\right]=0.
\end{equation}

When considering a higher-order power of $\mathcal{T}$ in the function, the field equations include products in the form of $\mathcal{T}^{n-1}\Theta_{ab}$ and powers of $\mathcal{T}^n$. For the linear case, we have shown that for $\mathcal{T}$ to be well-defined in the distributional formalism, the matching must be smooth, meaning there is no thin shell, i.e., $S_{\alpha \beta}=0$. This requirement ensures that $\mathcal{T}$ and the auxiliary tensor ($\Theta_{\alpha \beta}$), as defined in Equation \eqref{def_theta_2}, are completely regular, featuring only terms proportional to $\theta\left(l\right)$ and no terms proportional to $\delta\left(l\right)$. As a result, the products of $\mathcal{T}$ and $\Theta_{\alpha \beta}$, as well as the powers of $\mathcal{T}^n$, maintain this regularity. Therefore, introducing a higher-order power law of $\mathcal{T}$ in the function of $f(R,\mathcal{T})$ does not lead to additional junction conditions, as long as crossed $R\mathcal T$ terms are absent.

\paragraph{Matching with an Exterior Vacuum}

\textls[-35]{Let us now apply the junction conditions derived in \mbox{the previous {Section Junction Conditions}} 
 to match the interior wormhole spacetime with an exterior spherically symmetric and static vacuum solution. The metrics for the interior and exterior spacetimes are given by}
\begin{equation}\label{metrici}
ds_-^2=-C e^{\zeta_0\left(\frac{r_0}{r}\right)^\alpha}dt^2+\left[1-\left(\frac{r_0}{r}\right)^{\beta+1}\right]^{-1}dr^2+r^2d\Omega^2,
\end{equation}
\begin{equation}\label{metrice}
ds_+^2=-\left(1-\frac{2M}{r}\right)dt^2+\left(1-\frac{2M}{r}\right)^{-1}dr^2+r^2d\Omega^2,
\end{equation}
respectively. The metric in Equation \eqref{metrici} is derived from Equation \eqref{def_metric} using the suggested ansatz for the redshift and shape functions, as given in Equation \eqref{zbfunctions}, and $C$ is introduced for convenience to ensure that the time coordinates in both the interior and exterior metrics coincide. The metric in Equation \eqref{metrice} corresponds to the Schwarzschild solution with a mass of $M$ \cite{Schwarzschild:1916uq}
.

The analysis becomes more convenient if we begin with the second junction condition (Equation \eqref{junction}). Due to the spherical symmetry of the metrics under consideration, the extrinsic curvatures ($K_{ab}^\pm$) have only two independent components, namely $K_{00}$ \mbox{and $K_{\theta\theta} = K_{\phi\phi} \sin^2\theta$.} Thus, we derive two independent constraints on the matching, that is, $\left[K_{00}\right] = 0$ and $\left[K_{\theta\theta}\right] = 0$, which are given by
\begin{equation}\label{cond1}
\frac{\alpha\zeta_0}{2}\left(\frac{r_0}{r}\right)^\alpha\sqrt{1-\left(\frac{r_0}{r}\right)^{\beta+1}}+\frac{M}{r}\sqrt{\frac{r}{r-2M}}=0,
\end{equation}
\begin{equation}\label{cond2}
\sqrt{1-\left(\frac{r_0}{r}\right)^{\beta+1}}=\sqrt{1-\frac{2M}{r}},
\end{equation}
respectively. Solving the second of these conditions for the radius ($r$) using Equation \eqref{cond2}
, we obtain unique real solutions for $M>0$ and $r_0>0$, given by
\begin{equation}\label{rsigma}
r_\Sigma=\left(2M\right)^{-\frac{1}{\beta}}\left(r_0\right)^{1+\frac{1}{\beta}},
\end{equation}
which corresponds to the radius ($r_\Sigma$) at which the matching must be performed. The radius ($r_\Sigma$) must satisfy $r_\Sigma > 2M$ to prevent event horizons in the complete wormhole spacetime, which implies that the throat radius ($r_0$) should also satisfy $r_0>2M$ for any $\beta \geq 1$. Introducing the obtained solutions for $r_\Sigma$ back into the first condition (\mbox{Equation \eqref{cond1}}), we can solve it with respect to the value of $\zeta_0$, for which matching at the radius ($r=r_\Sigma$) is possible. Doing so, the expression for $\zeta_0$ becomes
\begin{equation}\label{zsigma}
\zeta_0=\frac{\left(2M\right)^{\frac{1-\alpha+\beta}{\beta}}\left(r_0\right)^{\frac{\alpha}{\beta}}}{\alpha\left[\left(2M\right)^{1+\frac{1}{\beta}}-\left(r_0\right)^{1+\frac{1}{\beta}}\right]}.
\end{equation}
Given that $r_0>2M$, as derived from the previous constraint, it follows that for any $\alpha\geq 1$ and $\beta\geq 1$, we have $\zeta_0<0$. This is expected because negative values of $\zeta_0$ ensure that the derivative of $g_{00}$ maintains consistent signs in both the interior and exterior metrics, which is necessary for a smooth matching.

Let us now consider the first junction condition given in Equation \eqref{junction}. Since the angular components of the metrics in Equations \eqref{metrici} and \eqref{metrice} coincide, the angular parts of the induced metric ($h_{\alpha\beta}$) are straightforwardly continuous. Therefore, the continuity condition ($\left[h_{00}\right]=0$) is analyzed independently and takes the following form:
\begin{equation}\label{cond3}
C e^{\zeta_0\left(\frac{r_0}{r}\right)^\alpha}=\left(1-\frac{2M}{r}\right).
\end{equation}

\textls[15]{We take the results derived from the second junction condition, namely the radius ($r_\Sigma$) from\mbox{ Equation \eqref{rsigma}}, where the matching occurs, and the corresponding value of $\zeta_0$ from \mbox{Equation \eqref{zsigma},} and substitute them into Equation \eqref{cond3} and solve for the constant \mbox{($C$), obtaining}}
\begin{equation}\label{csigma}
C=\left[1-\left(\frac{2M}{r_0}\right)^{1+\frac{1}{\beta}}\right]e^{-\alpha\left[\left(\frac{r_0}{2M}\right)^{1+\frac{1}{\beta}}-1\right]}.
\end{equation}

Since $r_0>2M$, the constant ($C$) remains strictly positive, regardless of the values of $\alpha \geq 1$ and $\beta \geq 1$, therefore preserving the correct metric signature.

To summarize, given $r_0>2M$, $\alpha\geq 1$, 
 and $\beta\geq 1$, the second junction condition ($\left[K_{\alpha\beta}\right]=0$) determines the radius ($r_\Sigma$) for the matching (see Equation \eqref{rsigma}) and the corresponding value of $\zeta_0$ (see Equation \eqref{zsigma}). Meanwhile, the first junction condition ($\left[h_{\alpha\beta}\right]$) sets the value of the constant ($C$), ensuring the continuity of the complete spacetime metric (see Equation \eqref{csigma}).

\begin{figure}[H]
    \begin{minipage}[b]{0.48\textwidth}
        \includegraphics[scale=0.7]{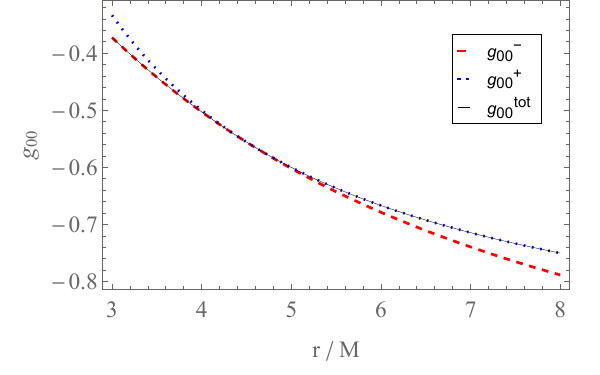}
    \end{minipage}
    \begin{minipage}[b]{0.48\textwidth}
        \includegraphics[scale=0.7]{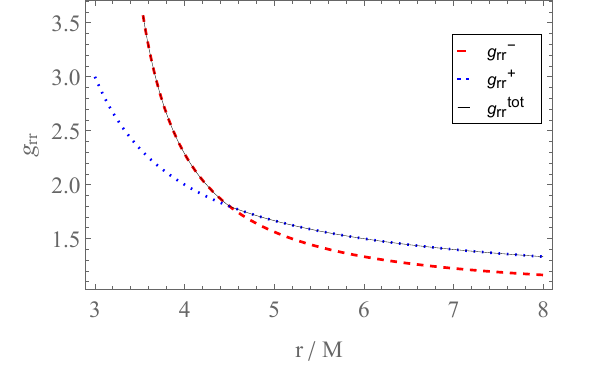}
    \end{minipage}
    \caption{The interior wormhole spacetime in Equation \eqref{metrici} (red dashed curve) and the exterior Schwarzschild spacetime in Equation \eqref{metrice} (blue dotted curve) for $\beta = 1$, $r_0 = 3M$, and $\alpha = 1$. The thin black line represents the solution obtained by matching the interior and exterior solutions at $r = r_\Sigma$ for the $g_{00}^{\text{tot}}$ component (\textbf{left panel}) and $g_{rr}^{\text{tot}}$ component (\textbf{right panel}).}
    \label{fig:matching}
\end{figure}

Let us provide the following example. Consider the specific case where $r_0=3M$, $\alpha=1$, and $\beta=1$. With these parameters, Equation \eqref{rsigma} gives the matching radius at $r_\Sigma=\frac{9}{2}M$, Equation \eqref{zsigma} gives $\zeta_0=-\frac{6}{5}$, and Equation \eqref{csigma} results in $C=\frac{5}{9}e^{\frac{4}{5}}$. The $g_{00}$ components of the interior, exterior, and matched metrics are plotted in the left panel of Figure \ref{fig:matching}. Notice that the $g_{00}$ component of the metric transitions smoothly from the interior to the exterior metric, ensuring the continuity of both the induced metric and the \mbox{extrinsic curvature.}

Analyzing the radial component of the $g_{rr}$ metric
, as shown in the right panel of \mbox{Figure \ref{fig:matching},} we observe that even though the radial component is not directly constrained by the junction conditions, as both the induced metric ($h_{ab}$) and the extrinsic curvature ($K_{ab}$) are three-dimensional tensors on the hypersurface ($\Sigma$), we find that $g_{rr}$ remains continuous, although not differentiable at $r = r_\Sigma$. This continuity of $g_{rr}$ is anticipated when considering its dependence on the mass function within a spherical hypersurface of radius $r$ ($m\left(r\right)$), as given by $g_{rr} = \left( 1 - \frac{2m(r)}{r} \right)^{-1}$, from which we obtain $m(r) = \frac{r_0}{2} \left( \frac{r_0}{r} \right)^\beta$ (see \mbox{Equations 
 \eqref{zbfunctions} and \eqref{metrici}).} Indeed, since the matching between the interior and exterior spacetimes are smooth, i.e., lacking a thin shell
 , it follows that the mass function ($m(r)$) is continuous at $r_\Sigma$, which, in turn, ensures the continuity of the $g_{rr}$ component of the metric.

\subsubsection{Summary}

In conclusion, this analysis of traversable wormhole spacetimes within the framework of $f\left(R,\mathcal T\right)$ gravity for a linear model on both $R$ and $\mathcal T$ demonstrates the existence of numerous traversable wormhole solutions that satisfy all energy conditions, therefore possessing significant physical relevance. An intriguing aspect of the obtained solutions is that, although the spacetime metrics are asymptotically flat, the $f\left(R,\mathcal T\right)$ theory permits matter distributions that are not asymptotically vacuum and, therefore, not localized. However, the localization of the solutions is achievable via the use of the theory's junction conditions. These conditions were derived, revealing that only smooth matching is allowed under this theory, as the scalar ($\mathcal T$) becomes singular in the presence of a thin shell. Consequently, the junction conditions for a smooth matching reduce to those of GR, i.e., the continuity of the induced metric and extrinsic curvature at the hypersurface separating the interior and exterior spacetime regions. Upon performing this matching, localized wormhole solutions satisfying all energy conditions throughout the entire spacetime were obtained, highlighting their particular astrophysical relevance. Moreover, the methods introduced in this work can be straightforwardly generalized to more complex dependencies of the function of $f\left(R,\mathcal T\right)$ in $\mathcal T$, provided that crossed terms between $R$ and $\mathcal T$ are absent. Furthermore, due to the requirement that the matching in this theory be smooth, the absence of these crossed terms also ensures that no additional junction conditions arise, allowing for effective localization of the solutions obtained in the same manner as in the linear \linebreak($\mathcal T$) 
counterpart.

\section{Conclusions}\label{sec:conclusion}

The exploration of EMSG presents a novel framework for extending our understanding of gravitational phenomena beyond the limits of GR. This comprehensive study encompasses the theoretical formalism, thermodynamic implications, cosmological models, and the nature of compact objects within EMSG, providing a robust foundation for future research and potential observational confirmations. In this work, we delved into the mathematical formulation of EMSG by presenting the action and field equations in the geometrical and scalar-tensor representations, laying the groundwork for the theory.
The thermodynamic aspects of EMSG are crucial in understanding the Universe's evolution. We also examined particle production in cosmology, highlighting how EMSG influences matter creation. The thermodynamic interpretation of irreversible matter creation provides insights into a deeper connection between thermodynamics and cosmological expansion. This motivated the analysis of the cosmological implications of EMSG. The generalized Friedmann equations were derived, incorporating the energy balance equation and the deceleration parameter, and various de Sitter expansion scenarios were analyzed, including self-interacting potentials and constant-density solutions. The exploration of general cosmological models through dimensionless and redshift representations offered a qualitative discussion of specific models and their observational prospects.

In fact, matter creation processes are believed to play a fundamental role in quantum field theoretical approaches to gravity, where these processes emerge naturally within the framework. A well-established result in quantum field theory, particularly in curved spacetimes, is the creation of quanta from minimally coupled scalar fields in an expanding Friedmann--Robertson--Walker (FRW) universe. This phenomenon, as rigorously demonstrated by Leonard Parker in his pioneering research, has become a cornerstone of our understanding of particle creation in cosmological settings \cite{Parker:2012at, Parker:1968mv,Parker:1969au,Parker:1971pt,Parker:1972kp}. Parker's work, beginning in the late 1960s, showed that the expansion of the Universe leads to the spontaneous creation of particles---an insight that has profound implications for both quantum mechanics and cosmology. His findings revealed that the dynamics of an expanding universe could result in the production of both fermions and bosons, establishing a deep connection between the geometry of spacetime and quantum field behavior.

Given the significance of these results, finding an equivalent microscopic quantum description of the matter creation processes discussed in the present work could provide critical insights into the physical mechanisms that govern particle generation through the coupling of gravity and matter geometry. Although this analysis lies outside the scope of the present work, by developing such a microscopic perspective, we could further illuminate the fundamental processes by which particles are created in the Universe, potentially bridging the gap between quantum field theory and gravitational phenomena. These models, which offer a synthesis of classical and quantum descriptions of the Universe, provide a robust platform for exploring how particles( both fermions and bosons) are generated in an expanding cosmological background. Parker's extensive body of work serves as a foundational reference for these explorations, offering both theoretical insights and analytical techniques necessary to delve into the complexities of particle \mbox{creation \cite{Parker:2012at, Parker:1968mv,Parker:1969au,Parker:1971pt,Parker:1972kp}.}
These seminal works provide the theoretical framework and foundational principles that continue to guide research on how the expanding Universe influences quantum fields, leading to the creation of matter as we observe it. Through these studies, the intricate relationship between gravity, quantum fields, and matter creation is progressively being unraveled, offering new perspectives on the nature of the Universe. Work along these lines is currently underway.

Furthermore, we investigated the impact of EMSG on compact objects. Indeed, EMSG reduces to GR in vacuum, making them indistinguishable without matter or energy. However, in the presence of an energy-momentum distribution, differences arise because EMSG includes higher-order corrections dependent on the square of the energy-momentum tensor. These corrections become significant in high-curvature regions, such as the dense cores of neutron stars, black holes, or wormholes, leading to noticeable deviations from GR. In this work, electrodynamics within the EMSG framework were explored, suggesting potential deviations in electromagnetic phenomena near massive objects. Wormhole geometries were also studied, with a focus on the metric and field equations, and specific wormhole solutions were obtained. Junction conditions and matching provide the criteria for viable physical models. Therefore, the dense cores of compact objects serve as ideal laboratories to test the predictions of EMSG against those of GR. By studying high-curvature regions, we can gain insights into the validity of EMSG and potentially identify observable phenomena that distinguish it from GR. As observational technology advances, particularly in the realms of gravitational wave astronomy and high-energy astrophysics, these deviations could provide crucial evidence supporting or refuting the modifications introduced \mbox{by EMSG. }

An interesting avenue for further exploration would be to consider the effects of non-spherical symmetry, particularly those induced by the distribution of angular momentum. When a compact object is set into slow rotation, as described in the works of \mbox{Hartle \cite{Hartle:1967he, Hartle:1968si},} both the geometry of the surrounding spacetime and the interior distribution of stress energy experience significant alterations. These modifications become even more pronounced in the presence of EMSG. In the context of EMSG, the slow rotation of a compact object could lead to complex changes in the spacetime geometry, potentially affecting the stability, structure, and observable properties of these objects. The inclusion of angular momentum distribution in such models is likely to yield new insights into the behavior of rotating compact objects, particularly in how they deviate from the predictions of GR. Given the complexity and importance of these considerations, a detailed analysis of the impact of angular momentum on non-spherical compact objects within the framework of EMSG would be a valuable contribution to the field. However, this investigation involves substantial theoretical and computational challenges, which we intend to address in future work. This future research will aim to provide a more comprehensive understanding of how rotation and EMSG interact to shape the properties of compact astrophysical objects.

In this context of compact objects, it is also of interest to define the mass of these geometries as viewed by a distant observer, which is, indeed, not a trivial matter. Here, one can, indeed, enumerate several concepts, such as the Komar mass \cite{Komar:1958wp,Wald:1984rg}, ADM mass (Arnowitt-Deser-Misner Mass) \cite{Arnowitt:1962hi,Misner:1973prb}, and Bondi mass \cite{Bondi:1962px,Sachs:1962wk}, which are not equivalent in GR but are related concepts that apply in different contexts. The Komar mass  is the mass associated with stationary spacetimes, specifically those with a time-like Killing vector field. The Komar mass is a measure of the total mass (including contributions from both matter and gravitational energy) within a certain region of spacetime. It is applicable primarily in spacetimes with certain symmetries (e.g., stationary or axisymmetric spacetimes) and is not defined in non-stationary situations. The ADM mass is a measure of the total mass (including gravitational energy) of an isolated system as seen from spatial infinity. It is defined in asymptotically flat spacetimes and represents the conserved quantity associated with the asymptotic time translation symmetry. The Bondi mass is defined at null infinity and measures the total mass of an isolated system as seen by a distant observer, accounting for any energy radiated away as gravitational waves. Unlike the ADM mass, which is defined in a spatial context, the Bondi mass is relevant in dynamical spacetimes where gravitational radiation is present.
In GR, these three concepts are not directly equivalent but can be related under specific conditions. For instance, in a stationary, asymptotically flat spacetime, the Komar mass and ADM mass can be shown to be equivalent. This is because both are designed to measure the total mass of a static or stationary system, although using different methods. The ADM mass and Bondi mass differ because the Bondi mass decreases as gravitational waves carry energy away from the system. In a stationary situation with no gravitational radiation, they would be equivalent, but in general, the Bondi mass is less than the ADM mass for a radiating system.

In modified theories of gravity, such as in EMSG, the relationship between these masses can differ significantly from GR due to changes in the underlying field equations and the nature of gravitational interaction. Here, the Komar mass typically relies on the Einstein field equations and the presence of a time-like Killing vector. In modified gravity theories, the field equations are altered, and the existence and properties of Killing vectors may also change, leading to different expressions or interpretations for Komar mass. The ADM mass is defined based on the Hamiltonian formulation of GR, which may be altered in modified theories of gravity. As a result, the ADM mass could take a different form or might not even be well-defined, depending on the asymptotic structure of the modified theory. Relative to the Bondi mass, in theories with different gravitational radiation properties or different treatments of null infinity, the definition of Bondi mass could also be modified. The relationship between energy carried away by gravitational waves and the reduction in Bondi mass could differ from those in GR.

In conclusion, the study of EMSG provides a compelling extension to existing gravitational theories, offering potential solutions to longstanding problems in cosmology and astrophysics. The formalism developed here lays a solid foundation for further theoretical exploration and potential empirical verification. By bridging the gap between gravity, thermodynamics, and quantum effects, EMSG emerges as a promising candidate for a more comprehensive theory of gravity, potentially furthering our understanding of \mbox{the Universe.}



\vspace{6pt}
\authorcontributions{Conceptualization, R.A.C.C. and F.S.N.L.; methodology, R.A.C.C., N.G., T.H., F.S.N.L., M.A.S.P. and J.L.R.; software, R.A.C.C., N.G., T.H., F.S.N.L., M.A.S.P. and J.L.R.; validation, R.A.C.C., N.G., T.H., F.S.N.L., M.A.S.P. and J.L.R.; formal analysis, R.A.C.C., N.G., T.H., F.S.N.L., M.A.S.P. and J.L.R.; investigation, R.A.C.C., N.G., T.H., F.S.N.L., M.A.S.P. and J.L.R.; resources, R.A.C.C., N.G., T.H., F.S.N.L., M.A.S.P. and J.L.R.; writing---original draft preparation, R.A.C.C. and F.S.N.L.; writing---review and editing, R.A.C.C., N.G., T.H., F.S.N.L., M.A.S.P. and J.L.R.; visualization, R.A.C.C., N.G., T.H., F.S.N.L., M.A.S.P. and J.L.R.; supervision, F.S.N.L.; project administration, F.S.N.L.; funding acquisition, R.A.C.C., N.G., F.S.N.L. and M.A.S.P. All authors have read and agreed to the published version of the manuscript.}

\funding{ This research was funded by the Fundação para a Ciência e a Tecnologia (FCT) under research grants UIDB/04434/2020, UIDP/04434/2020, and PTDC/FIS-AST/0054/2021.}



\dataavailability{Data are contained within the article.}




\acknowledgments{F.S.N.L. acknowledges support from the Fundação para a Ciência e a Tecnologia (FCT) Scientific Employment Stimulus contract with reference CEECINST/00032/2018.}

\conflictsofinterest{The authors declare no conflicts of interest.}



\begin{adjustwidth}{-\extralength}{0cm}
\printendnotes[custom]

\reftitle{References}

\PublishersNote{}
\end{adjustwidth}

\begin{thebibliography}{999}

%

\bibitem{SupernovaSearchTeam:1998fmf}
Riess, A.G.~et al. [Supernova Search Team].
Observational evidence from supernovae for an accelerating universe and a cosmological constant.
\emph{Astron. J.} \textbf{1998}, \emph{116}, 1009--1038. [\href{http://doi.org/10.1086/300499}{CrossRef}]


\bibitem{SupernovaCosmologyProject:1998vns}
Perlmutter, S.~et al. [Supernova Cosmology Project].
Measurements of $\Omega$ and $\Lambda$ from 42 High Redshift Supernovae.
\emph{Astrophys. J.} \textbf{1999}, \emph{517}, 565--586. [\href{http://dx.doi.org/10.1086/307221}{CrossRef}]

\bibitem{Capozziello:2002rd}
Capozziello, S.
Curvature quintessence.
\emph{Int. J. Mod. Phys. D} \textbf{2002}, \emph{11}, 483--492. [\href{http://dx.doi.org/10.1142/S0218271802002025}{CrossRef}]

\bibitem{Nojiri:2006ri}
Nojiri, S.; Odintsov, S.D.
Introduction to modified gravity and gravitational alternative for dark energy.
\emph{Int. J. Geom. Methods Mod. Phys. 
} \textbf{2007} \emph{4}, 115--145. [\href{http://dx.doi.org/10.1142/S0219887807001928}{CrossRef}]

\bibitem{Lobo:2008sg}
Lobo, F.S.N.
The Dark side of gravity: Modified theories of gravity. \emph{arXiv} \textbf{2008}, arXiv:0807.1640.

\bibitem{Sotiriou:2008rp}
Sotiriou, T.P.; Faraoni, V.
$f(R)$ Theories of Gravity.
\emph{Rev. Mod. Phys.} \textbf{2010}, \emph{82}, 451--497. [\href{http://dx.doi.org/10.1103/RevModPhys.82.451}{CrossRef}]

\bibitem{Nojiri:2010wj}
Nojiri, S.; Odintsov, S.D.
Unified cosmic history in modified gravity: From F(R) theory to Lorentz non-invariant models.
\emph{Phys. Rep.} \textbf{2011}, \emph{505}, 59--144. [\href{http://dx.doi.org/10.1016/j.physrep.2011.04.001}{CrossRef}]
%

\bibitem{Olmo:2011uz}
Olmo, G.J.
Palatini Approach to Modified Gravity: F(R) Theories and Beyond.
\emph{Int. J. Mod. Phys. D} \textbf{2011}, \emph{20}, 413--462. [\href{http://dx.doi.org/10.1142/S0218271811018925}{CrossRef}]

\bibitem{Capozziello:2011et}
Capozziello, S.; De Laurentis, M.
Extended Theories of Gravity.
\emph{Phys. Rep.} \textbf{2011}, \emph{509}, 167--321. [\href{http://dx.doi.org/10.1016/j.physrep.2011.09.003}{CrossRef}]

\bibitem{Clifton:2011jh}
Clifton, T.; Ferreira, P.G.; Padilla, A.; Skordis, C.
Modified Gravity and Cosmology.
\emph{Phys. Rep.} \textbf{2012}, \emph{513}, 1--189.

\bibitem{Harko:2018ayt}
Harko, T.; Lobo, F.S.N.
\emph{Extensions of $f(R)$ Gravity: Curvature-Matter Couplings and Hybrid Metric-Palatini Theory};
Cambridge University Press: Cambridge, UK,  2018;
ISBN 978-1-108-42874-3; 978-1-108-58457-9

\bibitem{CANTATA:2021ktz}
Saridakis, E.N. et al. [CANTATA].
\emph{Modified Gravity and Cosmology: An Update by the CANTATA Network};
Springer: Cham, Switzerland, 2021;
ISBN 978-3-030-83714-3; 978-3-030-83717-4; 978-3-030-83715-0

\bibitem{Copeland:2006wr}
Copeland, E.J.; Sami, M.; Tsujikawa, S.
Dynamics of dark energy.
\emph{Int. J. Mod. Phys. D} \textbf{2006}, \emph{15}, 1753--1936. [\href{http://dx.doi.org/10.1142/S021827180600942X}{CrossRef}]

\bibitem{Bertolami:2007gv}
Bertolami, O.; Boehmer, C.G.; Harko, T.; Lobo, F.S.N.
Extra force in f(R) modified theories of gravity.
\emph{Phys. Rev. D} \textbf{2007}, \emph{75}, 104016. [\href{http://dx.doi.org/10.1103/PhysRevD.75.104016}{CrossRef}]


\bibitem{Harko:2014sja}
Harko, T.; Lobo, F.S.N.; Otalora, G.; Saridakis, E.N.
Nonminimal torsion-matter coupling extension of f(T) gravity.
\emph{Phys. Rev. D} \textbf{2014}, \emph{89}, 124036. [\href{http://dx.doi.org/10.1103/PhysRevD.89.124036}{CrossRef}]

\bibitem{Harko:2014aja}
Harko, T.; Lobo, F.S.N.; Otalora, G.; Saridakis, E.N.
$f(T,\mathcal{T})$ gravity and cosmology.
\emph{J. Cosmol. Astropart. Phys.}
\textbf{2014}, \emph{12}, 021. [\href{http://dx.doi.org/10.1088/1475-7516/2014/12/021}{CrossRef}]

\bibitem{Harko:2014gwa}
Harko, T.; Lobo, F.S.N.
Generalized curvature-matter couplings in modified gravity.
\emph{Galaxies} \textbf{2014}, \emph{2}, 410--465. [\href{http://dx.doi.org/10.3390/galaxies2030410}{CrossRef}]

\bibitem{Harko:2018gxr}
Harko, T.; Koivisto, T.S.; Lobo, F.S.N.; Olmo, G.J.; Rubiera-Garcia, D.
Coupling matter in modified $Q$ gravity.
\emph{Phys. Rev. D} \textbf{2018}, \emph{98}, 084043. [\href{http://dx.doi.org/10.1103/PhysRevD.98.084043}{CrossRef}]

\bibitem{Harko:2010mv}
Harko, T.; Lobo, F.S.N.
$f(R,L_{m})$ gravity.
\emph{Eur. Phys. J. C} \textbf{2010}, \emph{70}, 373--379. [\href{http://dx.doi.org/10.1140/epjc/s10052-010-1467-3}{CrossRef}]

\bibitem{Harko:2011kv}
Harko, T.; Lobo, F.S.N.; Nojiri, S.; Odintsov, S.D.
$f(R,T)$ gravity.
\emph{Phys. Rev. D} \textbf{2011}, \emph{84}, 024020. [\href{http://dx.doi.org/10.1103/PhysRevD.84.024020}{CrossRef}]

\bibitem{Barrientos:2018cnx}
Barrientos, E.; Lobo, F.S.N.; Mendoza, S.; Olmo, G.J.; Rubiera-Garcia, D.
Metric-affine f(R,T) theories of gravity and their applications.
\emph{Phys. Rev. D} \textbf{2018}, \emph{97}, 104041. [\href{http://dx.doi.org/10.1103/PhysRevD.97.104041}{CrossRef}]



\bibitem{Katirci:2013okf}
Kat\i{}rc\i{}, N.; Kavuk, M.
$ f(R,T_{\mu\nu}T^{\mu\nu})$ gravity and Cardassian-like expansion as one of its consequences.
\emph{Eur. Phys. J. Plus} \mbox{\textbf{2014}, \emph{129}, 163.} [\href{http://dx.doi.org/10.1140/epjp/i2014-14163-6}{CrossRef}]

\bibitem{Roshan:2016mbt}
Roshan, M.; Shojai, F.
Energy-Momentum Squared Gravity.
\emph{Phys. Rev. D} \textbf{2016}, \emph{94}, 044002. [\href{http://dx.doi.org/10.1103/PhysRevD.94.044002}{CrossRef}]

\bibitem{Haghani:2013oma}
Haghani, Z.; Harko, T.; Lobo, F.S.N.; Sepangi, H.R.; Shahidi, S.
Further matters in space-time geometry: $f(R,T,R_{\mu\nu}T^{\mu\nu})$ gravity.
\emph{Phys. Rev. D}\textbf{2013}, \emph{88}, 044023. [\href{http://dx.doi.org/10.1103/PhysRevD.88.044023}{CrossRef}]

\bibitem{Odintsov:2013iba}
Odintsov, S.D.; S\'aez-G\'omez, D.
$f(R, T, R_{\mu\nu} T^{\mu\nu})$ gravity phenomenology and $\Lambda$CDM universe.
\emph{Phys. Lett. B} \textbf{2013}, \emph{725}, 437--444. [\href{http://dx.doi.org/10.1016/j.physletb.2013.07.026}{CrossRef}]


\bibitem{Bahamonde:2019urw}
Bahamonde, S.; Marciu, M.; Rudra, P.
Dynamical system analysis of generalized energy-momentum-squared gravity.
\emph{Phys. Rev. D} \textbf{2019}, \emph{100}, 083511. [\href{http://dx.doi.org/10.1103/PhysRevD.100.083511}{CrossRef}]

\bibitem{Akarsu:2018drb}
Akarsu, \"O.; Katirci, N.; Kumar, S.
Energy-momentum powered gravity and cosmic acceleration. In Proceedings of the PoS CORFU2017, Corfu, Greece, 2--28 September 2017; p. 105.

\bibitem{Akarsu:2020vii}
Akarsu, \"O.; Barrow, J.D.; Uzun, N.M.
Screening anisotropy via energy-momentum squared gravity: $\Lambda$CDM model with hidden anisotropy.
\emph{Phys. Rev. D} \textbf{2020}, \emph{102}, 124059. [\href{http://dx.doi.org/10.1103/PhysRevD.102.124059}{CrossRef}]

\bibitem{Board:2017ign}
Board, C.V.R.; Barrow, J.D.
Cosmological Models in Energy-Momentum-Squared Gravity.
\emph{Phys. Rev. D} \textbf{2017}, \emph{96}, 123517;
Erratum: \emph{Phys. Rev. D} \textbf{2018}, \emph{98}, 129902. [\href{http://dx.doi.org/10.1103/PhysRevD.96.123517}{CrossRef}]

\bibitem{Barbar:2019rfn}
Barbar, A.H.; Awad, A.M.; AlFiky, M.T.
Viability of bouncing cosmology in energy-momentum-squared gravity.
\emph{Phys. Rev. D} \textbf{2020}, \emph{101}, 044058. [\href{http://dx.doi.org/10.1103/PhysRevD.101.044058}{CrossRef}]

\bibitem{Akarsu:2017ohj}
Akarsu, \"O.; Kat\i{}rc\i{}, N.; Kumar, S.
Cosmic acceleration in a dust only universe via energy-momentum powered gravity.
\emph{Phys. Rev. D} \textbf{2018}, \emph{97}, 024011. [\href{http://dx.doi.org/10.1103/PhysRevD.97.024011}{CrossRef}]

\bibitem{Cipriano:2023yhv}
Cipriano, R.A.C.; Harko, T.; Lobo, F.S.N.; Pinto, M.A.S.; Rosa, J.L.
Gravitationally induced matter creation in scalar-tensor $f(R,T_{\mu\nu}T^{\mu\nu})$ gravity.
\emph{Phys. Dark Universe} \textbf{2024}, \emph{44}, 101463 [\href{http://dx.doi.org/10.1016/j.dark.2024.101463}{CrossRef}]

\bibitem{Sharif:2023uyv}
Sharif, M.; Zeeshan Gul, M.
Stability Analysis of the Inhomogeneous Perturbed Einstein Universe in Energy\textendash{}Momentum Squared Gravity.
\emph{Universe} \textbf{2023}, \emph{9}, 145. [\href{http://dx.doi.org/10.3390/universe9030145}{CrossRef}]

\bibitem{Canuto:2023gdv}
Canuto, \'A.J.C.; Santos, A.F.
G\"odel-type universes in energy\textendash{}momentum-squared gravity.
\emph{Eur. Phys. J. C }\textbf{2023}, \emph{83}, 404. [\href{http://dx.doi.org/10.1140/epjc/s10052-023-11570-3}{CrossRef}]
%

\bibitem{Shahidi:2021lqf}
Shahidi, S.
Non-minimal energy\textendash{}momentum squared gravity.
\emph{Eur. Phys. J. C} \textbf{2021}, \emph{81}, 274. [\href{http://dx.doi.org/10.1140/epjc/s10052-021-09082-z}{CrossRef}]
%




\bibitem{Akarsu:2018zxl}
Akarsu, \"O.; Barrow, J.D.; \c{C}\i{}k\i{}nto\u{g}lu, S.; Ek\c{s}i, K.Y.; Kat\i{}rc\i{}, N.
Constraint on energy-momentum squared gravity from neutron stars and its cosmological implications.
\emph{Phys. Rev. D} \textbf{2018}, \emph{97}, 124017. [\href{http://dx.doi.org/10.1103/PhysRevD.97.124017}{CrossRef}]
%

\bibitem{Chen:2019dip}
Chen, C.Y.; Chen, P.
Eikonal black hole ringings in generalized energy-momentum squared gravity.
\emph{Phys. Rev. D} \textbf{2020}, \emph{101}, 064021. [\href{http://dx.doi.org/10.1103/PhysRevD.101.064021}{CrossRef}]
%

\bibitem{Sharif:2021rck}
Sharif, M.; Zeeshan Gul, M.
Dynamics of charged anisotropic spherical collapse in energy-momentum squared gravity.
\emph{Chin. J. Phys.} \textbf{2021}, \emph{71}, 365--374. [\href{http://dx.doi.org/10.1016/j.cjph.2021.03.005}{CrossRef}]

\bibitem{Yousaf:2021ltg}
Yousaf, Z.; Bhatti, M.Z.; Farwa, U.
Evolution of axially and reflection symmetric source in energy\textendash{}momentum squared gravity.
\emph{Eur. Phys. J. Plus} \textbf{2022}, \emph{137}, 49. [\href{http://dx.doi.org/10.1140/epjp/s13360-021-02253-7}{CrossRef}]

\bibitem{Sharif:2022ibt}
Sharif, M.; Naz, S.
Stable charged gravastar model in ${f}(\mathfrak {R},\mathbf{T} ^{2})$ gravity with conformal motion.
\emph{Eur. Phys. J. Plus} \textbf{2022}, \emph{137}, 421. [\href{http://dx.doi.org/10.1140/epjp/s13360-022-02636-4}{CrossRef}]
%

\bibitem{Sharif:2022cud}
Sharif, M.; Naz, S.
Impact of charge on gravastars in ${f}(\mathfrak {R},\mathbf{T} ^{2})$ gravity.
\emph{Mod. Phys. Lett. A} \textbf{2022}, \emph{37}, 2250065. [\href{http://dx.doi.org/10.1142/S0217732322500651}{CrossRef}]
%

\bibitem{Sharif:2022prm}
Sharif, M.; Anjum, A.
Impact of charge on the complexity of static sphere in $f(R,\mathbf{T} ^{2})$ gravity.
\emph{Eur. Phys. J. Plus} \textbf{2022}, \emph{137}, 602. [\href{http://dx.doi.org/10.1140/epjp/s13360-022-02816-2}{CrossRef}]
%

\bibitem{Sharif:2022mdp}
Sharif, M.; Gul, M.Z.
Anisotropic compact stars with Karmarkar condition in energy-momentum squared gravity.
\emph{Gen. Relativ. Gravit.} \textbf{2023}, \emph{55}, 10. [\href{http://dx.doi.org/10.1007/s10714-022-03062-8}{CrossRef}]

\bibitem{ZeeshanGul:2023ysx}
Zeeshan Gul, M.; Sharif, M.
Traversable Wormhole Solutions Admitting Noether Symmetry in Theory. 
\emph{Symmetry} \textbf{2023}, \emph{15}, 684.

\bibitem{Sharif:2023uac}
Sharif, M.; Naz, S.
Study of decoupled gravastars in energy\textendash{}momentum squared gravity.
\emph{Ann. Phys.} \textbf{2023}, \emph{451}, 169240. [\href{http://dx.doi.org/10.1016/j.aop.2023.169240}{CrossRef}]
%

\bibitem{Sharif:2023gbl}
Sharif, M.; Naz, S.
Viable decoupled solutions in energy\textendash{}momentum squared gravity.
\emph{Pramana} \textbf{2023}, \emph{97}, 116. [\href{http://dx.doi.org/10.1007/s12043-023-02595-0}{CrossRef}]

\bibitem{Sharif:2023nrl}
Sharif, M.; Naz, S.
Stable gravastars with Krori\textendash{}Barua metric in f(R,T2)~gravity.
\emph{Ann. Phys.} \textbf{2023}, \emph{457}, 169426. [\href{http://dx.doi.org/10.1016/j.aop.2023.169426}{CrossRef}]

\bibitem{Sharif:2023ccr}
Sharif, M.; Gul, M.Z.
Compact stellar objects in ${f}(\mathfrak {R},\mathbf{T} ^{2})$ gravity.
\emph{Pramana} \textbf{2023}, \emph{97}, 122. [\href{http://dx.doi.org/10.1007/s12043-023-02598-x}{CrossRef}]

\bibitem{Kazemi:2020hep}
Kazemi, A.; Roshan, M.; De Martino, I.; De Laurentis, M.
Jeans analysis in energy\textendash{}momentum-squared gravity.
\emph{Eur. Phys. J. C} \textbf{2020}, \emph{80}, 150. [\href{http://dx.doi.org/10.1140/epjc/s10052-020-7662-y}{CrossRef}]
%



\bibitem{Sharif:2023rcd}
Sharif, M.; Manzoor, S.
Compact objects admitting Finch\textendash{}Skea symmetry in f(R,T2) gravity.
\emph{Ann. Phys.} \textbf{2023}, \emph{454}, 169337. [\href{http://dx.doi.org/10.1016/j.aop.2023.169337}{CrossRef}]
%

\bibitem{HosseiniMansoori:2023mqh}
Hosseini Mansoori, S.A.; Felegray, F.; Talebian, A.; Sami, M.
PBHs and GWs from \ensuremath{\mathbb{T}}$^{2}$-inflation and NANOGrav 15-year data.
\emph{J. Cosmol. Astropart. Phys.}
\textbf{2023}, \emph{08}, 067. [\href{http://dx.doi.org/10.1088/1475-7516/2023/08/067}{CrossRef}]
%

\bibitem{Sharif:2022aei}
Sharif, M.; Zeeshan Gul, M.
Noether Symmetries and Some Exact Solutions in f(R, T~$^{2}$) Theory.
\emph{J. Exp. Theor. Phys.} \textbf{2023}, \emph{136}, 436--445. [\href{http://dx.doi.org/10.1134/S106377612303007X}{CrossRef}]
%

\bibitem{Nasir:2023pzq}
Nasir, M.M.M.; Bhatti, M.Z.; Yousaf, Z.
Influence of EMSG on complex systems: Spherically symmetric, static case.
\emph{Int. J. Mod. Phys. D} \textbf{2023}, \emph{32}, 2350009. [\href{http://dx.doi.org/10.1142/S0218271823500098}{CrossRef}]

\bibitem{Pretel:2023avv}
Pretel, J.M.Z.; Tangphati, T.; Banerjee, A.
Relativistic structure of charged quark stars in energy\textendash{}momentum squared gravity.
\emph{Ann. Phys.} \textbf{2023}, \emph{458}, 169440. [\href{http://dx.doi.org/10.1016/j.aop.2023.169440}{CrossRef}]
%

\bibitem{Naz:2022vvn}
Naz, S.; Sharif, M.
Gravastars with Kuchowicz Metric in Energy-Momentum Squared Gravity.
\emph{Universe} \textbf{2022}, \emph{8}, 142. [\href{http://dx.doi.org/10.3390/universe8030142}{CrossRef}]

\bibitem{Sharif:2022cjv}
Sharif, M.; Naz, S.
Gravastars with Karmarkar condition in $f(R,T^2)$ gravity.
\emph{Int. J. Mod. Phys. D} \textbf{2022}, \emph{31}, 2240008. [\href{http://dx.doi.org/10.1142/S0218271822400089}{CrossRef}]

\bibitem{Rudra:2021ksp}
Rudra, P.
Energy\textendash{}momentum squared symmetric Teleparallel gravity: $f(Q,T_{\mu\nu}T^{\mu\nu})$ gravity.
\emph{Phys. Dark Universe} \textbf{2022}, \emph{37}, 101071. [\href{http://dx.doi.org/10.1016/j.dark.2022.101071}{CrossRef}]
%

\bibitem{Nazari:2022xhv}
Nazari, E.; Roshan, M.; De Martino, I.
Constraining energy-momentum-squared gravity by binary pulsar observations.
\emph{Phys. Rev. D} \textbf{2022}, \emph{105}, 044014. [\href{http://dx.doi.org/10.1103/PhysRevD.105.044014}{CrossRef}]



\bibitem{Ashtekar:2011ni}
Ashtekar, A.; Singh, P.
Loop Quantum Cosmology: A Status Report.
\emph{Class. Quantum Gravity}
\textbf{2011}, \emph{28}, 213001. [\href{http://dx.doi.org/10.1088/0264-9381/28/21/213001}{CrossRef}]
%

\bibitem{Maartens:2003tw}
Maartens, R.
Brane world gravity.
\emph{Living Rev. Relativ.} \textbf{2004}, \emph{7}, 7. [\href{http://dx.doi.org/10.12942/lrr-2004-7}{CrossRef}]
%

\bibitem{Maartens:2010ar}
Maartens, R.; Koyama, K.
Brane-World Gravity.
\emph{Living Rev. Relativ.} \textbf{2010}, \emph{13}, 5. [\href{http://dx.doi.org/10.12942/lrr-2010-5}{CrossRef}]
%


\bibitem{Brax:2003fv}
Brax, P.; van de Bruck, C.
Cosmology and brane worlds: A Review.
\emph{Class. Quantum Gravity} \textbf{2003}, \emph{20}, R201--R232. [\href{http://dx.doi.org/10.1088/0264-9381/20/9/202}{CrossRef}]
%

\bibitem{Armendariz-Picon:2000ulo}
Armendariz-Picon, C.; Mukhanov, V.F.; Steinhardt, P.J.
Essentials of k essence.
\emph{Phys. Rev. D} \textbf{2001}, \emph{63}, 103510. [\href{http://dx.doi.org/10.1103/PhysRevD.63.103510}{CrossRef}]
%
%



\bibitem{Nazari:2020gnu}
Nazari, E.; Sarvi, F.; Roshan, M.
Generalized Energy-Momentum-Squared Gravity in the Palatini Formalism.
\emph{Phys. Rev. D} \textbf{2020}, \emph{102}, 064016. [\href{http://dx.doi.org/10.1103/PhysRevD.102.064016}{CrossRef}]
%

\bibitem{Sharif:2021ptz}
Sharif, M.; Zeeshan Gul, M.
Viable wormhole solutions in energy\textendash{}momentum squared gravity. 
\emph{Eur. Phys. J. Plus} \textbf{2021}, \emph{136}, 503. [\href{http://dx.doi.org/10.1140/epjp/s13360-021-01512-x}{CrossRef}]

\bibitem{Woodard:2006nt}
Woodard, R.P.
Avoiding dark energy with 1/r modifications of gravity.
\emph{Lect. Notes Phys.} \textbf{2007}, \emph{720}, 403--433.
%

\bibitem{Ostrogradski}
Ostrogradski, M. \emph{Mémoires sur les Équations Différentielles, Relatives au Problème des Isopérimètres}; Mémoires de l’Académie Impériale des Sciences: St-Pétersbourg, Russia, 1850; Volume 4, p. 385.

\bibitem{Abedi:2022hzq}
Abedi, H.; Bajardi, F.; Capozziello, S.
Linearized field equations and extra force in f(R,T(n)) extended gravity.
\emph{Int. J. Mod. Phys. D }\textbf{2022}, \emph{31}, 2240015. [\href{http://dx.doi.org/10.1142/S0218271822400156}{CrossRef}]
%

\bibitem{Schrodinger:1939}
Schrödinger, E.
The proper vibrations of the expanding universe.
\emph{Physica } \textbf{1939}, \emph{6}, 899. [\href{http://dx.doi.org/10.1016/S0031-8914(39)90091-1}{CrossRef}]

\bibitem{Schrodinger:1940}
Schrödinger, E.
The General Theory of Relativity and Wave Mechanics.
\emph{Physica }\textbf{1940}, \emph{46}, 25.

\bibitem{Parker:2012at}
Parker, L.
Particle creation and particle number in an expanding universe.
\emph{J. Phys. A} \textbf{2012}, \emph{45}, 374023. [\href{http://dx.doi.org/10.1088/1751-8113/45/37/374023}{CrossRef}]

\bibitem{Parker:1968mv}
Parker, L.
Particle creation in expanding universes.
\emph{Phys. Rev. Lett.} \textbf{1968}, \emph{21}, 562--564. [\href{http://dx.doi.org/10.1103/PhysRevLett.21.562}{CrossRef}]

\bibitem{Parker:1969au}
Parker, L.
Quantized fields and particle creation in expanding universes. 1.
\emph{Phys. Rev.} \textbf{1969}, \emph{183}, 1057--1068. [\href{http://dx.doi.org/10.1103/PhysRev.183.1057}{CrossRef}]

\bibitem{Parker:1971pt}
Parker, L.
Quantized fields and particle creation in expanding universes. 2.
\emph{Phys. Rev. D} \textbf{1971}, \emph{3}, 346--356;
Erratum in \emph{Phys. Rev. D} \textbf{1971}, \emph{3}, 2546--2546. [\href{http://dx.doi.org/10.1103/PhysRevD.3.346}{CrossRef}]

\bibitem{Parker:1972kp}
Parker, L.
Particle creation in isotropic cosmologies.
\emph{Phys. Rev. Lett.} \textbf{1972}, \emph{28}, 705--708;
Erratum in \emph{Phys. Rev. Lett.} \textbf{1972}, \emph{28}, 1497. [\href{http://dx.doi.org/10.1103/PhysRevLett.28.705}{CrossRef}]

\bibitem{Prigogine:1989zz}
Prigogine,  I.; Geheniau, J.; Gunzig, E.; Nardone, P.
Thermodynamics and cosmology.
\emph{Gen. Relativ. Gravit.} \textbf{1989}, \emph{21}, 767--776. [\href{http://dx.doi.org/10.1007/BF00758981}{CrossRef}]

\bibitem{Prigogine:1986}
Prigogine, I.; Géhéniau, J.
Entropy, matter, and cosmology.
\emph{Proc. Nat. Acad. Sci. USA} \textbf{1986}, \emph{83}, 6245--6249. [\href{http://dx.doi.org/10.1073/pnas.83.17.6245}{CrossRef}] [\href{http://www.ncbi.nlm.nih.gov/pubmed/16593747}{PubMed}]


\bibitem{Prigogine:1988jax}
Prigogine, I.; Geheniau, J.; Gunzig, E.; Nardone, P.
Thermodynamics of cosmological matter creation.
\emph{Proc. Nat. Acad. Sci. USA} \textbf{1988}, \emph{85}, 7428. [\href{http://dx.doi.org/10.1073/pnas.85.20.7428}{CrossRef}]


\bibitem{Ford:1986sy}
Ford, L.H.
Gravitational Particle Creation and Inflation.
\emph{Phys. Rev. D} \textbf{1987}, \emph{35}, 2955. [\href{http://dx.doi.org/10.1103/PhysRevD.35.2955}{CrossRef}]

\bibitem{Brout:1979qe}
Brout, R.; Englert, F.; Spindel, P.
Cosmological Origin of the Grand Unification Mass Scale.
\emph{Phys. Rev. Lett.} \textbf{1979}, \emph{43}, 417;
Erratum in \emph{Phys. Rev. Lett.} \textbf{1979}, \emph{43}, 890. [\href{http://dx.doi.org/10.1103/PhysRevLett.43.417}{CrossRef}]

\bibitem{Grib:1976pw}
Grib, A.A.; Mamaev, S.G.; Mostepanenko, V.M.
Particle Creation from Vacuum in Homogeneous Isotropic Models of the Universe.
\emph{Gen. Relativ. Gravit.} \textbf{1976}, \emph{7}, 535--547 [\href{http://dx.doi.org/10.1007/BF00766413}{CrossRef}]


\bibitem{Harko:2014pqa}
Harko, T.
Thermodynamic interpretation of the generalized gravity models with geometry---Matter coupling.
\emph{Phys. Rev. D} \textbf{2014}, \emph{90}, 044067. [\href{http://dx.doi.org/10.1103/PhysRevD.90.044067}{CrossRef}]
%

\bibitem{Pinto:2022tlu}
Pinto, M.A.S.; Harko, T.; Lobo, F.S.N.
Gravitationally induced particle production in scalar-tensor f(R,T) gravity.
\emph{Phys. Rev. D} \textbf{2022}, \emph{106}, 044043. [\href{http://dx.doi.org/10.1103/PhysRevD.106.044043}{CrossRef}]
%

\bibitem{Harko:2015pma}
Harko, T.; Lobo, F.S.N.; Mimoso, J.P.; Pav\'on, D.
Gravitational induced particle production through a nonminimal curvature\textendash{}matter coupling.
\emph{Eur. Phys. J. C} \textbf{2015}, \emph{75}, 386. [\href{http://dx.doi.org/10.1140/epjc/s10052-015-3620-5}{CrossRef}]
%

\bibitem{Pinto:2023phl}
Pinto, M.A.S.; Harko, T.; Lobo, F.S.N.
Irreversible Geometrothermodynamics of Open Systems in Modified Gravity.
\emph{Entropy} \textbf{2023}, \emph{25}, 944. [\href{http://dx.doi.org/10.3390/e25060944}{CrossRef}] [\href{http://www.ncbi.nlm.nih.gov/pubmed/37372288}{PubMed}]
%

\bibitem{Chernodub:2023pwf}
Chernodub, M.N.
Conformal anomaly and gravitational pair production. \emph{arXiv} \textbf{2023}, arXiv:2306.03892.

\bibitem{Berezin:2022phu}
Berezin, V.A.; Dokuchaev, V.I.
Cosmological particle creation in Weyl geometry.
\emph{Class. Quantum Gravity} \textbf{2023}, \emph{40}, 015006. [\href{http://dx.doi.org/10.1088/1361-6382/aca57e}{CrossRef}]
%

\bibitem{Xue:2020tpf}
Xue, S.S.
Massive particle pair production and oscillation in Friedman Universe: Reheating energy and entropy, and cold dark matter.
\emph{Eur. Phys. J. C} \textbf{2023}, \emph{83}, 355. [\href{http://dx.doi.org/10.1140/epjc/s10052-023-11524-9}{CrossRef}]
%

\bibitem{Schutz} Schutz, F. Perfect Fluids in General Relativity: Velocity Potentials and a Variational Principle. \emph{Phys. Rev. D} {\bf 1970}, \emph{2}, 2762. [\href{http://dx.doi.org/10.1103/PhysRevD.2.2762}{CrossRef}]

\bibitem{Brown} Brown, J.D. Action functionals for relativistic perfect fluids.
\emph{Class. Quantum Gravity} {\bf 1993}, \emph{10}, 1579. [\href{http://dx.doi.org/10.1088/0264-9381/10/8/017}{CrossRef}]


\bibitem{Bertolami:2008ab}
Bertolami, O.; Lobo, F.S.N.; Paramos, J.
Non-minimum coupling of perfect fluids to curvature.
\emph{Phys. Rev. D} \textbf{2008}, \emph{78}, 064036. [\href{http://dx.doi.org/10.1103/PhysRevD.78.064036}{CrossRef}]
%

\bibitem{Dod} Dodelson, S. \emph{Modern Cosmology}; Academic Press: San Diego, CA, USA, 2003.

\bibitem{wi1} Berera, A.; Fang, L.Z. Thermally Induced Density Perturbations in the Inflation Era. \emph{Phys. Rev. Lett.} {\bf 1995}, \emph{74}, 1912. [\href{http://dx.doi.org/10.1103/PhysRevLett.74.1912}{CrossRef}]

\bibitem{wi2} Berera, A. Warm Inflation. \emph{Phys. Rev. Lett.} {\bf 1995}, \emph{75}, 3218. [\href{http://dx.doi.org/10.1103/PhysRevLett.75.3218}{CrossRef}]

\bibitem{wi3} Harko, T.; Sheikhahmadi, H. Irreversible thermodynamical description of warm inflationary cosmological models. \emph{Phys. Dark Universe} {\bf 2020}, \emph{28}, 100521. [\href{http://dx.doi.org/10.1016/j.dark.2020.100521}{CrossRef}]

\bibitem{wi4} Berera, A. The Warm Inflation Story. \emph{Universe} {\bf 2023}, \emph{9}, 272. [\href{http://dx.doi.org/10.3390/universe9060272}{CrossRef}]

\bibitem{wi5} Kamali, V.; Motaharfar, M.; Ramos, R.O. Recent Developments in Warm Inflation. \emph{Universe} {\bf 2023}, \emph{9}, 124. [\href{http://dx.doi.org/10.3390/universe9030124}{CrossRef}]
\bibitem{wi6} Matei, T.M.; Harko, T. Warm inflation in a Universe with a Weylian boundary. \emph{Phys. Dark Universe} {\bf 2024}, \emph{46}, 101578. [\href{http://dx.doi.org/10.1016/j.dark.2024.101578}{CrossRef}]



\bibitem{nari1}
Nari, N.; Roshan, M.
Compact stars in Energy-Momentum Squared Gravity.
\emph{Phys. Rev. D} \textbf{2018}, \emph{98}, 024031. [\href{http://dx.doi.org/10.1103/PhysRevD.98.024031}{CrossRef}]
%


\bibitem{singh1}
Singh, K.N.; Banerjee, A.; Maurya, S.K.; Rahaman, F.; Pradhan, A.
Color-flavor locked quark stars in energy\textendash{}momentum squared gravity.
\emph{Phys. Dark Universe }\textbf{2021}, \emph{31}, 100774. [\href{http://dx.doi.org/10.1016/j.dark.2021.100774}{CrossRef}]
%

\bibitem{sharif1}
Sharif, M.; Zeeshan Gul, M.
Dynamics of spherical collapse in energy\textendash{}momentum squared gravity.
\emph{Int. J. Mod. Phys. A} \mbox{\textbf{2021}, \emph{36}, 2150004.} [\href{http://dx.doi.org/10.1142/S0217751X21500044}{CrossRef}]

\bibitem{Rudra:2020rhs}
Rudra, P.; Pourhassan, B.
Thermodynamics of the apparent horizon in the generalized energy\textendash{}momentum-squared cosmology.
\emph{Phys. Dark Universe} \textbf{2021}, \emph{33}, 100849. [\href{http://dx.doi.org/10.1016/j.dark.2021.100849}{CrossRef}]
%

\bibitem{Khodadi:2022xtl}
Khodadi, M.; Firouzjaee, J.T.
A survey of strong cosmic censorship conjecture beyond Einstein\textquoteright{}s gravity.
\emph{Phys. Dark Universe} \textbf{2022}, \emph{37}, 101084. [\href{http://dx.doi.org/10.1016/j.dark.2022.101084}{CrossRef}]

\bibitem{Moraes:2017dbs}
Moraes, P.H.R.S.; Sahoo, P.K.
Nonexotic matter wormholes in a trace of the energy-momentum tensor squared gravity.
\emph{Phys. Rev. D} \textbf{2018}, \emph{97}, 024007. [\href{http://dx.doi.org/10.1103/PhysRevD.97.024007}{CrossRef}]
%



\bibitem{Rosa:2023guo}
Rosa, J.L.; Ganiyeva, N.; Lobo, F.S.N.
Non-exotic traversable wormholes in $f\left( R,T_{ab}T^{ab}\right) $ gravity.
\emph{Eur. Phys. J. C} \textbf{2023}, \emph{83}, 1040.
%


\bibitem{Rosa:2023tph}
Rosa, J.L.
Junction conditions in gravity theories with extra scalar degrees of freedom.
\emph{Phys. Rev. D} \textbf{2024}, \emph{109}, 064018. [\href{http://dx.doi.org/10.1103/PhysRevD.109.064018}{CrossRef}]

\bibitem{Lobo:2020ffi}
Lobo, F.S.N.; Rodrigues, M.E.; de Sousa Silva, M.V.; Simpson, A.; Visser, M.
Novel black-bounce spacetimes: Wormholes, regularity, energy conditions, and causal structure.
\emph{Phys. Rev. D} \textbf{2021}, \emph{103}, 084052. [\href{http://dx.doi.org/10.1103/PhysRevD.103.084052}{CrossRef}]
%

\bibitem{Rodrigues:2023vtm}
Rodrigues, M.E.; Silva, M.V.d.S.
Source of black bounces in general relativity.
\emph{Phys. Rev. D} \textbf{2023}, \emph{107}, 044064. [\href{http://dx.doi.org/10.1103/PhysRevD.107.044064}{CrossRef}]
%

\bibitem{Junior:2023qaq}
Junior, J.T.S.S.; Rodrigues, M.E.
Coincident $f(\mathbb {Q})$ gravity: Black holes, regular black holes, and black bounces.
\emph{Eur. Phys. J. C} \textbf{2023}, \emph{83}, 475. [\href{http://dx.doi.org/10.1140/epjc/s10052-023-11660-2}{CrossRef}]
%

\bibitem{Fabris:2023opv}
Fabris, J.C.; Junior, E.L.B.; Rodrigues, M.E.
Generalized models for black-bounce solutions in f(R) gravity.
\emph{Eur. Phys. J. C} \mbox{\textbf{2023}, \emph{83}, 884.} [\href{http://dx.doi.org/10.1140/epjc/s10052-023-12022-8}{CrossRef}]
%

\bibitem{Junior:2023ixh}
Junior, J.T.S.S.; Lobo, F.S.N.; Rodrigues, M.E.
(Regular) Black holes in conformal Killing gravity coupled to nonlinear electrodynamics and scalar fields.
\emph{Class. Quantum Gravity} \textbf{2024}, \emph{41}, 055012. [\href{http://dx.doi.org/10.1088/1361-6382/ad210e}{CrossRef}]
%

\bibitem{Junior:2024xmm}
Junior, J.T.S.S.; Lobo, F.S.N.; Rodrigues, M.E.
Black holes and regular black holes in coincident $f({\mathbb {Q}},{\mathbb {B}}_Q)$ gravity coupled to nonlinear electrodynamics.
\emph{Eur. Phys. J. C} \textbf{2024}, \emph{84}, 332. [\href{http://dx.doi.org/10.1140/epjc/s10052-024-12696-8}{CrossRef}]
%

\bibitem{Junior:2024vrv}
Junior, J.T.S.S.; Lobo, F.S.N.; Rodrigues, M.E.
Black bounces in conformal Killing gravity.
\emph{Eur. Phys. J. C} \textbf{2024}, \emph{84}, 557. [\href{http://dx.doi.org/10.1140/epjc/s10052-024-12922-3}{CrossRef}]
%

\bibitem{morris1}
Morris, M.S.; Thorne, K.S. Wormholes in spacetime and their use for interstellar travel: A tool for teaching general relativity. \emph{Am. J. Phys.} \textbf{1988}, \emph{56}, 395. [\href{http://dx.doi.org/10.1119/1.15620}{CrossRef}]

\bibitem{Morris:1988tu}
Morris, M.S.; Thorne, K.S.; Yurtsever, U.
Wormholes, Time Machines, and the Weak Energy Condition.
\emph{Phys. Rev. Lett.} \textbf{1988}, \emph{61}, 1446--1449. [\href{http://dx.doi.org/10.1103/PhysRevLett.61.1446}{CrossRef}]

\bibitem{visser1}
Visser, M. \textit{Lorentzian Wormholes: From Einstein to Hawking}; Springer: New York, NY, USA, 1996.

\bibitem{lemos1}
Lemos, J.P.S.; Lobo, F.S.N.; Quinet de Oliveira, S.
Morris-Thorne wormholes with a cosmological constant.
\emph{Phys. Rev. D} \mbox{\textbf{2003}, \emph{68}, 064004.} [\href{http://dx.doi.org/10.1103/PhysRevD.68.064004}{CrossRef}]
%


\bibitem{Visser:2003yf}
Visser, M.; Kar, S.; Dadhich, N.
Traversable wormholes with arbitrarily small energy condition violations.
\emph{Phys. Rev. Lett.} \mbox{\textbf{2003}, \emph{90}, 201102.} [\href{http://dx.doi.org/10.1103/PhysRevLett.90.201102}{CrossRef}]
%

\bibitem{Kar:1995ss}
Kar, S.; Sahdev, D.
Evolving Lorentzian wormholes.
\emph{Phys. Rev. D} \textbf{1996}, \emph{53}, 722--730. [\href{http://dx.doi.org/10.1103/PhysRevD.53.722}{CrossRef}] [\href{http://www.ncbi.nlm.nih.gov/pubmed/10020052}{PubMed}]
%

\bibitem{Hawking:1973uf}
Hawking, S.W.; Ellis, G.F.R.
\emph{The Large Scale Structure of Space-Time};
Cambridge University Press: Cambridge, UK, 1973.
ISBN 978-1-00-925316-1.

\bibitem{Sajadi:2016hko}
Sajadi, S.N.; Riazi, N.
Gravitational lensing by multi-polytropic wormholes.
\emph{Can. J. Phys.} \textbf{2020}, \emph{98}, 1046--1054. [\href{http://dx.doi.org/10.1139/cjp-2019-0524}{CrossRef}]
%

\bibitem{agnese1}
Agnese, A.G.; La Camera, M.
Wormholes in the Brans-Dicke theory of gravitation.
\emph{Phys. Rev. D} \textbf{1995}, \emph{51}, 2011--2013. [\href{http://dx.doi.org/10.1103/PhysRevD.51.2011}{CrossRef}]

\bibitem{nandi1}
Nandi, K.K.; Bhattacharjee, B.; Alam, S.M.K.; Evans, J.
Brans-Dicke wormholes in the Jordan and Einstein frames.
\emph{Phys. Rev. D} \textbf{1998}, \emph{57}, 823--828. [\href{http://dx.doi.org/10.1103/PhysRevD.57.823}{CrossRef}]
%

\bibitem{camera1}
La Camera, M.
Wormhole solutions in the Randall-Sundrum scenario. 
\emph{Phys. Lett. B} \textbf{2003}, \emph{573}, 27--32. [\href{http://dx.doi.org/10.1016/j.physletb.2003.08.042}{CrossRef}]
%


\bibitem{lobo1}
Lobo, F.S.N. Exotic solutions in general relativity: Traversable wormholes and `warp drive' spacetimes. In \textit{Classical and Quantum Gravity Research}; Christiansen, M.N., Rasmussen, T.K., Eds.; Nova Science Publishers: Hauppauge, NY, USA, 2008; p. 1.

\bibitem{garattini1}
Garattini, R.; Lobo, F.S.N.
Self sustained phantom wormholes in semi-classical gravity.
\emph{Class. Quantum Gravity} \textbf{2007}, \emph{24}, 2401--2413. [\href{http://dx.doi.org/10.1088/0264-9381/24/9/016}{CrossRef}]
%

\bibitem{lobo2}
Lobo, F.S.N.
General class of wormhole geometries in conformal Weyl gravity.
\emph{Class. Quantum Gravity} \textbf{2008}, \emph{25}, 175006. [\href{http://dx.doi.org/10.1088/0264-9381/25/17/175006}{CrossRef}]
%

\bibitem{garattini2}
Garattini, R.; Lobo, F.S.N.
Self-sustained traversable wormholes in noncommutative geometry.
\emph{Phys. Lett. B} \textbf{2009}, \emph{671}, 146--152. [\href{http://dx.doi.org/10.1016/j.physletb.2008.11.064}{CrossRef}]
%

\bibitem{lobo3}
Lobo, F.S.N.; Oliveira, M.A.
General class of vacuum Brans-Dicke wormholes.
\emph{Phys. Rev. D} \textbf{2010}, \emph{81}, 067501. [\href{http://dx.doi.org/10.1103/PhysRevD.81.067501}{CrossRef}]

\bibitem{MontelongoGarcia:2011ag}
Montelongo Garcia, N.; Lobo, F.S.N.
Exact solutions of Brans-Dicke wormholes in the presence of matter.
\emph{Mod. Phys. Lett. A} \textbf{2011}, \emph{40}, 3067--3076. [\href{http://dx.doi.org/10.1142/S021773231103739X}{CrossRef}]
%

\bibitem{garattini3}
Garattini, R.; Lobo, F.S.N.
Self-sustained wormholes in modified dispersion relations.
\emph{Phys. Rev. D} \textbf{2012}, \emph{85}, 024043. [\href{http://dx.doi.org/10.1103/PhysRevD.85.024043}{CrossRef}]
%

\bibitem{myrzakulov1}
Myrzakulov, R.; Sebastiani, L.; Vagnozzi, S.; Zerbini, S.
Static spherically symmetric solutions in mimetic gravity: Rotation curves and wormholes.
\emph{Class. Quantum Gravity} \textbf{2016}, \emph{33}, 125005. [\href{http://dx.doi.org/10.1088/0264-9381/33/12/125005}{CrossRef}]
%

\bibitem{lobo4}
Lobo, F.S.N. (Ed.) \emph{Wormholes, Warp Drives and Energy Conditions}; Fundamental Theories of Physics; Springer: \mbox{New York, NY, USA}, 2017; Volume 189.

\bibitem{lobo5}
Lobo, F.S.N.; Oliveira, M.A.
Wormhole geometries in f(R) modified theories of gravity.
\emph{Phys. Rev. D} \textbf{2009}, \emph{80}, 104012. [\href{http://dx.doi.org/10.1103/PhysRevD.80.104012}{CrossRef}]
%

\bibitem{capozziello1}
Capozziello, S.; Harko, T.; Koivisto, T.S.; Lobo, F.S.N.; Olmo, G.J. Wormholes supported by hybrid metric-Palatini gravity. \emph{Phys. Rev. D} \textbf{2012}, \emph{86}, 127504. [\href{http://dx.doi.org/10.1103/PhysRevD.86.127504}{CrossRef}]

\bibitem{rosa1}
Rosa, J.L.; Lemos, J.P.S.; Lobo, F.S.N. Wormholes in generalized hybrid metric-Palatini gravity obeying the matter null energy condition everywhere. \emph{Phys. Rev. D} \textbf{2018}, \emph{98}, 064054. [\href{http://dx.doi.org/10.1103/PhysRevD.98.064054}{CrossRef}]

\bibitem{rosa2}
Rosa, J.L. Double gravitational layer traversable wormholes in hybrid metric-Palatini gravity. \emph{Phys. Rev. D} \textbf{2021}, \emph{104}, 064002. [\href{http://dx.doi.org/10.1103/PhysRevD.104.064002}{CrossRef}]

\bibitem{rosalol}
Rosa, J.L.; Lemos, J.P.S. Junction conditions for generalized hybrid metric-Palatini gravity with applications. \emph{Phys. Rev. D} \textbf{2021}, \emph{104}, 124076. [\href{http://dx.doi.org/10.1103/PhysRevD.104.124076}{CrossRef}]

\bibitem{rosalol2}
Rosa, J.L.; Andr\'e, R.; Lemos, J.P.S. Traversable wormholes with double layer thin shells in quadratic gravity.
\emph{Gen. Relativ. Gravit.} \textbf{2023}, \emph{55}, 65. [\href{http://dx.doi.org/10.1007/s10714-023-03107-6}{CrossRef}]

\bibitem{kull1}
Rosa, J.L.; Kull, P.M.
Non-exotic traversable wormhole solutions in linear $f\left( R,T\right) $ gravity.
\emph{Eur. Phys. J. C} \textbf{2022}, \emph{82}, 1154. [\href{http://dx.doi.org/10.1140/epjc/s10052-022-11135-w}{CrossRef}]
%

\bibitem{garcia1}
Garcia, N.M.; Lobo, F.S.N.
Wormhole geometries supported by a nonminimal curvature-matter coupling.
\emph{Phys. Rev. D} \mbox{\textbf{2010}, \emph{82}, 104018.} [\href{http://dx.doi.org/10.1103/PhysRevD.82.104018}{CrossRef}]
%

\bibitem{garcia2}
Montelongo Garcia, N.; Lobo, F.S.N.
Nonminimal curvature-matter coupled wormholes with matter satisfying the null energy condition.
\emph{Class. Quantum Gravity }\textbf{2011}, \emph{28}, 085018. [\href{http://dx.doi.org/10.1088/0264-9381/28/8/085018}{CrossRef}]
%

\bibitem{harko1}
Harko, T.; Lobo, S.N.; Mak, M.K.; Sushkov, S.V.
Modified-gravity wormholes without exotic matter.
\emph{Phys. Rev. D} \textbf{2013}, \emph{87}, 067504. [\href{http://dx.doi.org/10.1103/PhysRevD.87.067504}{CrossRef}]
%

\bibitem{anchordoqui1}
Anchordoqui, L.A.; Perez Bergliaffa, S.E.; Torres, D.F.
Brans-Dicke wormholes in nonvacuum space-time.
\emph{Phys. Rev. D} \textbf{1997}, \emph{55}, 5226--5229. [\href{http://dx.doi.org/10.1103/PhysRevD.55.5226}{CrossRef}]
%

\bibitem{DiGrezia:2017daq}
Di Grezia, E.; Battista, E.; Manfredonia, M.; Miele, G.
Spin, torsion and violation of null energy condition in traversable wormholes.
\emph{Eur. Phys. J. Plus} \textbf{2017}, \emph{132}, 537. [\href{http://dx.doi.org/10.1140/epjp/i2017-11799-6}{CrossRef}]
%

\bibitem{bhawal1}
Bhawal, B.; Kar, S. Lorentzian wormholes in Einstein---Gauss-Bonnet theory. \emph{Phys. Rev. D} \textbf{1992}, \emph{46}, 2464. [\href{http://dx.doi.org/10.1103/PhysRevD.46.2464}{CrossRef}]

\bibitem{dotti1}
Dotti, G.; Oliva, J.; Troncoso, R.
Exact solutions for the Einstein-Gauss-Bonnet theory in five dimensions: Black holes, wormholes and spacetime horns.
\emph{Phys. Rev. D} \textbf{2007}, \emph{76}, 064038. [\href{http://dx.doi.org/10.1103/PhysRevD.76.064038}{CrossRef}]

\bibitem{mehdizadeh1}
Mehdizadeh, M.R.; Kord Zangeneh, M.; Lobo, F.S.N.
Einstein-Gauss-Bonnet traversable wormholes satisfying the weak energy condition.
\emph{Phys. Rev. D} \textbf{2015}, \emph{91},084004. [\href{http://dx.doi.org/10.1103/PhysRevD.91.084004}{CrossRef}]

\bibitem{bronnikov1}
Bronnikov, K.A.; Kim, S.W.
Possible wormholes in a brane world.
\emph{Phys. Rev. D} \textbf{2003}, \emph{67}, 064027. [\href{http://dx.doi.org/10.1103/PhysRevD.67.064027}{CrossRef}]

\bibitem{lobo6}
Lobo, F.S.N.
A General class of braneworld wormholes.
\emph{Phys. Rev. D} \textbf{2007}, \emph{75}, 064027. [\href{http://dx.doi.org/10.1103/PhysRevD.75.064027}{CrossRef}]

\bibitem{israel1}
Israel, W. Singular hypersurfaces and thin shells in general
relativity. \emph{Nuovo Cimento B} {\bf 1966}, \emph{44}, 1--14. [\href{http://dx.doi.org/10.1007/BF02710419}{CrossRef}]

\bibitem{visser2}
Visser, M. Traversable wormholes: Some simple examples. \emph{Phys. Rev. D} \textbf{1989}, \emph{39}, 3182. [\href{http://dx.doi.org/10.1103/PhysRevD.39.3182}{CrossRef}] [\href{http://www.ncbi.nlm.nih.gov/pubmed/9959561}{PubMed}]

\bibitem{visser3}
Visser, M. Traversable wormholes from surgically modified Schwarzschild spacetimes. \emph{Nucl. Phys. B} \textbf{1989}, \emph{328}, 203. [\href{http://dx.doi.org/10.1016/0550-3213(89)90100-4}{CrossRef}]

\bibitem{Lobo:2004rp}
Lobo, F.S.N.
Energy conditions, traversable wormholes and dust shells.
\emph{Gen. Relativ. Gravit.} \textbf{2005}, \emph{37}, 2023--2038. [\href{http://dx.doi.org/10.1007/s10714-005-0177-x}{CrossRef}]

\bibitem{Lobo:2004id}
Lobo, F.S.N.
Surface stresses on a thin shell surrounding a traversable wormhole.
\emph{Gen. Relativ. Gravit.} \textbf{2004}, \emph{21}, 4811--4832. [\href{http://dx.doi.org/10.1088/0264-9381/21/21/005}{CrossRef}]

\bibitem{Lobo:2005us}
Lobo, F.S.N.
Phantom energy traversable wormholes.
\emph{Phys. Rev. D} \textbf{2005}, \emph{71}, 084011. [\href{http://dx.doi.org/10.1103/PhysRevD.71.084011}{CrossRef}]

\bibitem{Lobo:2005yv}
Lobo, F.S.N.
Stability of phantom wormholes.
\emph{Phys. Rev. D} \textbf{2005}, \emph{71}, 124022. [\href{http://dx.doi.org/10.1103/PhysRevD.71.124022}{CrossRef}]
%

\bibitem{schwarzschild1}
Schwarzschild, K. \emph{Über das Gravitationsfeld Einer Kugel aus Inkompressibler Flüssigkeit Nach der Einsteinschen Theorie};  Sitzungsberichte der Königlich-Preussischen Akademie der Wissenschaften Berlin: Berlin, Germany, 1916; pp. 424--434.

\bibitem{rosafluid}
Rosa, J.L.; Pi\c{c}arra, P.
Existence and stability of relativistic fluid spheres supported by thin-shells.
\emph{Phys. Rev. D} \textbf{2020}, \emph{102}, 6. [\href{http://dx.doi.org/10.1103/PhysRevD.102.064009}{CrossRef}]

\bibitem{rosafluid2}
Rosa, J.L. Observational properties of relativistic fluid spheres with thin accretion disks. \emph{arXiv }\textbf{2023}, arXiv:2302.11915.

\bibitem{oppenheimer1}
Oppenheimer, J.R.; Snyder, H. On Continued Gravitational Contraction. \emph{Phys. Rev.} \textbf{1939}, \emph{56}, 455. [\href{http://dx.doi.org/10.1103/PhysRev.56.455}{CrossRef}]

\bibitem{rosa112}
Rosa, J.L.; Carloni, S.
Junction conditions for general LRS spacetimes in the $1+1+2$ covariant formalism. \emph{arXiv} \textbf{2023}, arXiv:2303.12457.


\bibitem{senovilla1}
Senovilla, J.M.M.
Junction conditions for F(R)-gravity and their consequences.
\emph{Phys. Rev. D} \textbf{2013}, \emph{88}, 064015. [\href{http://dx.doi.org/10.1103/PhysRevD.88.064015}{CrossRef}]

\bibitem{Vignolo:2018eco}
Vignolo, S.; Cianci, R.; Carloni, S.
On the junction conditions in $f(R)$-gravity with torsion.
\emph{Class. Quantum Gravity} \mbox{\textbf{2018}, \emph{35}, 095014.} [\href{http://dx.doi.org/10.1088/1361-6382/aab6fe}{CrossRef}]

\bibitem{Reina:2015gxa}
Reina, B.; Senovilla, J.M.M.; Vera, R.
Junction conditions in quadratic gravity: Thin shells and double layers.
\emph{Class. Quantum Gravity} \textbf{2016}, \emph{33}, 105008. [\href{http://dx.doi.org/10.1088/0264-9381/33/10/105008}{CrossRef}]

\bibitem{Deruelle:2007pt}
Deruelle, N.; Sasaki, M.; Sendouda, Y.
Junction conditions in f(R) theories of gravity.
\emph{Prog. Theor. Exp. Phys.} \textbf{2008}, \emph{119}, 237--251. [\href{http://dx.doi.org/10.1143/PTP.119.237}{CrossRef}]

\bibitem{Olmo:2020fri}
Olmo, G.J.; Rubiera-Garcia, D.
Junction conditions in Palatini $f(R)$ gravity.
\emph{Class. Quantum Gravity} \textbf{2020}, \emph{37}, 215002. [\href{http://dx.doi.org/10.1088/1361-6382/abb924}{CrossRef}]
%

\bibitem{rosafrt}
Rosa, J.L.
Junction conditions and thin shells in perfect-fluid $f(R,T)$ gravity.
\emph{Phys. Rev. D} \textbf{2021}, \emph{103}, 104069. [\href{http://dx.doi.org/10.1103/PhysRevD.103.104069}{CrossRef}]

\bibitem{rosafrt2}
Rosa, J.L.; Rubiera-Garcia, D.
Junction conditions of Palatini f(R,T) gravity.
\emph{Phys. Rev. D} \textbf{2022}, \emph{106}, 064007. [\href{http://dx.doi.org/10.1103/PhysRevD.106.064007}{CrossRef}]

\bibitem{suffern}
Suffern, K.G. Singular hypersurfaces in the Brans-Dicke theory of gravity. \emph{J. Phys. A Math. Gen.} \textbf{1982} \emph{15}, 1599. [\href{http://dx.doi.org/10.1088/0305-4470/15/5/021}{CrossRef}]

\bibitem{Barrabes:1997kk}
Barrabes, C.; Bressange, G.F. Singular hypersurfaces in scalar---Tensor theories of gravity. \emph{Class. Quantum Gravity} \mbox{\textbf{1997}, \emph{14}, 805--824.} [\href{http://dx.doi.org/10.1088/0264-9381/14/3/021}{CrossRef}]

\bibitem{Padilla:2012ze}
Padilla, A.; Sivanesan, V.
Boundary Terms and Junction Conditions for Generalized Scalar-Tensor Theories.
\emph{J. High Energy Phys.}
\textbf{2012}, \emph{08}, 122. [\href{http://dx.doi.org/10.1007/JHEP08(2012)122}{CrossRef}]

\bibitem{Casado-Turrion:2023omz}
Casado-Turri\'on, A.; de la Cruz-Dombriz, \'A.; Jim\'enez-Cano, A.; Maldonado Torralba, F.J.
Junction conditions in bi-scalar Poincar\'e gauge gravity.
\emph{J. Cosmol. Astropart. Phys.} \textbf{2023}, \emph{07}, 023. [\href{http://dx.doi.org/10.1088/1475-7516/2023/07/023}{CrossRef}]

\bibitem{delaCruz-Dombriz:2014zaa}
de la Cruz-Dombriz, \'A.; Dunsby, P.K.S.; Saez-Gomez, D.
Junction conditions in extended Teleparallel gravities.
\emph{J. Cosmol. Astropart. Phys.} \textbf{2014}, \emph{12}, 048. [\href{http://dx.doi.org/10.1088/1475-7516/2014/12/048}{CrossRef}]

\bibitem{Arkuszewski:1975fz}
Arkuszewski, W.; Kopczynski, W.; Ponomarev, V.N.
Matching Conditions in the Einstein-Cartan Theory of Gravitation.
\emph{Commun. Math. Phys.} \textbf{1975}, \emph{45}, 183--190. [\href{http://dx.doi.org/10.1007/BF01629248}{CrossRef}]

\bibitem{amacias}
Macias, A.; Lammerzahl, C.; Pimentel, L.O.
Matching conditions in metric affine gravity.
\emph{Phys. Rev. D} \textbf{2002}, \emph{66}, 104013. [\href{http://dx.doi.org/10.1103/PhysRevD.66.104013}{CrossRef}]

\bibitem{Schwarzschild:1916uq}
Schwarzschild, K.
On the gravitational field of a mass point according to Einstein's theory.
\emph{Sitzungsber. Preuss. Akad. Wiss. Berlin Math. Phys.} \textbf{1916}, \emph{1916}, 189--196.



\bibitem{Hartle:1967he}
Hartle, J.B.
Slowly rotating relativistic stars. 1. Equations of structure.
\emph{Astrophys. J.} \textbf{1967}, \emph{150}, 1005--1029. [\href{http://dx.doi.org/10.1086/149400}{CrossRef}]

\bibitem{Hartle:1968si}
Hartle, J.B.; Thorne, K.S.
Slowly Rotating Relativistic Stars. II. Models for Neutron Stars and Supermassive Stars.
\emph{Astrophys. J.} \textbf{1968}, \emph{153}, 807. [\href{http://dx.doi.org/10.1086/149707}{CrossRef}]


\bibitem{Komar:1958wp}
Komar, A.
Covariant conservation laws in general relativity.
\emph{Phys. Rev.} \textbf{1959}, \emph{113}, 934--936. [\href{http://dx.doi.org/10.1103/PhysRev.113.934}{CrossRef}]

\bibitem{Wald:1984rg}
Wald, R.M.
\emph{General Relativity};
The University of Chicago Press: Chicago, IL, USA, 1984.


\bibitem{Arnowitt:1962hi}
Arnowitt, R.L.; Deser, S.; Misner, C.W.
The Dynamics of general relativity.
\emph{Gen. Relativ. Gravit.} \textbf{2008}, \emph{40}, 1997--2027. [\href{http://dx.doi.org/10.1007/s10714-008-0661-1}{CrossRef}]


\bibitem{Misner:1973prb}
Misner, C.W.; Thorne, K.S.; Wheeler, J.A.
\emph{Gravitation};
W. H. Freeman: New York, NY, USA, 1973;
ISBN 978--0-7167-0344-0; 978-0-691-17779-3.


\bibitem{Bondi:1962px}
Bondi, H.; van der Burg, M.G.J.; Metzner, A.W.K.
Gravitational waves in general relativity. 7. Waves from axisymmetric isolated systems.
\emph{Proc. R. Soc. Lond. A} \textbf{1962}, \emph{269}, 21--52.

\bibitem{Sachs:1962wk}
Sachs, R.K.
Gravitational waves in general relativity. 8. Waves in asymptotically flat space-times.
\emph{Proc. R. Soc. Lond. A }\mbox{\textbf{1962}, \emph{270}, 103--126.}




\end{thebibliography}
\end{document}